\documentclass{egpubl}
\usepackage{eurovis2026}
\usepackage{booktabs}
\usepackage{subcaption}
\usepackage{xstring}
\usepackage{float}
\usepackage{siunitx}

\usepackage{ccicons} 

\preprint

\CGFpreprintccby

\usepackage[T1]{fontenc}
\usepackage{dfadobe} 
\usepackage{booktabs}
\usepackage{array}
\usepackage{multirow}
\usepackage{sidecap}
\usepackage{booktabs}
\usepackage{tabularx}
\usepackage{multirow}

\usepackage{cite}  
\BibtexOrBiblatex
\electronicVersion
\PrintedOrElectronic
\ifpdf \usepackage[pdftex]{graphicx} \pdfcompresslevel=9
\else \usepackage[dvips]{graphicx} \fi

\usepackage{egweblnk}

\usepackage{cuted} 

\usepackage[table]{xcolor}
\definecolor{best}{RGB}{240,245,255} 
\newcommand{\inlinevis}[1]{%
  \IfEqCase{#1}{%
    {oms}{\def\iconpath{figs/encodings_cell/oms_cell.pdf}\def\iconname{OMS~}}%
    {oma}{\def\iconpath{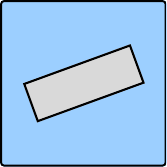}\def\iconname{OMA~}}%
    {b}{\def\iconpath{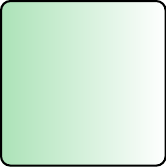}\def\iconname{BiC~}}%
    {bc}{\def\iconpath{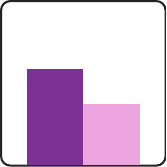}\def\iconname{Bar~}}%
    {alper_square}{\def\iconpath{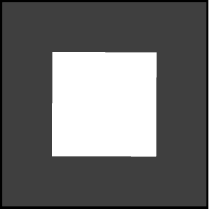}\def\iconname{}}%
    {alper_circle}{\def\iconpath{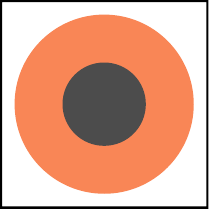}\def\iconname{}}%
    {alper_split}{\def\iconpath{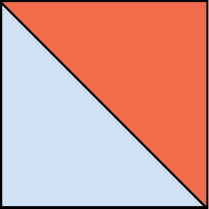}\def\iconname{}}%
    {alper_bars}{\def\iconpath{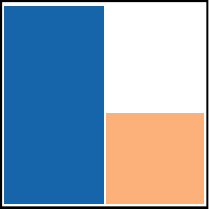}\def\iconname{}}%
  }[\def\iconpath{figs/encodings_cell/oms_cell.pdf}\def\iconname{OMS}]%
  {\sffamily{\iconname}}%
  \raisebox{-0.2\height}{\includegraphics[height=1em]{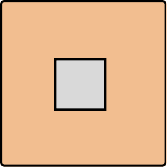}}%
}

\newcommand{\bic}{\protect\inlinevis{b}}
\newcommand{\oms}{\protect\inlinevis{oms}}
\newcommand{\oma}{\protect\inlinevis{oma}}
\newcommand{\barc}{\protect\inlinevis{bc}}

\newcommand{\inlinevisicon}[1]{%
  \IfEqCase{#1}{%
    {oms}{\def\iconpath{figs/encodings_cell/oms_cell.pdf}}%
    {oma}{\def\iconpath{figs/encodings_cell/oma_cell.pdf}}%
    {b}{\def\iconpath{figs/encodings_cell/b_cell.pdf}}%
    {bc}{\def\iconpath{figs/encodings_cell/bc_cell.pdf}}%
  }[\def\iconpath{figs/encodings_cell/oms_cell.pdf}]%
  \raisebox{-0.2\height}{\includegraphics[height=1em]{\iconpath}}%
}

\newcommand{\linkfont}{\sffamily\color{blue}}

\let\originalhref\href
\renewcommand{\href}[2]{\originalhref{#1}{{\linkfont#2}}}

\newif\ifshowappendix
\showappendixtrue

\newcommand{\papertitle}{Evaluating Encodings for Bivariate Edges in Adjacency Matrices}
\title[Evaluating Encodings for Bivariate Edges in AMs]%
      {\papertitle}

\author[J. Acosta-Hernández, A. Lex \& T. He]
{\parbox{\textwidth}{
\centering
J. Acosta-Hernández$^{1,2}$\orcid{0009-0003-4445-4119},
A. Lex$^{3,4}$\orcid{0000-0001-6930-5468}, and
T. He$^{3}$\orcid{0000-0002-9670-5587}
}
\\
{\parbox{\textwidth}{
\centering
$^1$Universidad Politécnica de Madrid, Spain \\
$^2$Center for Computational Simulation, Spain \\
$^3$Graz University of Technology, Austria \\
$^4$University of Utah, USA
\\
}}}

\DeclareRobustCommand{\gobblefour}[5]{}

\begin{document}




\maketitle

\begin{abstract}
    We present the first empirical evaluation of techniques for encoding distributions of quantitative edge values within adjacency matrices. In many real-world networks, edges represent not a single value but a set of measurements. While adjacency matrices preserve structural clarity, their compact cells limit the simultaneous display of multiple values. To address this, we explore edge encodings that represent distributions by two values: a measure of central tendency (mean, median, mode) and a measure of dispersion (standard deviation, variance, IQR). We select four possible encodings for evaluation that prior work has suggested are suitable for the limited space available in matrices: a bivariate color palette, embedded bar charts, and two overlaid-mark designs mapping the primary attribute to color and the secondary attribute to area or angle. In a preregistered crowdsourced study with 156 participants, we assessed performance of these encodings across eight analytical tasks and collected readability and aesthetic ratings. Results reveal clear performance regimes:
    area-based overlaid marks and bar charts achieved the highest overall performance;
    angle-based marks show moderate but less stable performance,
    and bivariate color consistently underperforms these alternatives.
    These findings clarify how visual channels behave under strict constraints and delineate the strengths and limitations of key design choices for multivariate edge visualization.
    \\

\ccsdesc[500]{Human-centered computing~Visualization techniques}
\ccsdesc[500]{Human-centered computing~Empirical studies in visualization}
\ccsdesc[300]{Human-centered computing~Visualization design and evaluation methods}

\printccsdesc   
\end{abstract}  


\section{Introduction}

Multivariate networks (MVNs)~\cite{kerren2014} arise across domains such as transportation logistics, neuroscience, or social networks, where edges often carry multiple quantitative descriptors~\cite{kaluza2010complex, bach2015small, rolls2021brain}. Including edge attributes in an analysis of a network provides critical context and enables meaningful analysis of complex relationships~\cite{kivela2014}. As in classical networks, MVNs can be sparse or dense depending on the domain and the generative process~\cite{krackhardt1987, bader2007temporal,barigozzi2010trade, matteo2013combinatorial, battinson2014structural }. 
When network density increases, adjacency matrices (AMs) become advantageous relative to node-link diagrams (NLs) because they avoid edge crossings and maintain structural clarity, unlike NLs where crossings and visual clutter quickly escalate~\cite{ghoniem2004comparison,alper2013weighted, nobre2019state}.
Contemporary examples of matrices used for dense attribute-rich networks include spatial genome data~\cite{kerpedjiev2018}, brain connectivity data~\cite{tamburro2024}, or mineral co-occurrence networks~\cite{que2024}.
In addition, AMs continue to be comparatively evaluated against alternative network representations and investigated from a perceptual perspective in recent visualization research~\cite{nobre2020evaluating,fuchs2024exploring,bae2025}.

Visualizing multiple edge attributes simultaneously is essential, as many patterns (e.g., correlations between interaction types, co-occurrences of extreme values, or attribute-driven clusters) emerge only when attributes are considered jointly~\cite{nobre2019state,mcgee2019state,kale2023}. Prior work has explored two main strategies for encoding multiple attributes in AMs: One common approach is to select a single representation of the underlying edge data (e.g., a mean, a specific attribute, or a derived network measure) and combine it with interaction techniques, including attribute switching, zooming, filtering, or focus+context operations to reveal more information~\cite{bach2014visualizing,bach2015small,safarli2019tamax,horak2020responsive, yang2022pattern}. While these interactions enhance analytical flexibility, they often hinder users from maintaining a cohesive overview and thus make it challenging to perceive patterns that emerge only when multiple attributes are considered together.
A second approach encodes all attributes directly within the matrix cells~\cite{alper2013weighted,nobre2020evaluating,safarli2019tamax,vogogias2020visual,filipov2024,fuchs2024exploring}. However, this approach is limited due to the relatively small size of matrix cells, which restrict how much information can be displayed simultaneously, making larger numbers of attributes challenging to represent.

\begin{figure*}[t]

\begin{subfigure}[b]{0.48\textwidth}
    \raggedright
    (a)\hfill\includegraphics[width=0.9\textwidth]{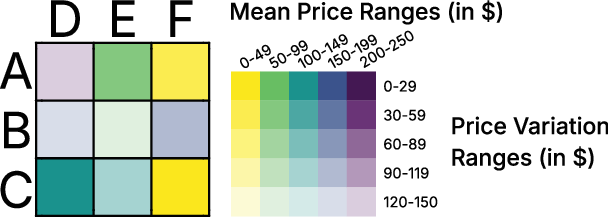}%
    \phantomcaption
    \label{fig:bic_legend}
\end{subfigure}
\hfill
\begin{subfigure}[b]{0.48\textwidth}
    \raggedright
    (b)\hfill\includegraphics[width=0.9\textwidth]{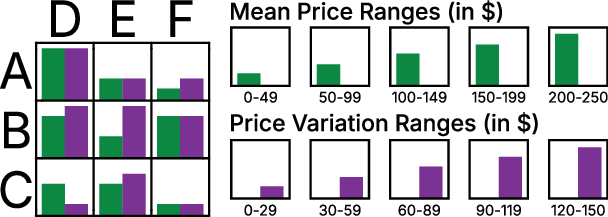}%
    \phantomcaption
    \label{fig:bar_legend}
\end{subfigure}

\par\vspace{1em}  

\begin{subfigure}[b]{0.48\textwidth}
    \raggedright
    (c)\hfill\includegraphics[width=0.9\textwidth]{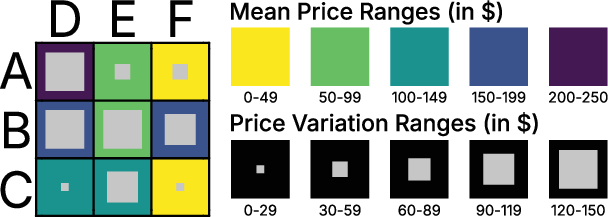}%
    \phantomcaption
    \label{fig:oms_legend}
\end{subfigure}
\hfill
\begin{subfigure}[b]{0.48\textwidth}
    \raggedright
    (d)\hfill\includegraphics[width=0.9\textwidth]{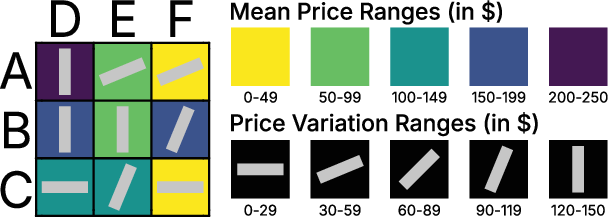}%
    \phantomcaption
    \label{fig:oma_legend}
\end{subfigure}

\caption{Four representative techniques illustrated using example 3×3 AMs with the same dataset. Each encoding represents two measures: central tendency (CT) and dispersion (D). (a) Bivariate color: hue encodes CT, lightness encodes D. (b) Embedded bar charts: the green bar encodes CT, the purple bar encodes D. (c) Overlaid mark size: hue encodes CT, the size of the overlaid square encodes D. (d) Overlaid mark angle: hue encodes CT, the angle of the overlaid bar encodes D.}
\label{fig:encodings}
\vspace{-6mm}
\end{figure*}

In this work, we therefore explore a middle ground. We summarize the distribution of an arbitrary number of edge attributes using two measures: one of central tendency (e.g., mean, median, mode) and one of dispersion (e.g., standard deviation, variance, range, IQR). This approach balances the need to capture information from multiple attributes with the practical constraints of limited cell space in AMs.

Yet, even the effectiveness of bivariate encoding in matrices remains largely untested, leaving designers without empirical guidance on when and how these encodings should be applied.
To address this gap, we present the first empirical evaluation of techniques for jointly encoding two quantitative edge attributes in AMs. In this study, we illustrate this setting by using the mean and standard deviation as representative attributes.

We first characterized techniques for mapping two quantitative edge attributes within AMs and selected four representative techniques for testing: (1) a bivariate color palette combining hue and lightness (\bic, \autoref{fig:bic_legend}), (2) embedded bar charts (\barc, \autoref{fig:bar_legend}), and two overlaid-mark designs that encode the secondary attribute using (3) area (\oms,  \autoref{fig:oms_legend}) or (4) angle (\oma,  \autoref{fig:oma_legend}). Although the first three techniques have been previously proposed and even compared against alternative layouts (e.g., AMs vs BioFabric both with embedded bar charts, AMs overlaid-mark size vs NL color and/or edge thickness)~\cite{alper2013weighted,nobre2020evaluating, fuchs2024exploring}, they have not yet been systematically compared with one another. The fourth design, \oma, is a novel variant that we derived from our design characterization as an alternative to \oms. It is intended to reduce interference between the overlaid mark and the perception of the attribute encoded in the matrix cell’s fill color~\cite{ware2010information}.

To evaluate these techniques, we conducted a crowdsourced study with 156 participants, using a network of flights and associated airline fares and eight representative analytic tasks covering structural-based, attribute-based, and estimation tasks (see \autoref{tab:tasks}) based on an established MVN task taxonomy~\cite{pretorius2014tasks}. We then compared accuracy, completion time, and perceived readability and aesthetics of these techniques.

In summary, this work contributes: (1) a systematic characterization of techniques for jointly encoding two edge attributes in AMs; (2) a controlled empirical evaluation comparing four representative encodings across a diverse set of analytical tasks; and (3) an analysis of objective performance and subjective preference that provides evidence-based guidance for designing multivariate AM visualizations.


\section{Related Work}
In this section, we first review foundational and recent MVN visualization approaches, then focus specifically on techniques for multivariate edge visualization, and finally discuss research on bivariate visualization.

\subsection{Multivariate Network Visualization (MNV)}

Multivariate network visualization (MNV) has been studied extensively, with research spanning layout algorithms (e.g., \cite{battista1998,guo2009,fuchs2025}),
interaction techniques (e.g., \cite{shneiderman2006, dunne2012, elzen2014}),
perceptual evaluations (e.g., \cite{alper2013weighted, nobre2020evaluating, fuchs2024exploring}),
and the design of representations capable of scaling to large, attribute-rich datasets (e.g., \cite{dunne2012,peysakhovich2015,domenico2014}). 

Node-link diagrams (NLs) and adjacency matrices (AMs) are the two most common techniques for network visualization. Classical works on network visualization have established their respective strengths and limitations.
Ghoniem et al.~\cite{ghoniem2004comparison} demonstrate that AMs outperform NLs for tasks involving dense graphs and edge-centric queries.
Okoe et al.~\cite{okoe2018, okoe2019} highlight tradeoffs between these two representations: AMs are suited to analyze neighborhoods and clusters but perform poorly when analyzing paths.
Beyond these two well known layouts (i.e., AM, NL) other works propose new approaches. For example, Bezerianos et al.~\cite{bezerianos2010} introduce Quilts specifically for large genealogies.
Longabaugh~\cite{longabaugh2012} introduces BioFabric for visualizing large biological networks by replacing nodes with horizontal lines and edges with vertical segments, producing clutter-free layouts.
Although neither layout is designed for multivariate data, the layout can be repurposed for such tasks, although with limitations in dense networks~\cite{nobre2019state}.
In summary, NLs provide intuitive depiction of topology, but AMs offer substantial advantages for dense networks, precise edge comparison, and attribute encoding, which aligns with the purposes of this work. Therefore, we focus on evaluating different edge encoding techniques in AMs rather than NLs or alternative techniques.

Recent surveys examine MVN from different angles.
McGee et al.~\cite{mcgee2019state} offer a perspective situating MVN visualization within the broader multilayer network (MLN) framework of Kivelä et al.~\cite{kivela2014}. They emphasize that MVNs should be conceptualized not merely as attribute-rich graphs, but as structurally heterogeneous systems whose layers, inter-layer relationships, and aspects should jointly shape visualization requirements. 
Kale et al.~\cite{kale2023} examine dynamic MVNs from a temporal perspective, showing that temporal evolution further amplifies scalability issues and that edge attributes continue to receive limited visual support. 
Nobre et al.~\cite{nobre2019state} offer a comprehensive taxonomy of layouts and operations for MVNs, noting that most existing techniques are based on node–link diagrams, despite their limited suitability for dense networks. They also observe that attribute encoding tends to focus on nodes rather than edges, even though matrix representations provide substantial expressive potential for edge attributes. Their taxonomy is informed by Pretorius et al.’s task analysis~\cite{pretorius2014tasks}, one of the most detailed frameworks for reasoning about MVN tasks. Together, these works motivate our focus on scalable encodings of edge attributes within AMs and inform the task set used in our evaluation.

\subsection{Multivariate Edge Visualization}

Researchers explore different representations of edge attributes in AMs.
Alper et al.~\cite{alper2013weighted} design on-edge encodings for two attributes and apply these techniques for weighted graph comparison.
Schöffel et al.~\cite{schoffel2016user} introduce on-edge bar charts within node–link diagrams, while Fuchs et al.~\cite{fuchs2024exploring} extend these concepts to AMs.
Following their previous work, Fuchs et al.~\cite{fuchs2025} investigate the effects of node and edge ordering on BioFabric, and show that edge ordering has a stronger influence on revealing structural patterns on this layout.
Vogogias et al.~\cite{vogogias2020visual} evaluate visual variables such as color and orientation for representing multiple edge types, and apply their insights to a Bayesian network analysis tool.

Some prior work investigates different techniques to visualize multiple edge attributes in AMs.
Early work explores summarization as a way to reduce cognitive load.
Bach et al.\ develop Small MultiPiles~\cite{bach2015small}, and introduce the piling metaphor to summarize dynamic networks. They later propose MatrixCube~\cite{bach2014visualizing}, which employs aggregation and 3D stacking to compactly represent evolving structures.
These tools demonstrate that structural and temporal patterns can be retained even when fine-grained detail is compressed. More recent research focuses on interaction as a complementary strategy.
Safarli and Lex~\cite{safarli2019tamax} introduce TaMax, a table–matrix technique that visualizes multiple node and edge attributes in dense graphs by juxtaposing a node-attribute table with an AM. 
Horak et al.~\cite{horak2020responsive} show how responsive matrix interfaces adapt to screen space and user intent, supporting scalable exploration when data density is high.
Building on this direction, Yang et al.~\cite{yang2022pattern} demonstrate that focus+context interactions substantially improve users’ ability to locate, compare, and contextualize information in large matrices.
However, no prior work systematically explores the design space of jointly encoding two edge attributes in AMs.

Researchers also conduct empirical evaluations to provide further guidance in designing effective MVNs visualizations. Alper et al.~\cite{alper2013weighted} compare NL and AM representations, showing the only empirical evaluation for encoding two edge attributes.
Nobre et al.~\cite{nobre2020evaluating} compare NL diagrams with on-node encodings against AMs with juxtaposed attribute tables, and demonstrate how study design, task selection, and crowdsourced evaluation can inform visualization methodology.
Filipov et al.~\cite{filipov2024} compare encodings and interaction techniques, including animation, in node–link and AMs for dynamic networks.
Fuchs et al.~\cite{fuchs2024exploring} evaluate AMs against BioFabric, and provide empirical evidence of the performance of each layout under different encodings.
These studies evaluate techniques across different layouts but do not compare different attribute-encoding techniques within AMs.

\subsection{Bivariate Visualization}

Bivariate visualization techniques aim to represent two data variables simultaneously in a single display, which have been studied by many cartographers. One of the most common approaches is the bivariate choropleth map, which uses color blending to encode two values. Trumbo~\cite{Trumbo:1981:Theory} introduced principles for constructing two-dimensional color legends for bivariate maps. Many bivariate color schemes have since been proposed and empirically evaluated (e.g., \cite{Robertson:1986:Generation, Teuling:2011:Bivariate, reimer:2011:squaring, karim2019, Strode:2020:Operationalizing}). Another technique is to overlay texture patterns on top of a colored map. Ware~\cite{Ware:2009:Quantitative} investigated the use of quantitative texture sequences layered on a color map. Ware~\cite{ware2010information} also notes that careful design is required to minimize perceptual interference when pattern and color encodings occupy the same space. Retchless and Brewer~\cite{retchless2016} evaluated eight different bivariate map designs and found that adding a patterned overlay to a color choropleth was the most preferred design by users. 

Researchers have also explored bivariate glyphs, where a single symbol encodes two data values. This idea dates back to Bertin \cite{Bertin:1998:SG, bertin:1983:semiology}, who proposed that certain visual channel combinations allow viewers to isolate each data dimension easily. Later studies further characterize the bivariate map symbol design (e.g., \cite{Elmer:2012:Symbol}) and empirically validate some variables are separated (e.g., hue and size \cite{Nelson:1999:Using, Nelson:2000:Designing}). In addition, there is growing interest in novel map techniques like bivariate cartograms (e.g., \cite{Nusrat:2018:Cartogram, Tran:2019:jointly}), where one variable is encoded by distorting the areas of regions and another variable is shown through the regions' color. Since our proposed design inherently maps two variables at once, we draw inspiration from the above bivariate visualization techniques in formulating our AM design.


\section{Design Characterization of Techniques for Encoding Two Edge Attributes Jointly in AMs}
\label{sec:characterization}

\begin{figure*}[t]

\centering
\begin{subfigure}[b]{0.12\textwidth}
  \raggedright
  (a)\hspace{0.3em}%
  \includegraphics[width=0.75\textwidth]{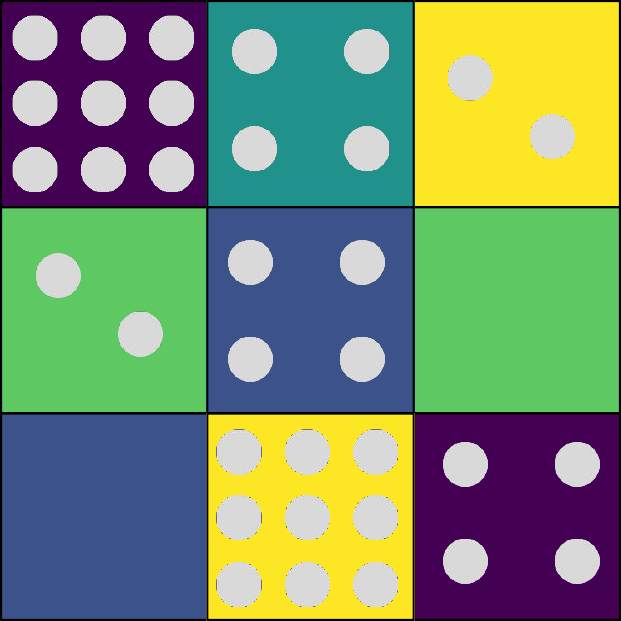}
  \phantomcaption
  \label{fig:example_hue_pattern}
\end{subfigure}
\hfill
\begin{subfigure}[b]{0.12\textwidth}
  \raggedright
  (b)\hspace{0.3em}%
  \includegraphics[width=0.75\textwidth]{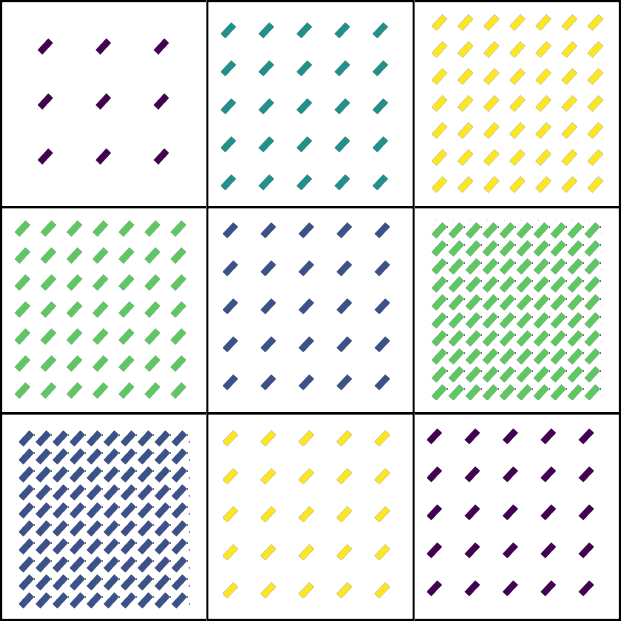}
  \phantomcaption
  \label{fig:example_orientation_hue_pattern}
\end{subfigure}
\hfill
\begin{subfigure}[b]{0.12\textwidth}
  \raggedright
  (c)\hspace{0.3em}%
  \includegraphics[width=0.75\textwidth]{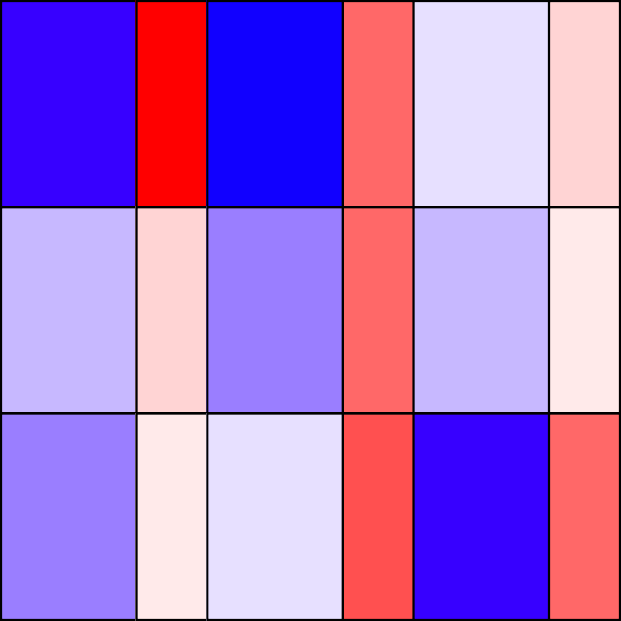}
  \phantomcaption
  \label{fig:example_asymm}
\end{subfigure}
\hfill
\begin{subfigure}[b]{0.12\textwidth}
  \raggedright
  (d)\hspace{0.3em}%
  \includegraphics[width=0.75\textwidth]{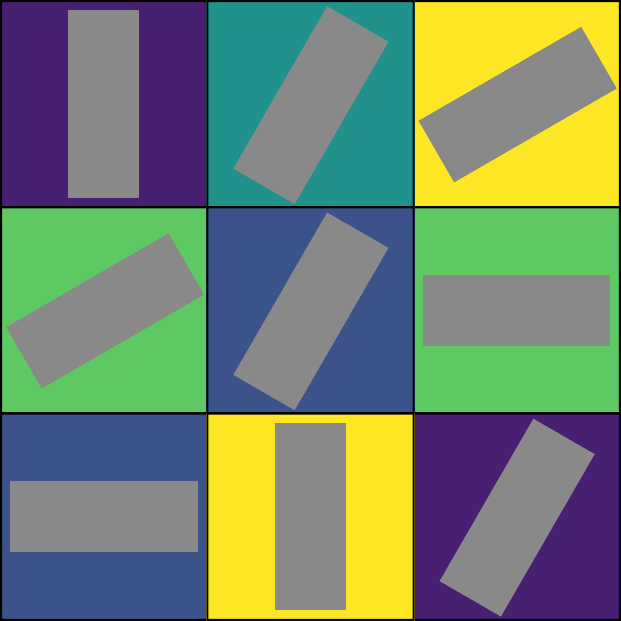}
  \phantomcaption
  \label{fig:example_oma}
\end{subfigure}
\hfill
\begin{subfigure}[b]{0.12\textwidth}
  \raggedright
  (e)\hspace{0.3em}%
  \includegraphics[width=0.75\textwidth]{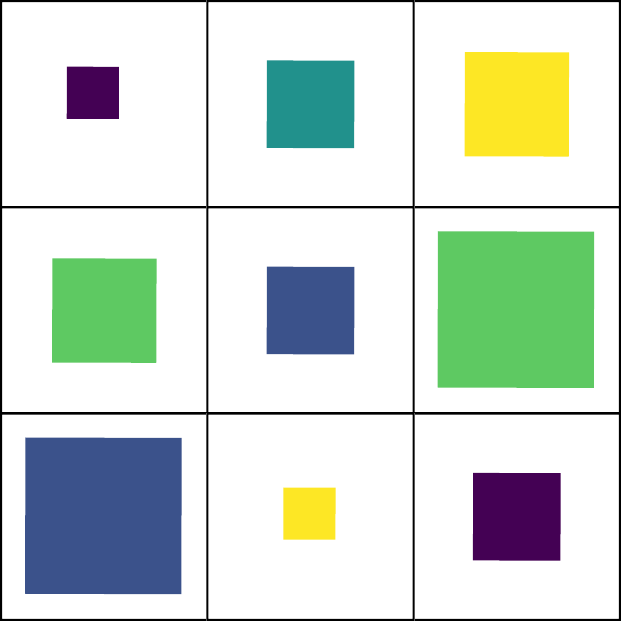}
  \phantomcaption
  \label{fig:example_hue_size}
\end{subfigure}
\hfill
\begin{subfigure}[b]{0.12\textwidth}
  \raggedright
  (f)\hspace{0.3em}%
  \includegraphics[width=0.75\textwidth]{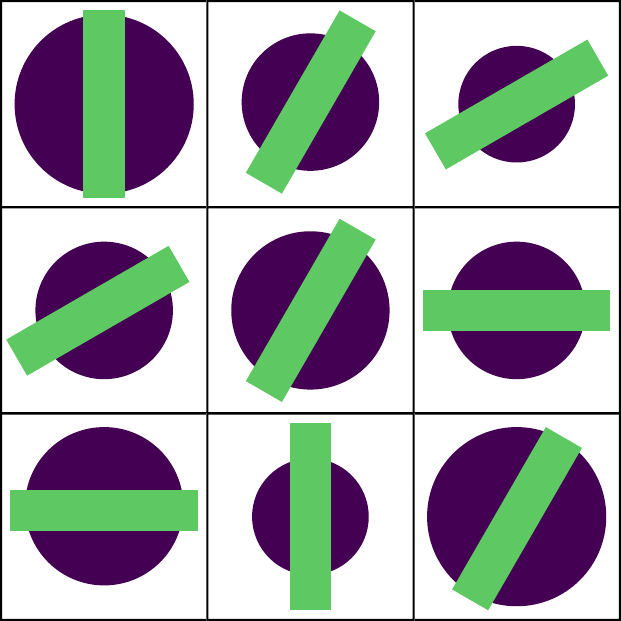}
  \phantomcaption
  \label{fig:example_size_ori}
\end{subfigure}
\hfill
\begin{subfigure}[b]{0.12\textwidth}
  \raggedright
  (g)\hspace{0.3em}%
  \includegraphics[width=0.75\textwidth]{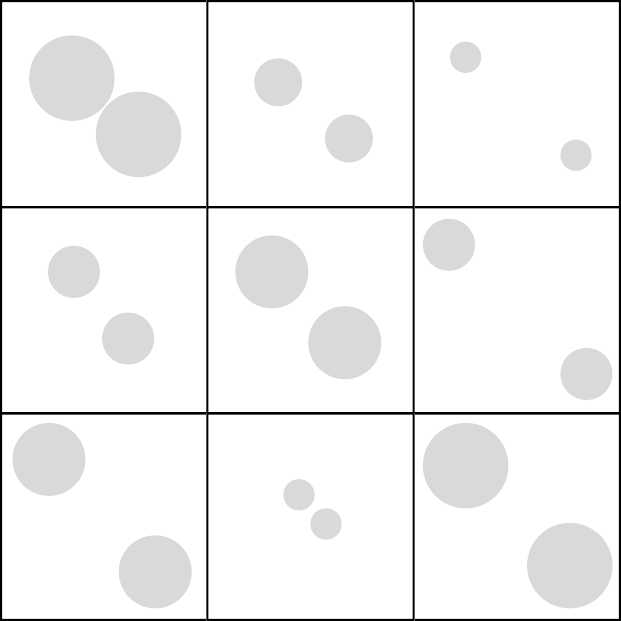}
  \phantomcaption
  \label{fig:example_dot_pattern}
\end{subfigure}

\caption{Examples of possible designs for jointly encoding two edge attributes in AM cells, beyond the four designs proposed by Alper et al.~\cite{alper2013weighted}. Designs (a)–(c) are discussed in \autoref{subsec:cell-only}, (d) and (e) in \autoref{subsec:one-mark}, and (f) and (g) in \autoref{subsubsec:two-marks}.}
\label{fig:example_encodings}

\end{figure*}

In this section, we characterize techniques for encoding two edge attributes per cell in AMs. To our knowledge, the only prior exploration of this problem is by Alper et al.~\cite{alper2013weighted}, who proposed four designs (\inlinevis{alper_split}, \inlinevis{alper_square}, \inlinevis{alper_circle}, and \inlinevis{alper_bars}). We extend their work by systematically examining the design space.

Encoding two edge attributes requires mapping two quantitative values to visual channels within each cell. We organize the design space based on which graphical elements serve as the carriers (marks) of these channels.
To support the selection and allow experimentation with key parameters (e.g., dataset, encoding, color scale), we developed a publicly accessible prototype \href{https://olmo.datsi.fi.upm.es/apache/~jacosta/prototypes/MVNV/}{olmo.datsi.fi.upm.es/apache/\textasciitilde jacosta/prototypes/MVNV/}.
We select representative techniques from each category to evaluate.

\subsection{The Cell as a Unique Mark}
\label{subsec:cell-only}

In this first category of techniques, the matrix cell itself is the only mark. The AM layout constrains several of the cell's visual channels: all cells share the same shape (typically squares) and size. As a consequence, these channels cannot vary at the cell level to encode edge attributes. The remaining degrees of freedom are channels related to the cell fill. We can encode both attributes directly by varying two visual channels of cells' fill, such as color (hue, saturation, lightness) or pattern parameters (e.g., \autoref{fig:example_hue_pattern} or \autoref{fig:example_orientation_hue_pattern}). 

In addition, we can split the cell into two regions, as proposed by Alper et al.~\cite{alper2013weighted} (\inlinevis{alper_split}). Splitting can be viewed as turning the cell into two submarks. However, because these subregions must collectively fill the cell, they inherit the same geometric constraints (fixed overall size, shape, and orientation), so we still characterize this technique within this category. Alper et al.~\cite{alper2013weighted} considered splitting the cell into two equal-area regions using three simple dividing boundaries (vertical, horizontal, or diagonal).
The two regions also need not have equal area~\cite{alper2013weighted} (e.g., \autoref{fig:example_asymm}). Moreover, one attribute can be assigned more area when it is conceptually the ``primary'' value or has more levels.

\paragraph*{Representative Technique Selection}
For this category, we did not select cell splitting because it introduces high-frequency boundaries that can create global artifacts and increase clutter in dense matrices~\cite{alper2013weighted}.
We therefore selected a bivariate color palette as the representative technique in our study  (\bic, \autoref{fig:bic_legend}). Such palettes are well-established in cartography (e.g., \cite{kaye2012,retchless2016}), yet remain underexplored for AMs. We map central tendency to hue and edge dispersion to color value, following the empirical results that support using lightness to convey uncertainty or variability~\cite{MacEachren:2012:Uncertainty,kaye2012,retchless2016}.

\subsection{Introducing One Additional Mark}
\label{subsec:one-mark}

In this category, each matrix cell contains one additional mark beyond the cell itself. Unlike the cell itself, the additional marks can take any shape, position and orientation within the cell, and all of their visual channels are, in principle, available for encoding. 

Two basic mapping strategies are possible.
First, one edge attribute can be mapped to a visual channel of the matrix cell (typically its fill), and the second attribute to a visual channel of the additional mark (e.g., \autoref{fig:example_oma}). Alper et al.'s inner--outer square design (\inlinevis{alper_square}) is an example of this strategy: it encodes the weight of one graph in the cell's outer region and the weight of the other graph in an overlaid inner square, mapping both weights to color brightness (lightness).
Second, both edge attributes can be mapped to different visual channels of the additional mark, with the cell background serving purely as a neutral container (e.g., \autoref{fig:example_hue_size}).

\paragraph*{Representative Technique Selection}

We adopt the inner–outer squares design by Alper et al.~\cite{alper2013weighted} for encoding two edge attributes in AMs (\oms, \autoref{fig:oms_legend}). In our adaptation, central tendency is mapped to cell hue and dispersion to the size of the overlaid mark.
Mapping dispersion to mark size provides a metric channel better aligned with magnitude judgments.
We further extend this design with a variant that maps edge dispersion to the angle of the overlaid mark (\oma, \autoref{fig:oma_legend}).
This choice keeps the overlaid mark area constant across dispersion values, potentially reducing interference with the perception of the cell fill \cite{ ware2010information, He:2026:Reframing}.
In addition, given that angle estimation has been shown to rely on an orthogonal internal reference frame~\cite{xu2018human}, the rectangular grids may reinforce orientation-based judgments.

\subsection{Introducing Two Additional Marks}
\label{subsubsec:two-marks}

In this category, each matrix cell contains two additional marks, with one edge attribute assigned to each mark. The cell itself is kept neutral and acts only as a container. Alper et al.'s study includes two examples of this category: embedded concentric circles (\inlinevis{alper_circle}) and embedded bar charts (\inlinevis{alper_bars}).
Given the flexibility of the additional marks, many other combinations are possible in terms of mark shape, position, and the visual channels used to encode edge attributes (e.g., \autoref{fig:example_size_ori} and \autoref{fig:example_dot_pattern}).

\paragraph*{Representative Technique Selection}
We selected embedded bar charts as the representative technique for this category (\barc, \autoref{fig:bar_legend}) because length is a highly accurate channel for magnitude and side-by-side bars offer good mark separability; bar-based designs are also well established in prior matrix and multi-scale views (e.g.,~\cite{elmqvist2008,horak2020responsive,fuchs2024exploring}), but have not been directly studied for encoding two edge attributes in AMs.


\section{Evaluation Study Design}

To evaluate the four representative encodings we selected, we conducted a controlled, online crowdsourced experiment with a between-subjects design to understand their effectiveness across core visual analytic tasks of MVNs, as well as their aesthetics and perceived readability. 
Our study was preregistered on OSF (\href{https://osf.io/vpmdk}{osf.io/vpmdk}) and received IRB approval from the
University of Utah (\textnumero\ IRB\_00191801). The study was implemented with reVISit~\cite{cutler2025revisit}.
All four conditions and study design can be seen at \href{https://jorgeacostaupm.github.io/revisit/}{jorgeacostaupm.github.io/revisit/}; the study code is available on \href{https://github.com/jorgeacostaupm/revisit}{github.com/jorgeacostaupm/revisit}.

\begin{figure*}[t]

\centering
\includegraphics[width=\textwidth]{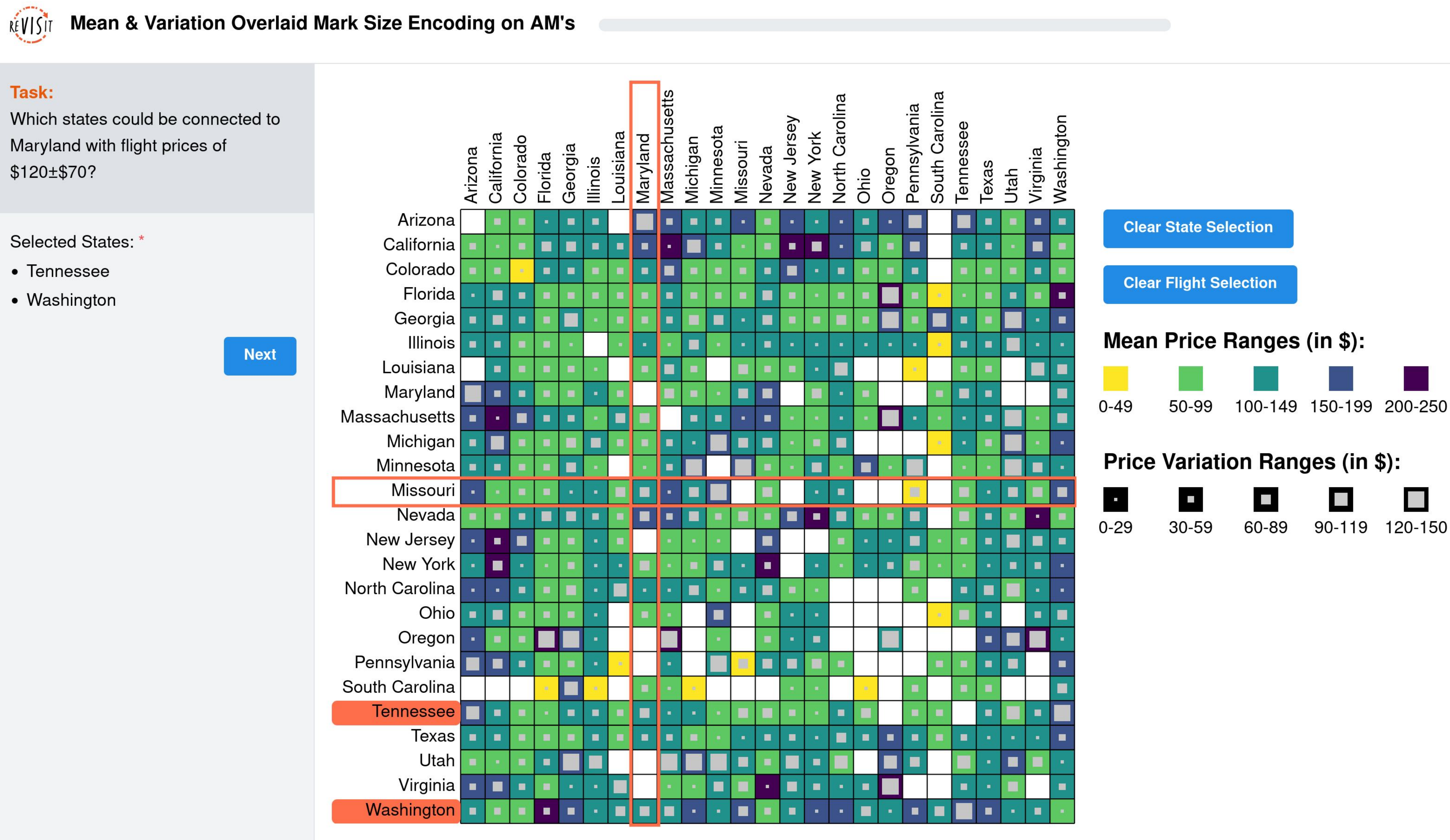}
\caption{Study interface for an example Adjacency by Attribute Combination task. The task prompt appears in the upper-left corner, with the selected nodes shown below it as the user’s responses. In this example, the participant used three interaction mechanisms, indicated by the orange elements: (1) the Maryland column is highlighted as a result of a previous click on its column label; (2) the Missouri row is highlighted on hover, facilitating inspection of node connectivity and selection; and (3) Tennessee and Washington have been selected as responses by clicking their corresponding row labels in the adjacency matrix.
}
\label{fig:task_example}

\end{figure*}

\subsection{Dataset}

For the study, we selected a dataset focused on domestic airline ticket fares using the U.S.
Bureau of Transportation Statistics DB1B Market dataset~\cite{bureauoftransportationstatistics_origin_2025}, which provides a 10\% sample of U.S.\ domestic airline ticket transactions. We selected this dataset based on the following three criteria: 
(1) Real-world data: While creating a synthetic dataset would have offered greater control over its characteristics, we prioritized the use of real-world datasets to enhance the ecological validity of our design and study. 
(2) Suitable network properties: We require a dense network that contains a sufficient number of link attributes, allowing their aggregation into mean and standard deviation, thus providing an MVN perspective.
(3) Familiar topic: this facilitates task comprehension and reduces cognitive load.

To prepare the data, we filtered incomplete records, excluded non-continental U.S. routes, and removed connections with fewer than 15 passengers to avoid distortions associated with bulk or contract fares.
We also removed outlier fares (i.e., standard deviation above \$150) to maintain interpretable linear scaling.
Because our study focuses on attribute encodings rather than directionality, origin–destination pairs were treated as undirected.
For each pair, we averaged fares across operators and computed the mean and standard deviation, resulting in a single quantitative attribute per edge.
Based on these aggregated values, we generated two networks: a smaller training network (12 nodes, density 0.55) and a larger study network (25 nodes, density 0.7). We detail our data processing process in \autoref{sec:data_proc}.

\subsection{Stimuli and Study Interface}

We designed the four stimuli based on the encoding strategies described in \autoref{sec:characterization}.
Across all conditions, we used the Viridis color scale, a perceptually uniform multihue colormap that is robust under common forms of color-vision deficiency.
Detailed information on color usage is provided in \autoref{sec:color-usage}.

Matrix ordering was fixed per task and identical across all four encodings. Non-cluster tasks used alphabetical row/column order. Cluster-identification tasks used reorder.js~\cite{fekete2015} with (1) Optimal Leaf Ordering (Euclidean distance, complete linkage) and (2) PCA-based ordering on either mean or standard deviation weights.
Details of the matrix-ordering procedures are given in \autoref{sec:matrix-ordering}.

We set the cell size to $23 \times 23$\,px to ensure legible labels, as label readability is a practical constraint in typical AMs use cases.

We implemented our study interface with the reVISit platform \cite{ding2023revisit, cutler2025revisit}.  \autoref{fig:task_example} illustrates a sample task interface, where the orange elements represent interaction feedback. To facilitate participants’ tasks and reduce fatigue, we implemented three interaction techniques: (1) row-column highlighting on cell hover; (2) column highlighting on top-node label click, which also enables row highlighting when hovering over the corresponding node label row; and (3) cell highlighting on individual cell click.
Depending on the task type, participants responded either by clicking on row node labels to select a node as their answer or by selecting a radio button.

\subsection{Tasks}
\label{sec:task}
We selected eight task types from a multivariate network (MVN) analysis taxonomy~\cite{pretorius2014tasks}, which emphasizes integrating topology and attribute information. The taxonomy defines four categories: Structure-Based, Attribute-Based, Estimation, and Browsing tasks; browsing tasks were excluded after the first round pilot, because their open-ended nature makes them unsuitable targets for controlled evaluation. \autoref{tab:tasks} shows the eight task types (T1–T8) we selected with example task prompts. See \autoref{sec:tasks-screenshots} for full task and stimuli descriptions.
 
T1 is a basic structural adjacency task (illustrated in \autoref{fig:task_example}). T2–T5 involve attribute-based analyses such as central tendency, dispersion, attribute combinations, and extreme-value identification. T6–T8 are high-level estimation tasks assessing global network properties; two of which focus on edge attributes. We include structure-only tasks to ensure that the edge encodings do not interfere with such tasks.

T1 was shown once due to its simplicity, while T2–T8 were repeated with different instances to ensure reliable performance estimates. Node-selection tasks (T1–T6) were scored using F1-scores, and multiple-choice tasks (T7–T8) using accuracy; scores were averaged across repetitions. Task instructions were refined via pilot studies for clarity, and full descriptions with visual examples are provided in the supplementary material.

\begin{table*}[hbt]

  \begin{tabular}{l l l p{9.5cm}}
  
        \toprule
        \textbf{ID} & \textbf{Category} & \textbf{Task Name} & \textbf{Example Task Prompt} \\
        \midrule
        \textbf{T1} & Structural & Structural Adjacency & Which states are connected to Oregon? \\
        \midrule
        \textbf{T2} & Attribute & Adjacency by Central Tendency & Which states are connected to Nevada with a mean price higher than \$150? \\
        \textbf{T3} & Attribute & Adjacency by Dispersion & Which states are connected to Missouri with a price variation higher than \$90? \\
        \textbf{T4} & Attribute & Adjacency by Attribute Combination & Which states are connected to Washington with flight prices of \$150±\$50? \\
        \textbf{T5} & Attribute & Extreme Detection & When flying from Utah, which destination states have the highest and lowest ticket prices? \\
        \midrule
        \textbf{T6} & Estimation & Classification by Dispersion & Assign each state to a category. \\
        \textbf{T7} & Estimation & Cluster Central Tendency Estimation & Which cluster has the highest mean price? \\
        \textbf{T8} & Estimation & Cluster Dispersion Estimation & Which cluster shows the highest price variation? \\
        \bottomrule
    
    \end{tabular}
    \caption{List of tasks grouped by category, presenting each task’s high-level objective,  and instructions as provided to participants. These tasks are based on Pretorius et al.~\cite{pretorius2014tasks} task taxonomy for MVNs. See the supplementary material for additional details.}
    \label{tab:tasks}
    
\end{table*}

In addition to analytical tasks, we also assessed subjective experiences using two validated instruments: PREVis~\cite{cabouat2024previs} for readability—capturing layout, understanding, data readability, and feature readability—and BeauVis~\cite{he2022beauvis} for aesthetic pleasure.

\subsection{Research Questions and Hypotheses}

Our primary objective is to assess the effect of encoding decisions on task performance (accuracy and completion) and subjective experience (perceived readability and aesthetics). Below we list our research questions and associated hypotheses.

\paragraph*{RQ1: How do the selected edge encoding techniques compare in task performance on common MVN tasks?}

\begin{itemize}

    \item 
    \textbf{H1.1}: Performance in structural adjacency tasks will be invariant across encoding conditions. \emph{Rationale:} Structural adjacency is determined solely by binary cell occupancy (i.e., which grid locations are filled); because all conditions preserve cell occupancy, the resulting adjacency structure is identical, and performance should not differ across conditions.
    
    \item
    \textbf{H1.2}: Overlaid mark encodings (i.e., \oms, \oma) will yield higher accuracy and faster completion times in attribute and estimation tasks than both \bic and \barc. \emph{Rationale:} A single additional mark can separate attributes while preserving a coherent base, reducing visual complexity and supporting magnitude comparison. In contrast, cell-only designs force both attributes into the same fill, increasing perceptual coupling, and dual-mark designs fragment the cell into competing visual elements~\cite{ware2010information}.
    
    \item
    \textbf{H1.3}: Within Overlaid mark encodings (i.e., \oms, \oma), a size-based mark will yield higher accuracy and faster completion times than an angle-based mark. \emph{Rationale:} Although the matrix grid provides external reference axes that may facilitate angle interpretation~\cite{xu2018human}, classical perceptual theory ranks size as a more precise channel than angle for quantitative magnitude estimation~\cite{ware2010information}.

\end{itemize}

\paragraph*{RQ2: How do the selected edge encoding techniques compare in perceived readability under small cell-size constraints?}

\begin{itemize}
    \item
    \textbf{H2}: Participants will rate \oms as the most readable encoding among all encoding types. \emph{Rationale:} A size-based mark is typically more interpretable than an angle-based mark~\cite{ware2010information}. Adding a single mark can improve separability without the clutter and fragmentation introduced by multiple marks or the reduced separability of cell-only encodings~\cite{Pelli2008,cleveland1985}.
\end{itemize}

\paragraph*{RQ3: How do the selected edge encoding techniques compare in perceived aesthetics under small cell-size constraints?}

\begin{itemize}
    \item
    \textbf{H3}: Encodings with fewer within-cell marks will receive higher aesthetic ratings than encodings introducing additional marks. \emph{Rationale:} Prior work in empirical aesthetics and visualization suggests that visual simplicity and reduced internal complexity are positively associated with aesthetic pleasure~\cite{he2022beauvis}.
\end{itemize}

\subsection{Power Analysis}

We conducted an a priori power analysis using G*Power~\cite{faul2007g}, targeting 0.8 power to detect medium effect sizes at the standard 0.05 alpha level. 
The analysis indicated that 180 participants would be required. These estimates are based on expected medium-sized effects.

\subsection{Participants}

We recruited a total of 184 participants through Prolific. Eligibility criteria required participants to be fluent English speakers and at least 18 years old. Participants were recruited through Prolific and compensated \$7 for an estimated 35 minutes of participation at an  hourly rate of \$12. 

In accordance with our preregistered exclusion criteria, we excluded 15 participants who failed the attention check. In addition, 13 participants were removed because they repeated the study due to a Prolific misconfiguration. The final sample thus consisted of 156 participants (age: $M = 35.24$, $SD = 11.33$; gender: 71 female, 78 male, 4 not disclosed; education: 28 High School, 81 Bachelor's, 38 Master's, 5 PhD, 2 Other, 2 Missing).
For additional demographic information (e.g., AM familiarity, country of origin, browser used, age distribution), please refer to \autoref{sec:demo}.

\subsection{Study Procedure}

We employed a between-subjects design to mitigate learning effects and participant fatigue. 
The eligibility criteria required participants to use a desktop or laptop device with a minimum display resolution of 1400 × 800 pixels. All participants were presented with the same AM layout. After giving their informed consent, participants went through four phases:

\paragraph*{Training} Participants received explanations of the AM visualization and its encoding. This included a guided walkthrough of the available interaction techniques—highlighting columns and cells—to facilitate task completion. A practice block of ten simple tasks ensured a baseline understanding~\cite{nobre2020evaluating}. Participants who failed to complete this block within three attempts were excluded from the study for failing to demonstrate comprehension of the core concepts.

\paragraph*{Analytic Tasks} Participants completed 15 trials (eight task types, randomized order, with repeats as described in \autoref{sec:task}). Responses were collected via direct node selection or multiple-choice options. An attention check ensured engagement and data quality.

\paragraph*{Subjective Ratings} Participants rated perceived readability and aesthetic appeal using validated instruments PREVis \cite{cabouat2024previs} and BeauVis \cite{he2022beauvis} after completing all analytical tasks.

\paragraph*{Demographics and User Feedback} Participants answered questions about their educational background and prior exposure to network visualizations. At the end of the study, participants had the option to provide qualitative feedback in the form of open-ended written responses.

\subsection{Pilot studies}

We conducted two pilot rounds to refine the study design and estimate task duration. The first, run in-lab with 8 participants (2 per encoding), lasted about 60 minutes and led to clarifying instructions, balancing task difficulty, removing browsing tasks, and verifying that the implementation was functioning correctly and free of bugs. These data were not included in the final analysis. A second pilot on Prolific with 12 participants (3 per encoding), averaging around 35 minutes, confirmed the clarity of the refined instructions and examples and the estimation of the study duration. Because this round matched the final procedure, we included its data in the formal analysis.


\section{Results}

We recruited 184 participants from Prolific for this study in total. A quarter of the participants were assigned to each encoding technique. A total of 156 participants were included in the analysis (\bic\ 42, \barc\ 42, \oms\ 37, \oma\ 35). All study data, results, and analysis scripts are available at \href{https://osf.io/2nrbc/}{osf.io/2nrbc}.

For each task we display F1-Score (T1-T6) or Accuracy (T7 and T8), for readability and aesthetics we report scores for each subscale, all in the form of violin plots, which approximate the density distribution of the selected measure. We superimpose the mean value with a 95\% confidence interval error bar to facilitate comparison. We applied the Kruskal–Wallis test due to the non-normal distribution of our results. Therefore, we report p-values and epsilon squared. For pairwise effect sizes we report Cohen's d due to simplicity and interpretability.
Completion time did not differ significantly across techniques; therefore, we do not report them here.
We summarize the results in two tables in the supplementary material: task performance (see \autoref{tab:task-results}) and perceived readability together with aesthetics (see \autoref{tab:perceived-results}).

\begin{figure*}[tb]

    \centering
    \includegraphics[width=\textwidth]{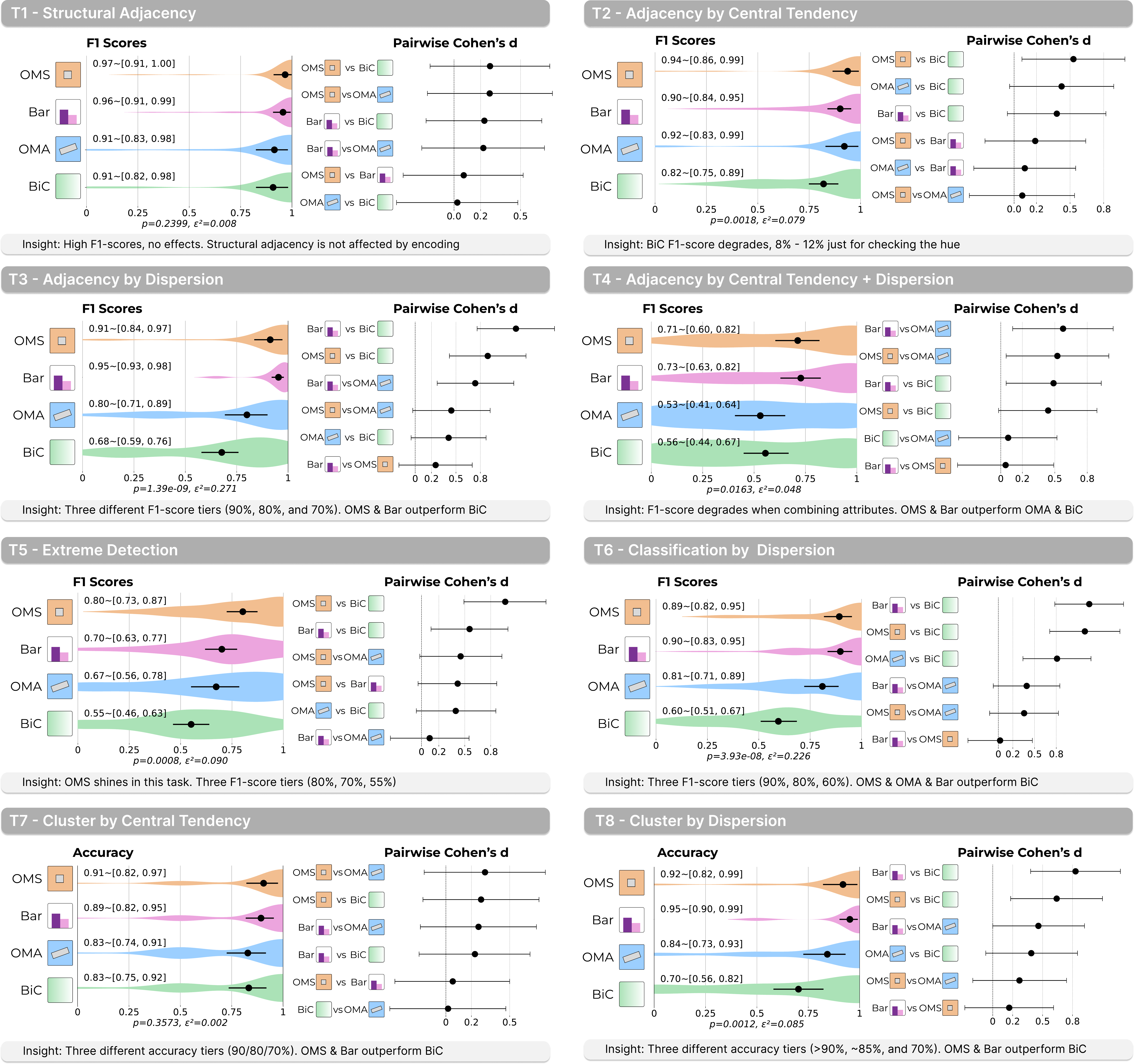}
     \caption{For each of the eight tasks (T1–T8), the corresponding performance metrics—F1-score or accuracy (higher is better), depending on the task—are shown using violin plots to illustrate the distribution of participant performance across conditions. The accompanying forest plots summarize the pairwise effects among the four encoding techniques (\protect\oms, \protect\barc, \protect\oma, \protect\bic), including effect size estimates (Cohen's d) and 95\%-confidence intervals. Together, these visualizations provide both distribution-level and comparative insights into how each encoding supports performance across the different task types. Additional details can be found in the supplementary material.}
    \label{fig:tasks-results}    
    
\end{figure*}

\subsection{Performance}

\begin{figure*}[t]

    \centering
    \includegraphics[width=\textwidth]{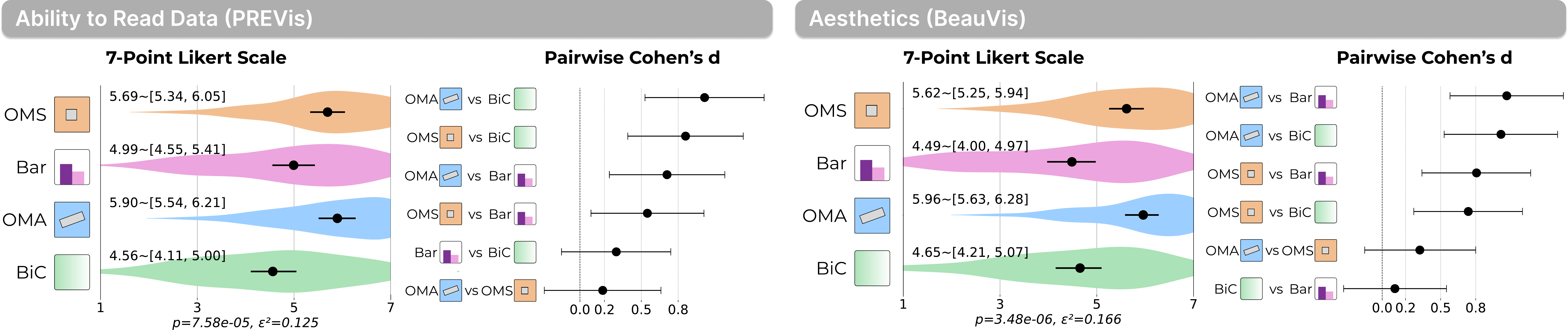}
    \caption{Comparative evaluation of perceived data reading ability (PREVis DataRead subscale~\cite{cabouat2024previs}) and perceived aesthetics (BeauVis~\cite{he2022beauvis}) across the four tested encoding techniques. Results for the remaining PREVis subscales are reported in the supplementary material.}
    \label{fig:results_reda_aes}   
    
\end{figure*}

For the structural adjacency task (T1), no significant differences were observed across encodings ($p = .24$, $\epsilon^2 = .01$).
All conditions achieved high accuracy ($\geq 91\%$). Therefore, \textbf{H1.1 is supported.}

Significant omnibus effects were observed in all attribute-based tasks (T2–T5) and in the dispersion-focused estimation tasks (T6, T8).
\oms\ and \barc\ tie as the top performers, each achieving the highest mean accuracy in 4/8 tasks and performing comparably (i.e., with less than a 4 percentage point difference) on all tasks except T5.
\oma\ performed a step below in most tasks except for T1 and T2 and showed significant differences against top performers in T4.
\bic\ consistently yielded the lowest mean accuracy and showed medium-high differences against top performers in T2, T3, T5, T6 and T8.
However, the expected performance hierarchy among the remaining techniques did not emerge (see \autoref{fig:tasks-results}).
Consequently, while single-mark encodings clearly outperform the cell-only approach in attribute-based tasks, they do not demonstrate a systematic advantage over the dual-mark design.
Therefore, \textbf{H1.2 is not supported.}

Across all attribute and estimation tasks (T2--T8), \oms\ outperformed \oma.
Significant pairwise difference was observed in T4  (see \autoref{fig:tasks-results}).
However, no significant differences in completion time were observed between these techniques for any task.
Therefore, \textbf{H1.3 is partially supported}: the size-based encoding outperformed the angle-based encoding in accuracy, but not in time.

\subsection{Perceived Readability}

We next report perceived readability using PREVis by subscale (i.e., \emph{Data Readability}, \emph{Understanding}, \emph{Layout}, and \emph{Feature Readability}) as recommended by Cabouat et al.~\cite{cabouat2024previs}.
Encoding condition had a significant effect on all four subscales (Kruskal--Wallis, all $p \le .01$, $\epsilon^2 = 0.06$--$0.13$).
\autoref{fig:results_reda_aes} reports the Data Readability subscale, which showed the largest effect ($\epsilon^2=0.13$); the remaining subscales and post-hoc statistics are provided in the supplementary material (see \autoref{tab:perceived-results}, \autoref{fig:readability_app}--\autoref{fig:features_app}).
\oma\ received the highest ratings, closely followed by \oms; however, the difference between them was not statistically significant. In contrast, \bic\ and \barc\ were consistently rated lower, with significant differences relative to the overlaid-mark encodings.
Therefore, \textbf{H2 is not supported}.

\subsection{Aesthetics}

Finally, we report aesthetic pleasure using the BeauVis scale; following He et al.~\cite{he2022beauvis}, we compute a single BeauVis score as the mean of its five items (i.e., enjoyable, likable, pleasing, nice, appealing).
Significant differences in aggregated aesthetic ratings were observed across encoding conditions (Kruskal--Wallis, $p < .001$, $\epsilon^2 = .17$).
Contrary to expectations, \oms\ achieved the highest mean aesthetic score, followed closely by \oma, whereas \barc\ and \bic\ received lower ratings and showed high significant differences against the first two as can be seen in \autoref{fig:results_reda_aes}.
Therefore, \textbf{H3 is not supported.}


\section{Discussion}

Our work investigated how to encode two quantitative edge attributes per cell in medium-sized dense AMs.
Based on this characterization, we selected four encodings for evaluation: \bic\ (cell-only), \oms\ and \oma\ (single-mark), and \barc\ (dual-mark).

Across eight MVN tasks grounded in a well-established task taxonomy~\cite{pretorius2014tasks}, we observe three consistent patterns. First, \bic\ produces the lowest performance in most tasks.
Second, \oms\ and \oma\ encodings, particularly the area-based mark encoding, achieve robust accuracy while also receiving strong subjective ratings.
Third, \barc remains unexpectedly competitive despite adding multiple within-cell elements.
Finally, completion time does not differ significantly across encodings, indicating that technique choice primarily affects correctness rather than speed within the tested spatial regime.

\paragraph*{RQ1}
Our results support \textbf{H1.1}: the choice of bivariate encoding did not affect basic adjacency perception, as participants detected the presence of a relation with comparable accuracy across conditions.

However, \textbf{H1.2} was not supported—the anticipated performance ranking based on mark multiplicity did not emerge.
Despite introducing additional graphical elements, \barc\ performed on par with \oms\ for most attribute-driven tasks and surpassed \oma\ in the majority of cases. In contrast, the cell-only \bic\ design consistently underperformed, aligning with our prior observations.

Finally, \textbf{H1.3} was partially supported: \oms\ was more accurate than \oma\ across attribute-reading and estimation tasks, with significant differences in several tasks, while completion times remained comparable.

These findings indicate that mark multiplicity alone does not determine analytical effectiveness.  
Instead, performance appears to depend on both perceptual separability and the precision of the visual channel used for magnitude estimation \cite{ware2010information}.
\barc\ preserves strong separability and relies on length, a visual variable known to support accurate quantitative comparison \cite{cleveland1985}.
\oms\ also benefits from spatial separation and uses area, which provides reasonably precise magnitude judgments. In contrast, \oma, despite spatial separation, encodes dispersion through angle,  
a channel associated with lower quantitative precision.
\bic\, although visually simple, couples two quantitative variables within a single color space,  
reducing separability and increasing perceptual interference.

\paragraph*{RQ2}
Participants’ readability ratings do not support \textbf{H2}, as \oms\ was not rated significantly higher than \oma, although both were consistently rated higher than \bic\ and \barc.
A key takeaway is the dissociation between perceived readability and task performance.
Although single-mark encodings are considered more readable, this does not systematically translate into higher accuracy, a disconnect also reported by Cabouat et al.~\cite{cabouat2024previs}, who caution against using performance as a direct proxy for readability.
In our study, \barc\ illustrates this gap: despite being criticized for within-cell visual complexity, it often achieved higher accuracy than \oma, reinforcing the idea that subjective impressions of clarity or clutter do not reliably predict performance.

\paragraph*{RQ3}
Aesthetic pleasure ratings do not support \textbf{H3}; the cell-only design is not preferred aesthetically. Instead, the single-mark overlays receive the highest aesthetic ratings.
From a design perspective, aesthetic preference does not appear to be driven by minimizing the number of marks, but rather by perceived visual order and clarity of mapping, and is therefore directly correlated with perceived readability.

\paragraph*{Spatial Regime}
As cell sizes decrease, encodings relying on geometric magnitude judgments can degrade due to pixel quantization, reduced spatial resolution, and visual crowding~\cite{Pelli2008,cleveland1985}.
We focus on cell sizes near the lower bound of label legibility; differences may become most pronounced only as cells approach a few pixels, where labels are no longer readable.

\paragraph*{Multiattribute Reasoning}
A clear performance bottleneck emerges in tasks that require integrating multiple attribute cues rather than reading a single attribute in isolation.
In our setting tasks T4 and T5 yield the lowest peak accuracies, with the best-performing encodings reaching only 73\% and 80\%, respectively (see \autoref{tab:task-results}).

Taken together, these results suggest that even with two encoded attributes, tasks that require multiattribute reasoning can be more error-prone in AMs size-constrained cells.
This pattern motivates the hypothesis that encoding higher-dimensional attribute sets directly within AM cells may further amplify errors unless part of the integration is externalized—e.g., via interaction~\cite{safarli2019tamax,horak2020responsive, yang2022pattern} and/or alternative layouts~\cite{fuchs2024exploring, fuchs2025} that reduce within-cell cognitive integration demands.
Importantly, this interpretation is conditioned on our task framing (e.g., the specific conjunctive constraints in T4 and neighborhood-based extremes in T5), the evaluation protocol, and the limitations of our sample and stimuli.

\paragraph*{Design Implications}
Our results translate into the following practical guidelines for encoding two quantitative edge attributes in AM cells under tight spatial constraints:

\begin{itemize}

\item
Use \oms\ as the default choice.
It offers the most reliable trade-off between accuracy and user acceptance across tasks, and is particularly effective for spotting extremes/outliers, likely due to efficient preattentive scanning with a hue and a size cue.

\item
Consider \barc\ as a strong alternative. Despite its higher within-cell complexity, it performs competitively; however, participants often perceived it as less readable, suggesting a risk of increased cognitive load.

\item
Reserve \oma\ for cases where only coarse differences are needed. It achieves acceptable performance overall and was rated as the most readable and appealing technique, making it a reasonable choice when perceived clarity is prioritized over fine-grained accuracy.

\item
Avoid \bic\ for quantitative comparisons.
Although this encoding is visually compact, mapping two quantitative variables into a single color space reduces perceptual separability, which makes magnitude- and dispersion-driven comparisons less reliable in small cells.
 
\end{itemize}




\section{Limitations and Future Work}

Our findings are bounded to the experimental regime we evaluated.
First, we tested dense AMs of moderate size and a fixed cell resolution; performance may differ for larger/sparser networks or smaller cells.
Second, we relied on a single real-world dataset and simplified the data to an undirected network; other domains, distributions, and directed networks may yield different trade-offs.
Third, we did not separately optimize each encoding for optimal parameters; better per-encoding optimization could change performance.
Fourth, we focus on two quantitative edge attributes summarized as central tendency and dispersion (i.e., mean/std); results may not transfer to other summary pairs (e.g., median/IQR), more than two attributes, or non-quantitative attributes.
Finally, our task set is representative but not exhaustive; additional analytical tasks could reveal different results.

Future work should extend this benchmark across matrix sizes/densities and cell resolutions, include directed networks and additional datasets/domains, test alternative distribution summaries (e.g., median/IQR) and more-than-bivariate attributes, systematically tune each encoding, and broaden the task set.


\section{Conclusions}

We presented the first systematic evaluation of techniques for encoding two quantitative edge attributes in AMs. Our results show that bivariate color scales perform consistently worse than alternatives under the spatial constraints of matrix cells. Overlaid mark designs, particularly area-based marks, provide the highest overall accuracy across attribute- and estimation-based tasks. Embedded bar charts—despite their visual complexity—offer competitive performance, highlighting the strength of familiar visual channels such as length. Angle-based marks, while perceived as more readable and aesthetically pleasing, do not translate into superior analytical performance. Together, these findings provide empirical guidance for choosing effective dual-attribute encodings in dense multivariate networks.


\section*{Acknowledgments}

This work was supported by the \textit{Programa Propio de Investigación, Innovación y Doctorado of the Universidad Politécnica de Madrid (UPM)} and the National Science Foundation (2213756).


\bibliographystyle{eg-alpha-doi} 
\bibliography{abbreviations,bibtex-output}

\ifshowappendix

\appendix 

\clearpage

\begin{strip} 
\noindent\begin{minipage}{\textwidth}
\makeatletter
\centering%
\sffamily\bfseries\fontsize{15}{16.5}\selectfont
\papertitle \\[.5em]
\large Supplementary Material \\[1.5em]
\makeatother
\normalfont\rmfamily\normalsize\noindent\raggedright In this supplementary material, we provide additional figures that informed our designs, the complete demographic data of our participant pool, and an overview of the tasks we evaluated. We include this material here because it extends beyond what could be presented in the main paper, either due to space constraints or because it was not essential for explaining our work.
\end{minipage}
\end{strip}

\section{Stimuli}
\label{sec:stimuli-images}

In this section, we present the stimuli used in our evaluation study.

\begin{center}
    \begin{minipage}{\textwidth}
        \centering
    
        \begin{minipage}{0.43\textwidth}
            \centering
            \includegraphics[width=\linewidth]{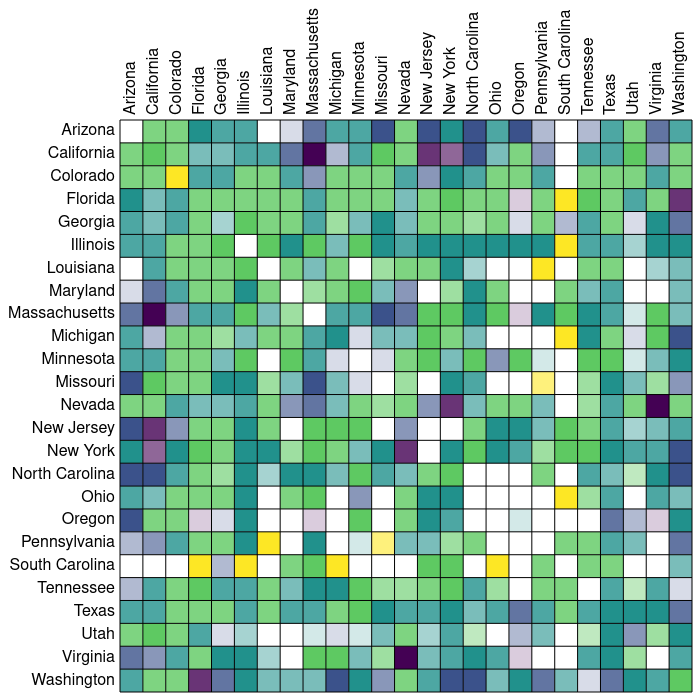}\\
            (a)
        \end{minipage}%
        \hfill
        \begin{minipage}{0.43\textwidth}
            \centering
            \includegraphics[width=\linewidth]{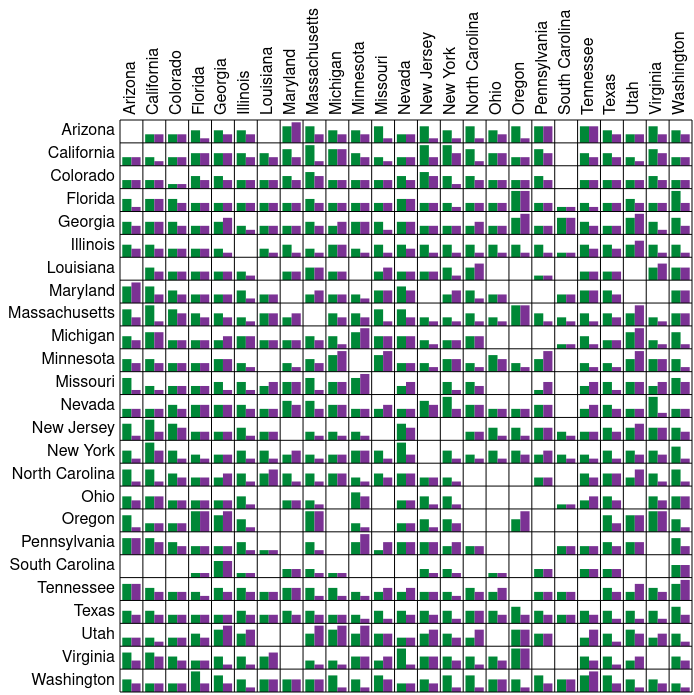}\\
            (b)
        \end{minipage}
    
        \vspace{0.5em}
    
        \begin{minipage}{0.43\textwidth}
            \centering
            \includegraphics[width=\linewidth]{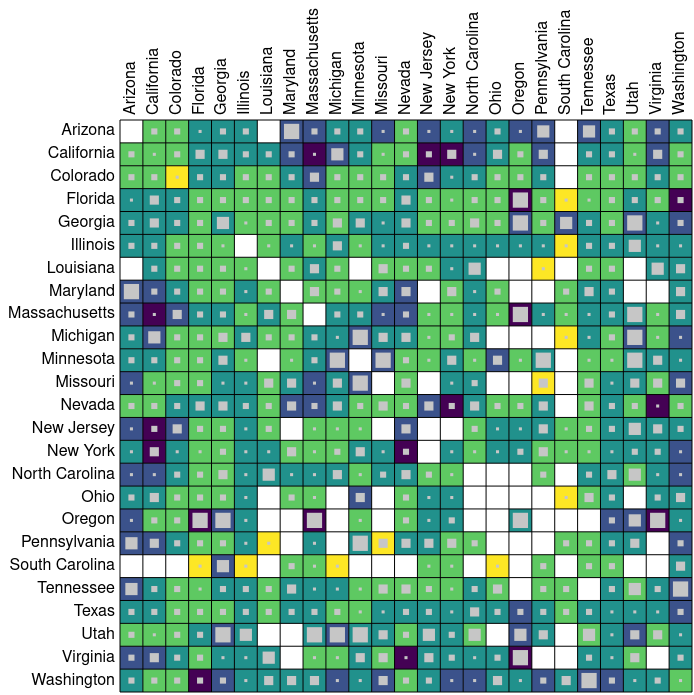}\\
            (c)
        \end{minipage}%
        \hfill
        \begin{minipage}{0.43\textwidth}
            \centering
            \includegraphics[width=\linewidth]{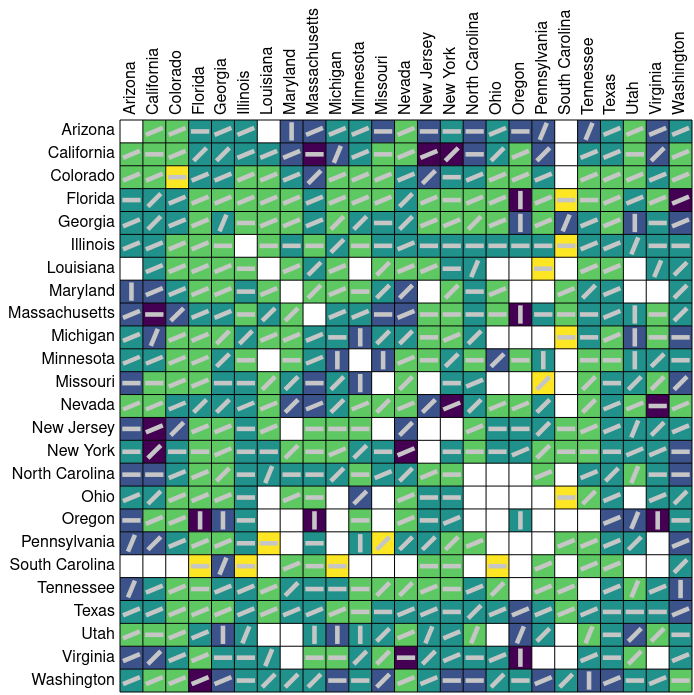}\\
            (d)
        \end{minipage}
    
        \captionof{figure}{Stimuli used in the study. 
        (a) Bivariate color palette encoding. 
        (b) Embedded bar chart encoding. 
        (c) Overlaid mark size encoding. 
        (d) Overlaid mark angle encoding.}
        \label{fig:stimuli-training}
    \end{minipage}
\end{center}

\clearpage

\section{Data Processing}
\label{sec:data_proc}

We first removed incomplete records, defined as entries with missing origin, destination, carrier, or fare information. We also excluded non-continental U.S. routes. Connections with fewer than 15 passengers were discarded, as these often correspond to bulk or contract fares reported as zero or implausibly high placeholder values. In addition, we removed infrequent carrier–route combinations ($\leq 5 $ weekly occurrences) and excluded routes whose aggregated fare standard deviation exceeded 150 USD, treating them as outliers to prevent distortions in the visual encodings and to preserve the interpretability of linear scaling without logarithmic or power-law transformations.

Our study focuses on the perceptual implications of two attribute encodings rather than flow direction. Therefore, we did not distinguish between origin and destination, making the AMs undirected and symmetrical. This reduces visual complexity allowing participants to focus on attribute analysis.

For each origin--destination pair, we first averaged fares across all operators serving that route, and then computed the mean and standard deviation of these aggregated values. This procedure emulates having a quantitative edge attribute per route—rather than treating each individual flight as an independent attribute—while producing the same effective outcome: a distribution of values associated with each edge. Based on these aggregated edge attributes, we constructed two networks: a smaller network for the Training phase (12 nodes, density 0.55) and a larger network for the main Study phase (25 nodes, density 0.7).

The complete preprocessing and aggregation pipeline is provided in the supplementary material (\texttt{Supplementary Material/dataGeneration/states.ipynb}), together with the dataset used to derive the training and study networks (\texttt{Origin\_and\_Destination\_Survey\_DB1BMarket\_2018\_1}).

\section{Color Usage}
\label{sec:color-usage}

This appendix documents the exact color usage in the study stimuli:
(1) the Viridis samples used to encode central tendency (CT) (with hex values),
(2) the white-mixing method used to encode dispersion (D) in the bivariate color condition (BiC),
and (3) the gray-mark color formula used in the overlaid-mark conditions (OMS/OMA).

\subsection{CT $\rightarrow$ Hue (Viridis)}
\label{app:color:ct}

Across all conditions, CT was mapped to hue using a quantized scale (d3 \texttt{scaleQuantize})
with domain $[\mathrm{meanMin}, \mathrm{meanMax}]$ and a discrete range of $m=5$ colors.
The colors were sampled from Viridis at normalized positions
\[
t_i = \frac{i}{m-1}, \qquad i \in \{0,\dots,m-1\},
\]
using \texttt{d3.interpolateViridis}.

For $m=5$, this yields:
\[
t = \{0,\,0.25,\,0.5,\,0.75,\,1\}.
\]

Table~\ref{tab:viridis-samples} lists the resulting hex values.

\begin{table}[t]
\centering

\begin{tabular}{c c c}
\hline
CT Range  & Hex & Color \\
\hline
0-49    & \#FDE725 & \colorbox[HTML]{FDE725}{\strut\hspace{1.6em}} \\
50-99   & \#5EC962 & \colorbox[HTML]{5EC962}{\strut\hspace{1.6em}} \\
100-149 & \#21918C & \colorbox[HTML]{21918C}{\strut\hspace{1.6em}} \\
150-199 & \#3B528B & \colorbox[HTML]{3B528B}{\strut\hspace{1.6em}} \\
200-250 & \#440154 & \colorbox[HTML]{440154}{\strut\hspace{1.6em}} \\
\hline
\end{tabular}

\caption{CT ranges with their associated Viridis hex and color}
\label{tab:viridis-samples}

\end{table}

\subsection{BiC: D $\rightarrow$ Lightness via White Mixing}
\label{app:color:bic}

In the bivariate color condition (BiC), dispersion D was mapped to lightness by \emph{white mixing}
implemented as an opacity-modulated overlay on a white base.

D was mapped with a quantized scale (d3 \texttt{scaleQuantize}) with domain
$[\mathrm{devMin}, \mathrm{devMax}]$ and a discrete set of $k=5$ opacity values
\[
\alpha_j = \alpha_{\max} - \frac{j}{k-1}\left(\alpha_{\max}-\alpha_{\min}\right),
\qquad j \in \{0,\dots,k-1\},
\]
with
\[
\alpha_{\min} = 0.2,
\qquad
\alpha_{\max} = 1.0.
\]

For $k=5$, this yields:
\[
\alpha \in \{1.0,\,0.8,\,0.6,\,0.4,\,0.2\}.
\]

Rendering was performed by drawing two stacked rectangles per cell:
(i) a white base, and (ii) a CT-colored rectangle with opacity $\alpha$.
This is equivalent to linear interpolation (mixing) in sRGB:
\[
\mathbf{c}_{out} = \alpha \cdot \mathbf{c}_{CT} + (1-\alpha)\cdot \mathbf{c}_{white},
\qquad \mathbf{c}_{white}=(1,1,1).
\]

Thus, BiC encodes CT by hue (Viridis) and D by the whitened lightness of that hue.

\subsection{OMS/OMA: Gray Overlaid Mark Color (Contrast-Aware)}
\label{app:color:marks}

In the overlaid-mark conditions (OMS/OMA), the cell background encodes CT using the same
Viridis mapping as in Appendix~\ref{app:color:ct}. The overlaid mark encodes D (by size in OMS,
by angle in OMA) and is drawn in a gray whose lightness is computed from the background to
maintain legible contrast without introducing additional hue.

Given the CT background color $\mathbf{c}_{CT}$, we convert it to CIELAB and extract its lightness
$L \in [0,100]$. We then compute:
\[
L_{\text{inv}} = 100 - L,
\qquad
L_{\text{gray}} = 80 - c + \frac{L_{\text{inv}}\cdot c}{100},
\]
where the interface parameter is set to
\[
c = 0.
\]

With $c=0$, this simplifies to:
\[
L_{\text{gray}} = 80,
\]
yielding a constant neutral gray:
\[
\mathrm{Lab}(80,\,0,\,0),
\]

\section{Matrix Ordering}
\label{sec:matrix-ordering}

Matrix ordering was controlled per task and kept identical across encoding conditions. 

For cluster-identification tasks, we computed task-relevant orderings using the \texttt{reorder.js} library~\cite{fekete2015}. Specifically, we applied (1) Optimal Leaf Ordering (OLO) with Euclidean distance and complete linkage, and (2) PCA-based ordering computed over either the mean or the standard deviation of link weights. These configurations resulted in four distinct orderings (e.g., OLO over mean values, PCA over standard deviation), depending on the structural property emphasized by the task.

For all remaining (non-cluster) tasks, we used a fixed alphabetical ordering of rows and columns. This ensured consistency across conditions and avoided introducing ordering-related confounds in comparisons between encoding techniques.

\section{\href{https://revisit.dev/}{reVISit}}
\label{sec:revisit}

The reVISit platform allows us to offer a completely transparent view of our study, all conditions, collected data and even participants’ interactions can be seen on the four studies that we created at: \href{https://jorgeacostaupm.github.io/revisit/}{jorgeacostaupm.github.io/revisit/}.

Beyond transparency, reVISit greatly facilitated the creation, management, and dissemination of our studies. Its intuitive interface streamlines study setup, data collection, and visualization of participant behavior, making it an invaluable tool for replicable and open research.

\section{Multivariate AM designs}
\label{sec:supp_figures}

In this section, we present several multivariate AM designs from prior work. While many of the studies discussed in the related work section have inspired us, here we include only a few of the most closely related ones.

\begin{figure}[htb]
    \centering
    \includegraphics[width=\columnwidth]{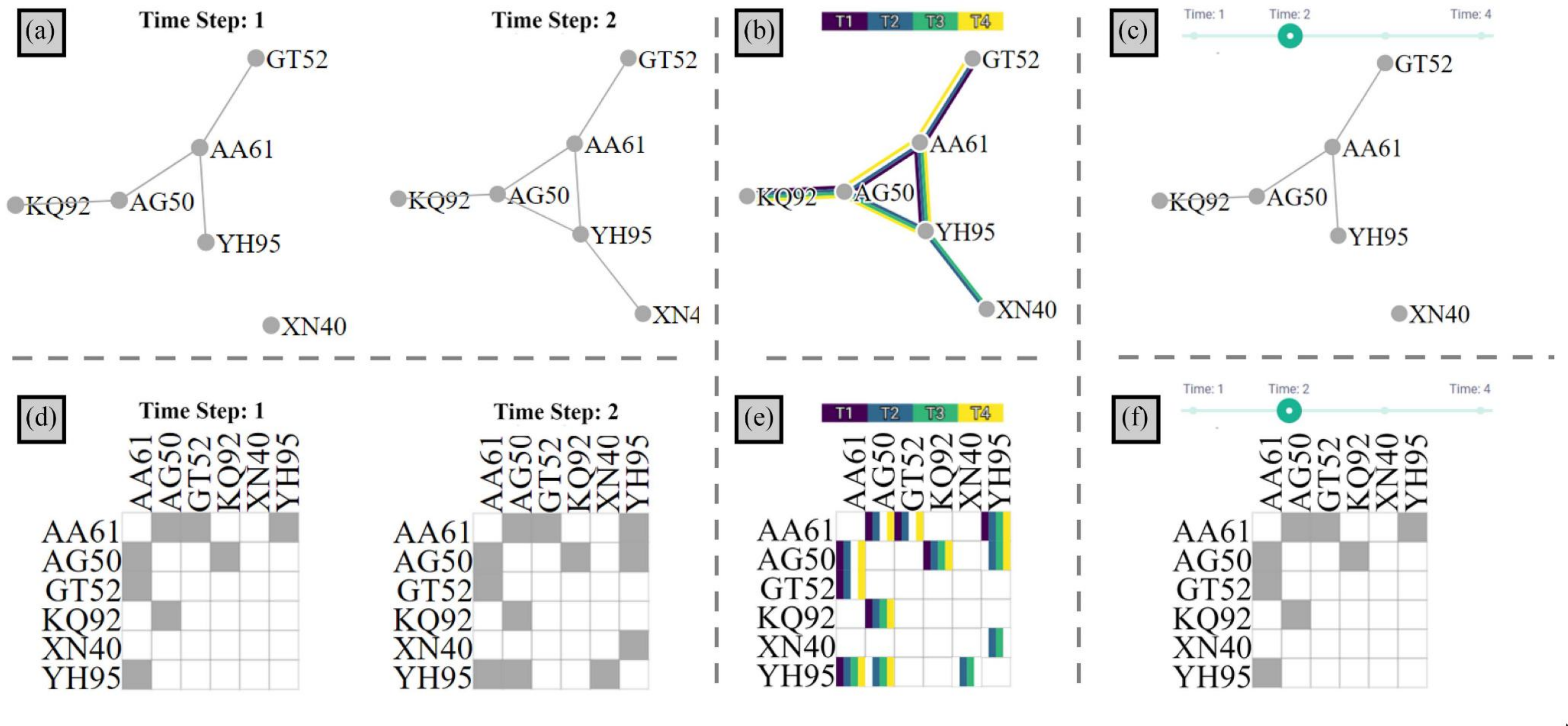}
    \caption{Encoding techniques for network data in the context of dynamic networks by Filipov et al. Image reproduced from \cite[Fig.~1]{filipov2024}.
    }
    \label{fig:Filipov}
\end{figure}

\begin{figure}[htb]
    \centering
    \includegraphics[width=\columnwidth]{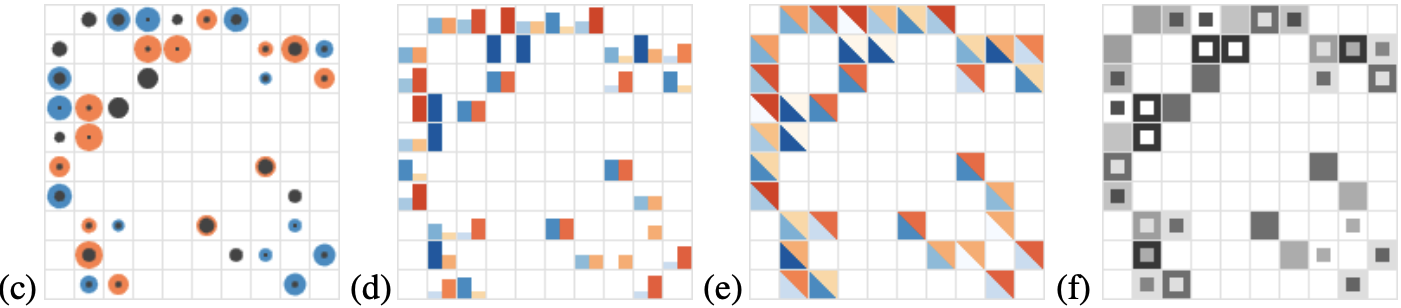}
    \caption{Encoding techniques for bivariate AMs by Alper et al. Image reproduced from \cite[Fig.~3]{alper2013weighted}}
    \label{fig:Alper}
\end{figure}

\begin{figure}[htb]
    \centering
    \includegraphics[width=\columnwidth]{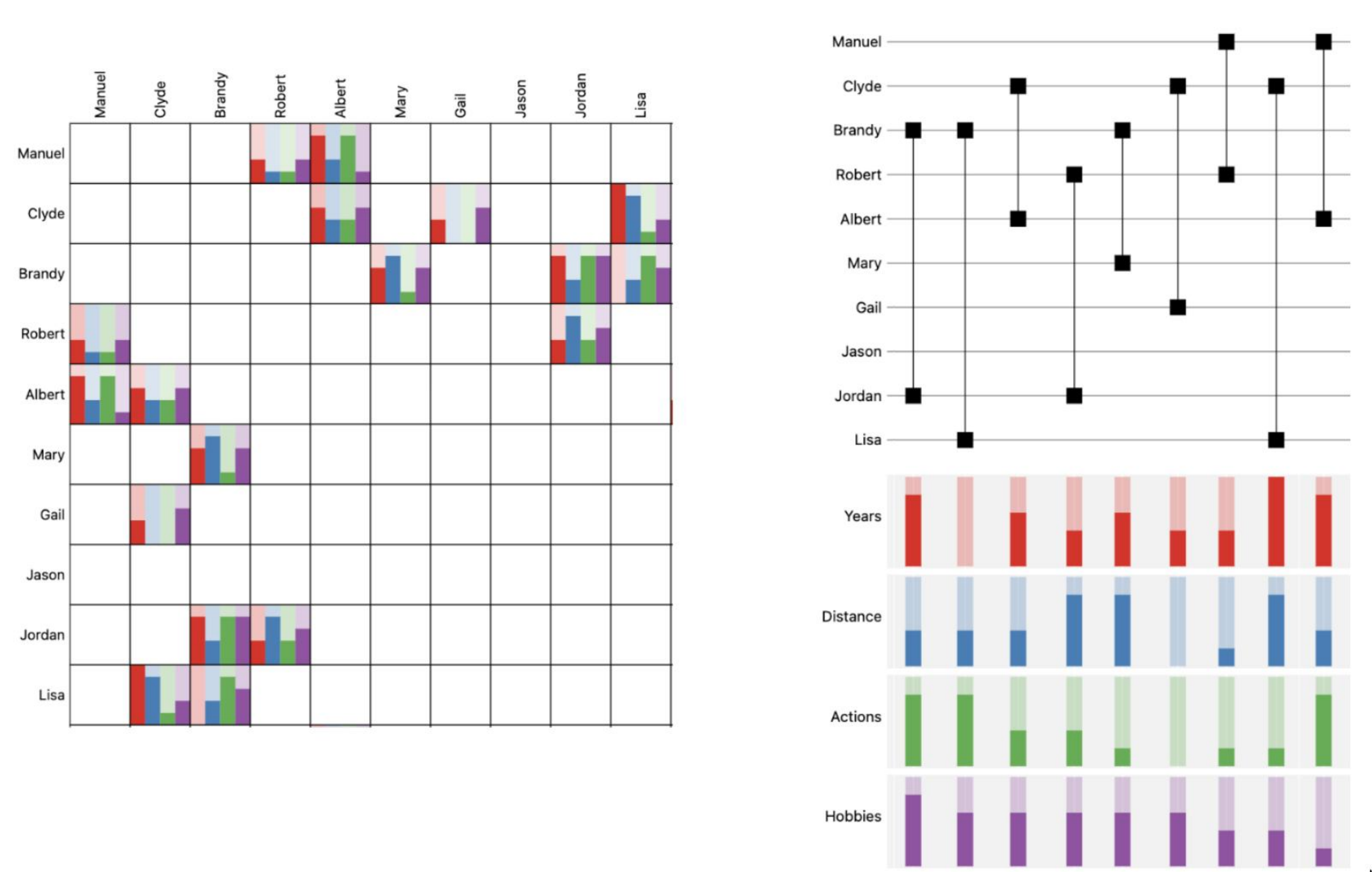}
    \caption{Bar Chart encoding tested by Fuchs et al. against a similar encoding on the BioFabric Layout. Image reproduced from \cite[Fig.~8]{fuchs2024exploring}.
    }
    \label{fig:Fuchs}
\end{figure}

\begin{figure}[htb]
    \centering
    \includegraphics[width=\columnwidth]{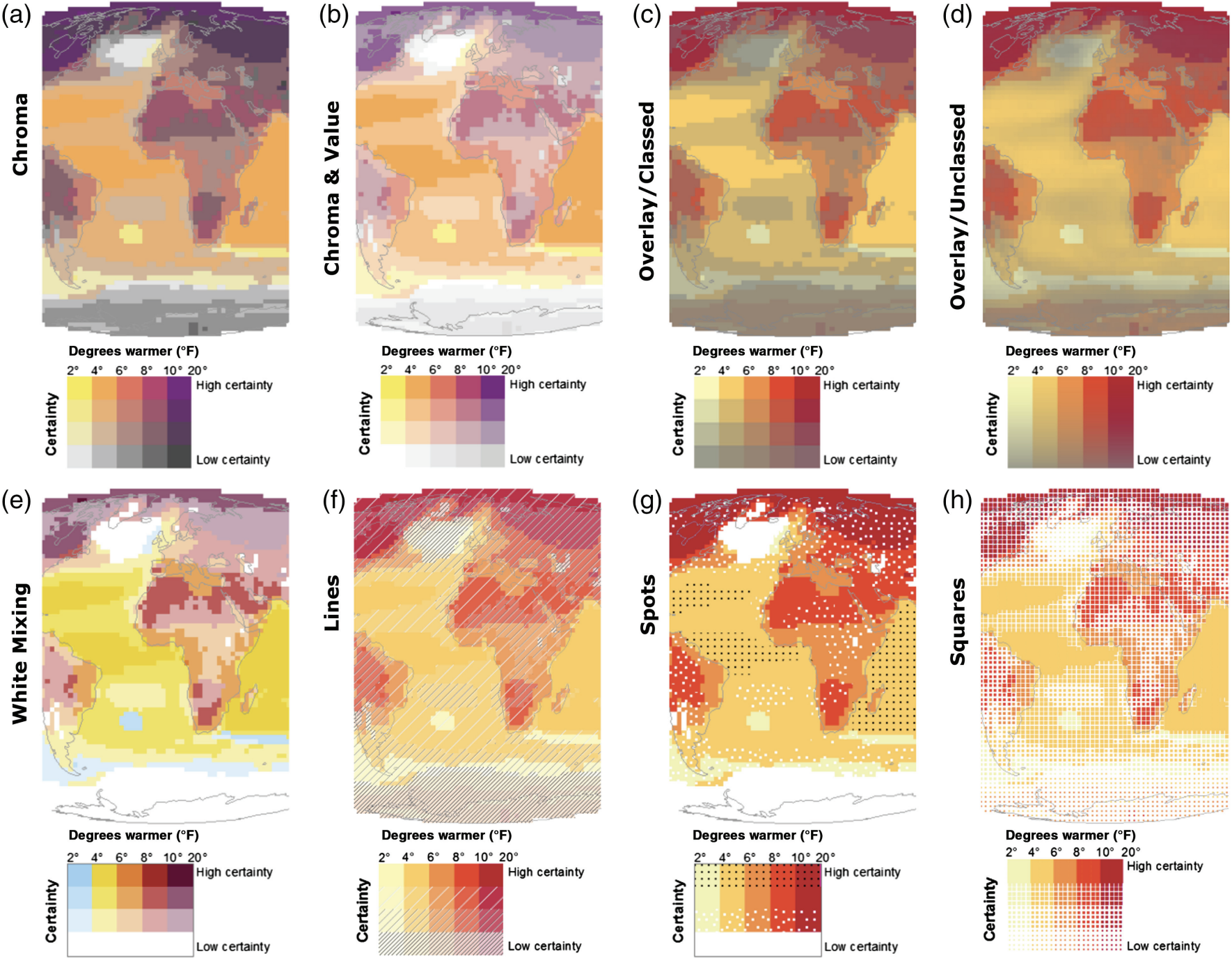}
    \caption{Comparison of eight bivariate representations by Retchless and Brewer. Image reproduced from \cite[Fig.~11]{retchless2016}.
    }
    \label{fig:Retchless}
\end{figure}

\begin{figure}[htb]
    \centering
    \includegraphics[width=\columnwidth]{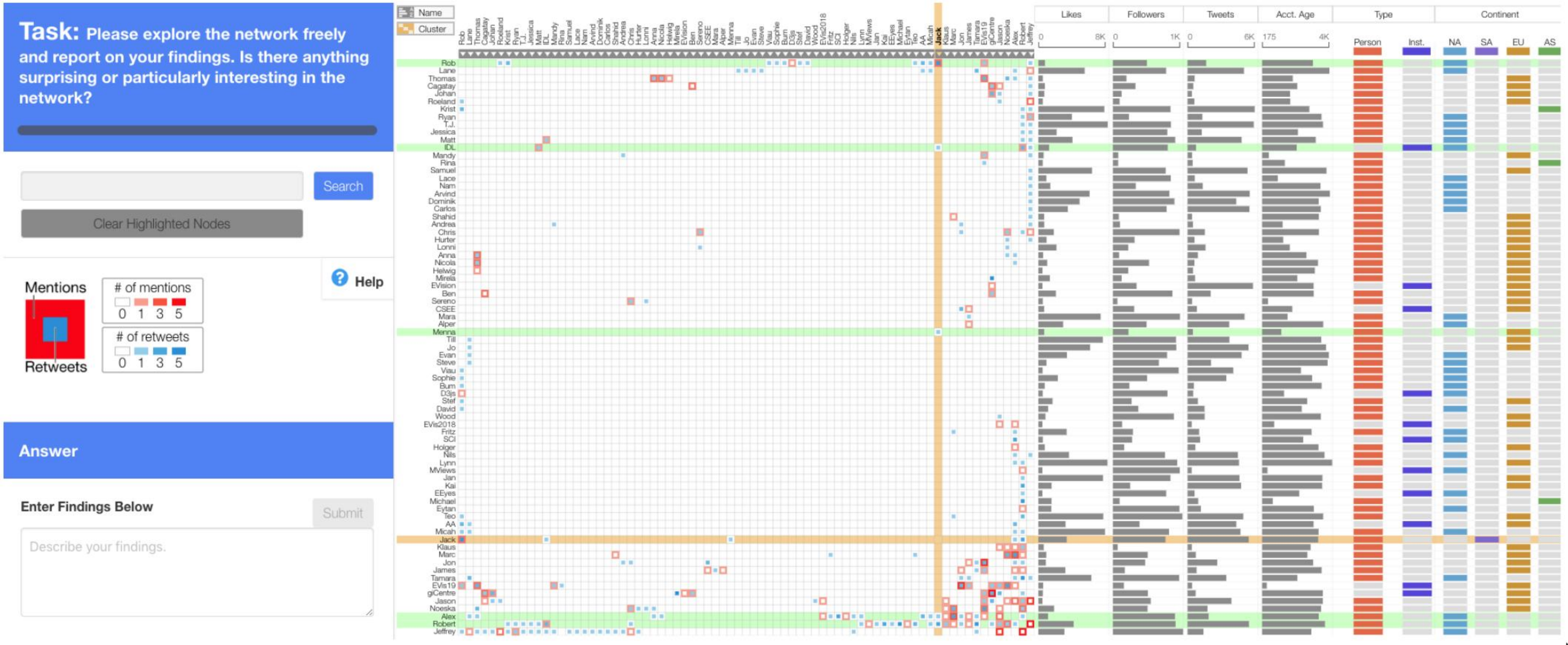}
    \caption{A task testing one of Alper et al. encodings against NL by Nobre et al. Image reproduced from \cite[Fig.~2]{nobre2020evaluating}.
    }
    \label{fig:Nobre}
\end{figure}

\begin{figure}[htb]
    \centering
    \includegraphics[width=\columnwidth]{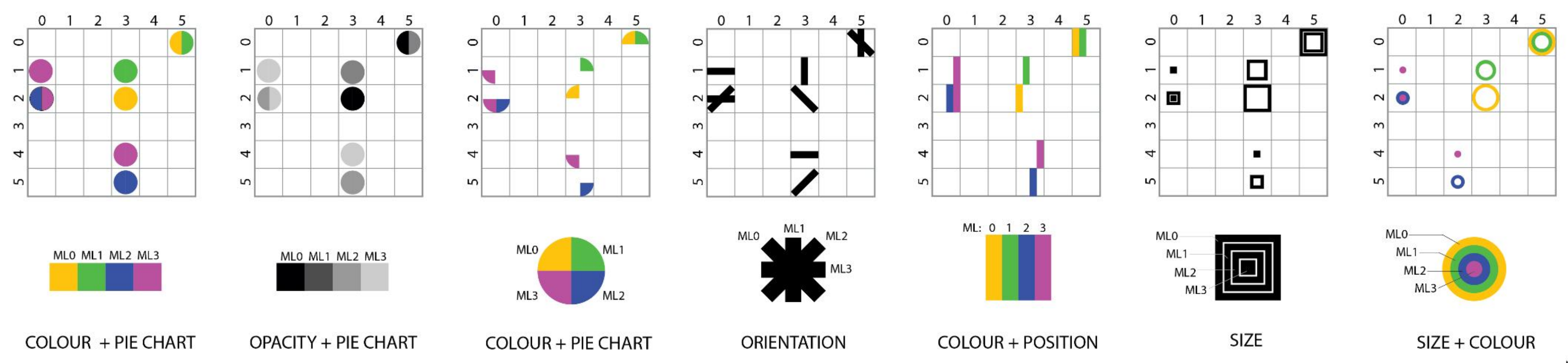}
    \caption{Proposed encodings for multivariate AMs by Vogogias et al. Image reproduced from \cite[Fig.~1]{vogogias2020visual}.
    }
    \label{fig:Vogogias}
\end{figure}

\clearpage

\section{Participant Demographics}
\label{sec:demo}
This section presents comprehensive demographic information for our participant pool (N=156). 
The distributions cover basic demographics including \textbf{age} and \textbf{sex}, 
educational background through \textbf{highest degree achieved}, 
\textbf{used browser}, expertise by \textbf{self-assessed visualization proficiency}, \textbf{number of participants per encoding technique} and \textbf{nationality}.

\begin{figure}[htbp]
    \centering
    \includegraphics[width=\linewidth]{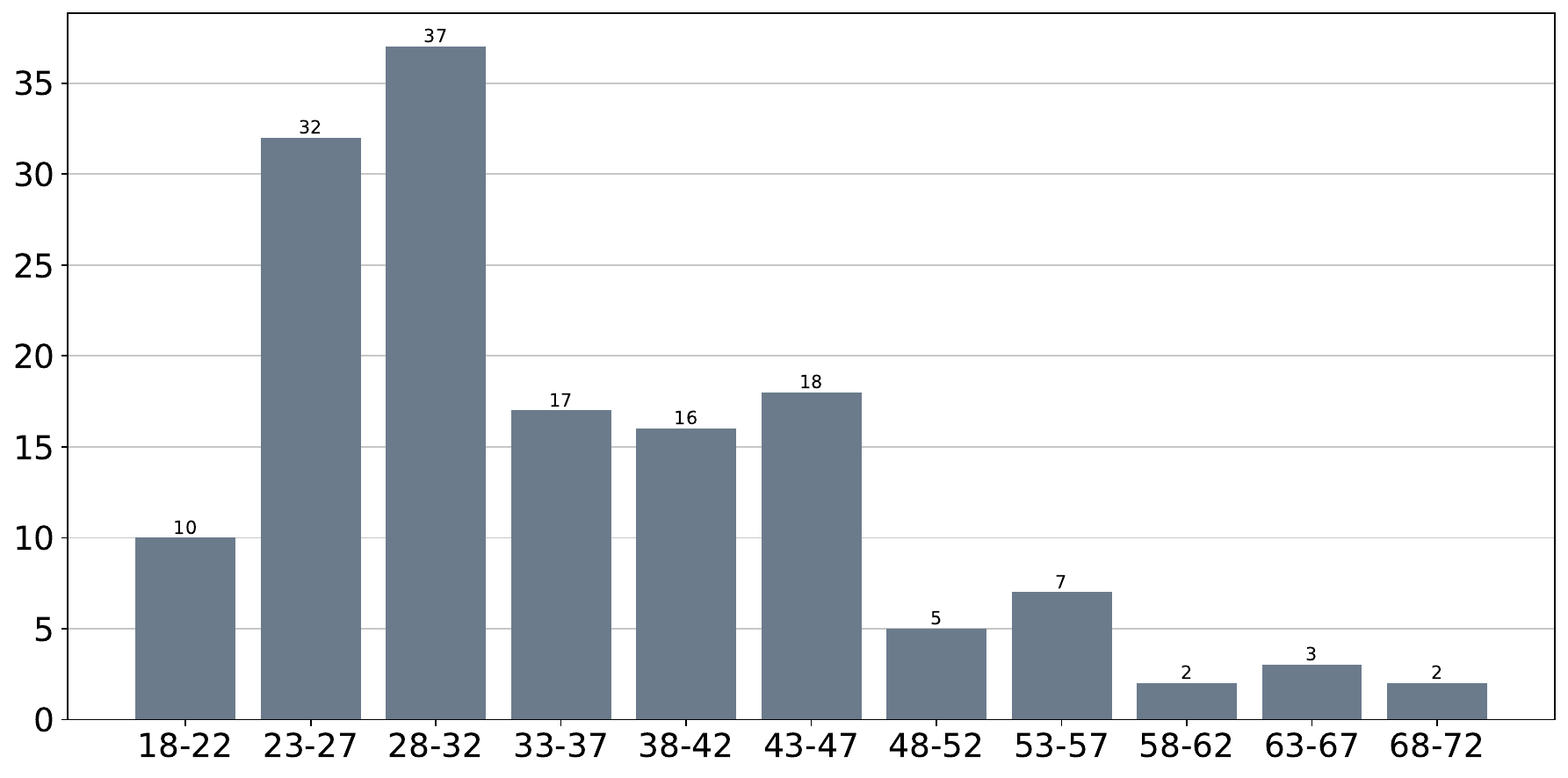}
    \caption{Age Distribution}
    \label{fig:age_distr}
\end{figure}

\begin{figure}[htbp]
    \centering
    \includegraphics[width=\linewidth]{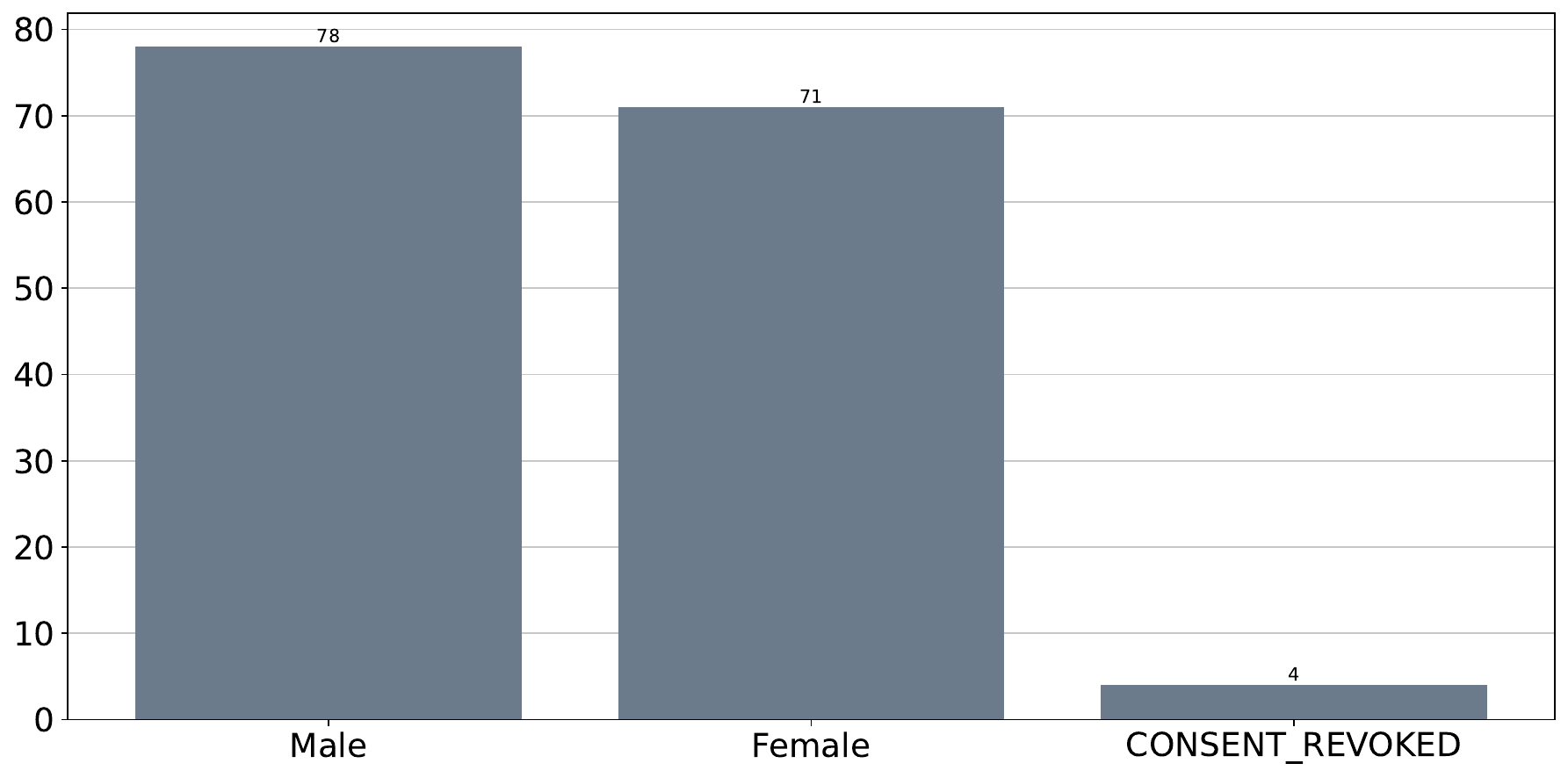}
    \caption{Sex Distribution}
    \label{fig:sex_distr}
\end{figure}

\begin{figure}[htbp]
    \centering
    \includegraphics[width=\linewidth]{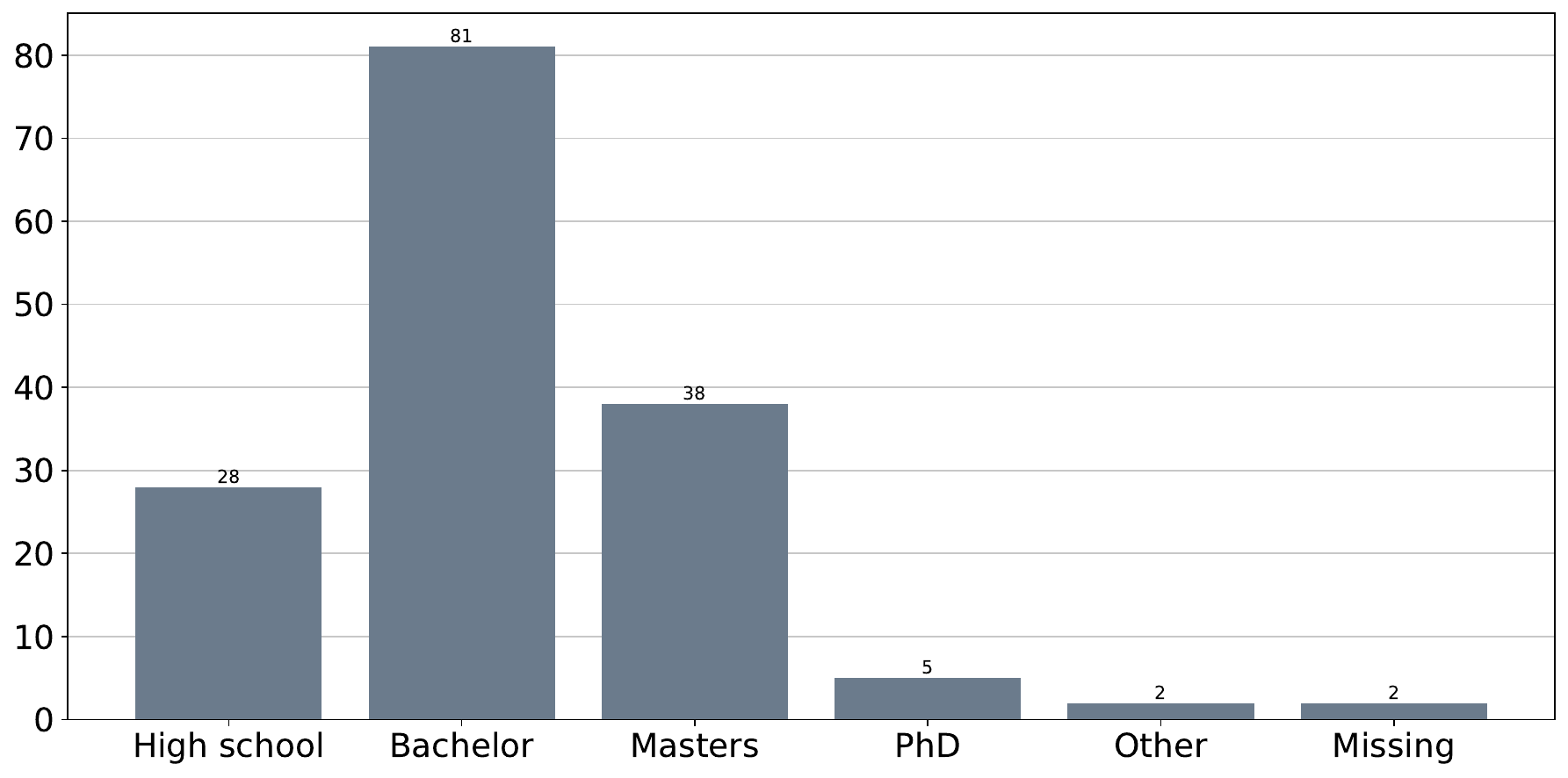}
    \caption{Education Level Distribution}
    \label{fig:education_distr}
\end{figure}

\begin{figure}[htbp]
    \centering
    \includegraphics[width=\linewidth]{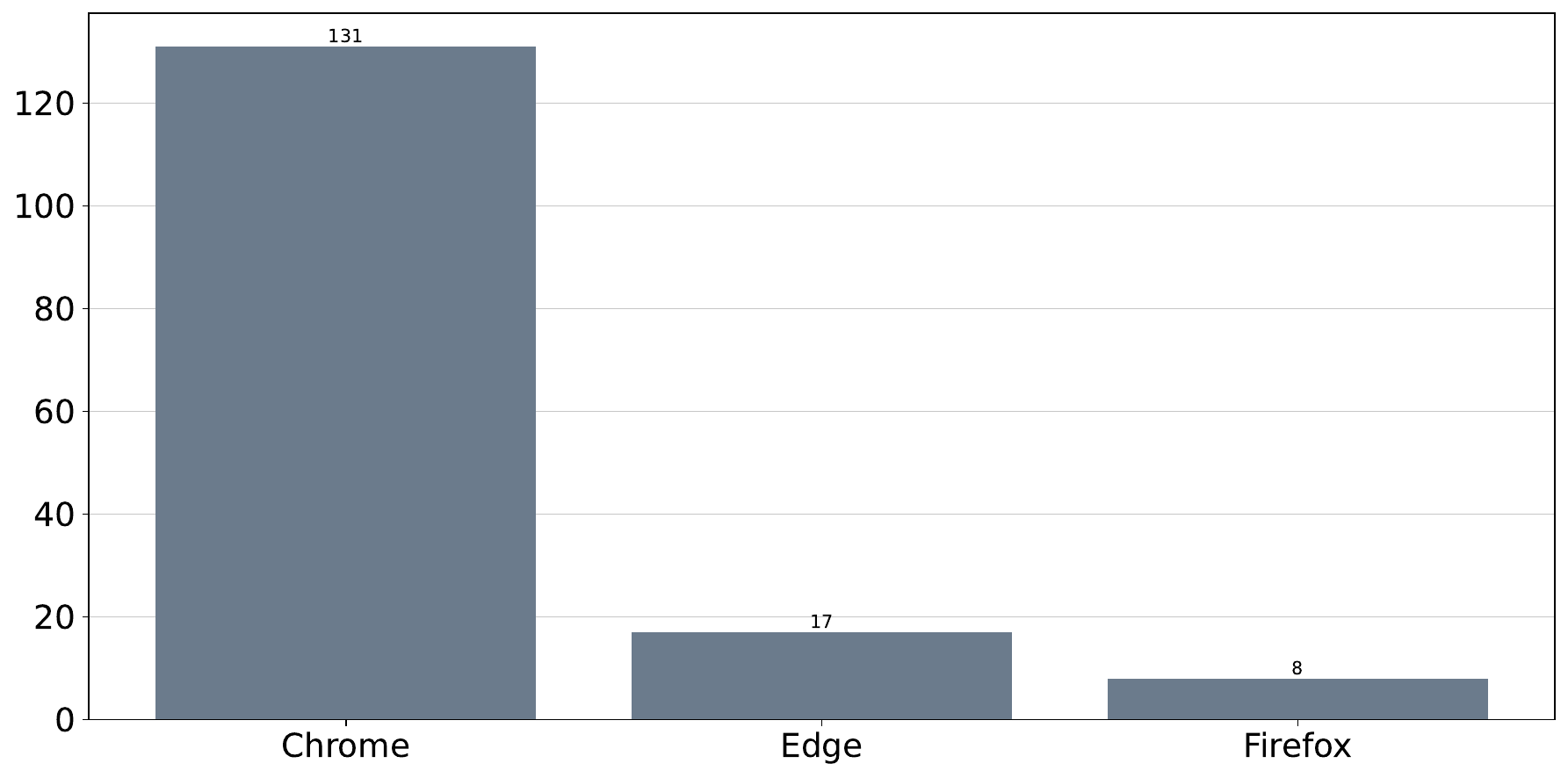}
    \caption{Browser Usage Distribution}
    \label{fig:browser_distr}
\end{figure}

\begin{figure}[htbp]
    \centering
    \includegraphics[width=\linewidth]{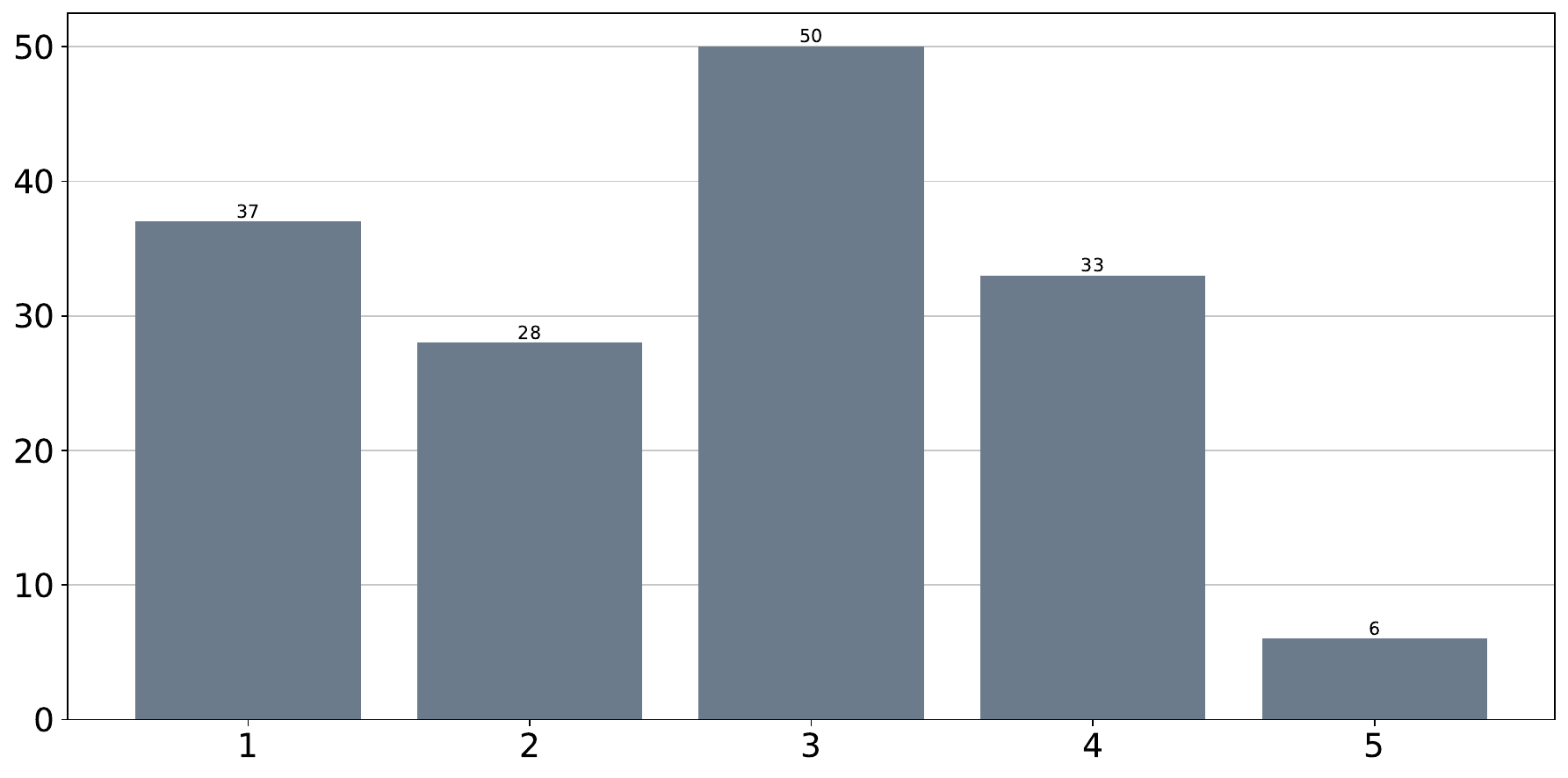}
    \caption{Matrix Experience Scores}
    \label{fig:am_experience_distr}
\end{figure}

\begin{figure}[htbp]
    \centering
    \includegraphics[width=\linewidth]{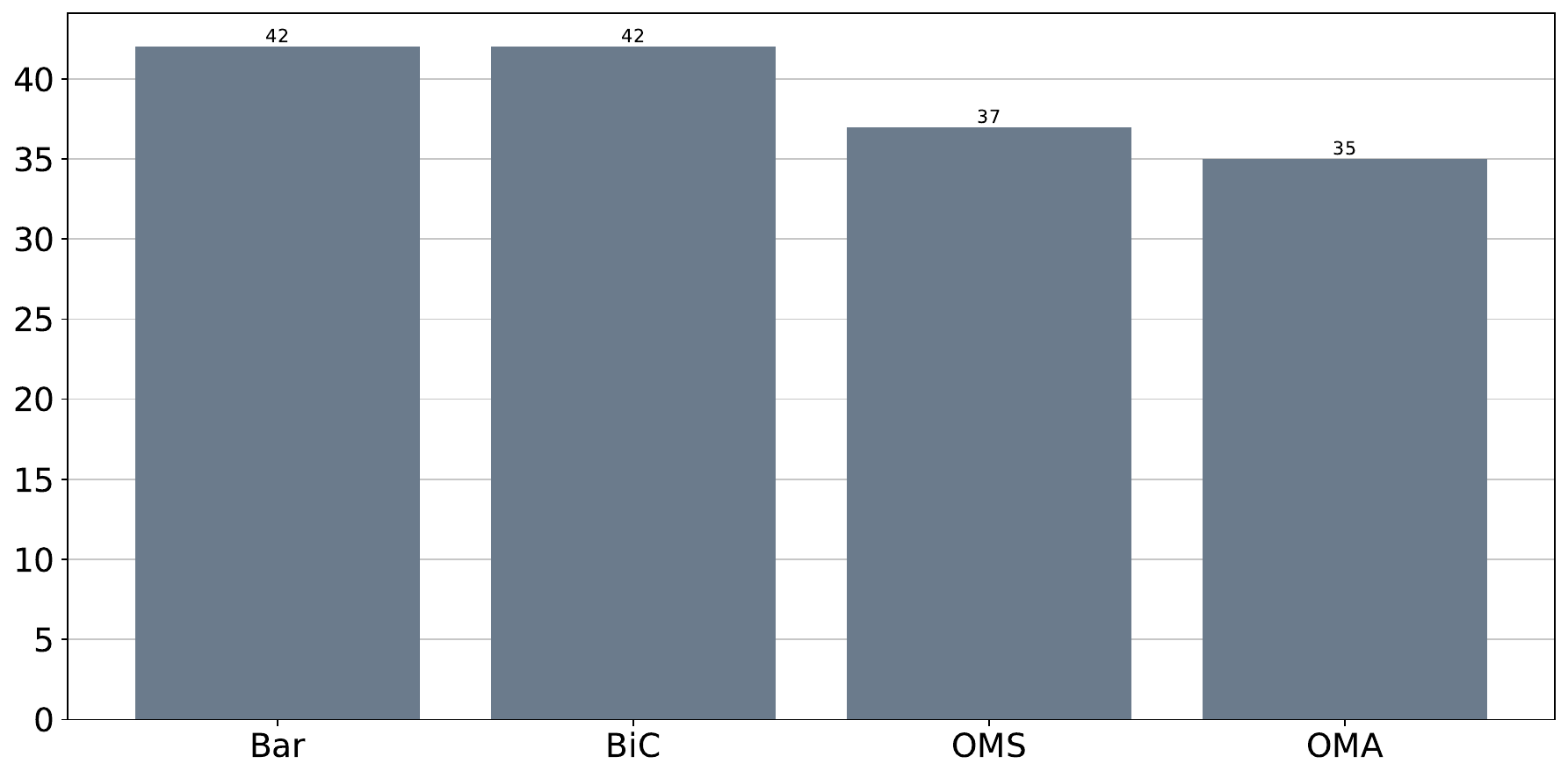}
    \caption{Encoding Distribution}
    \label{fig:encodings_distr}
\end{figure}

\begin{figure}[htbp]
    \centering
    \includegraphics[width=\linewidth]{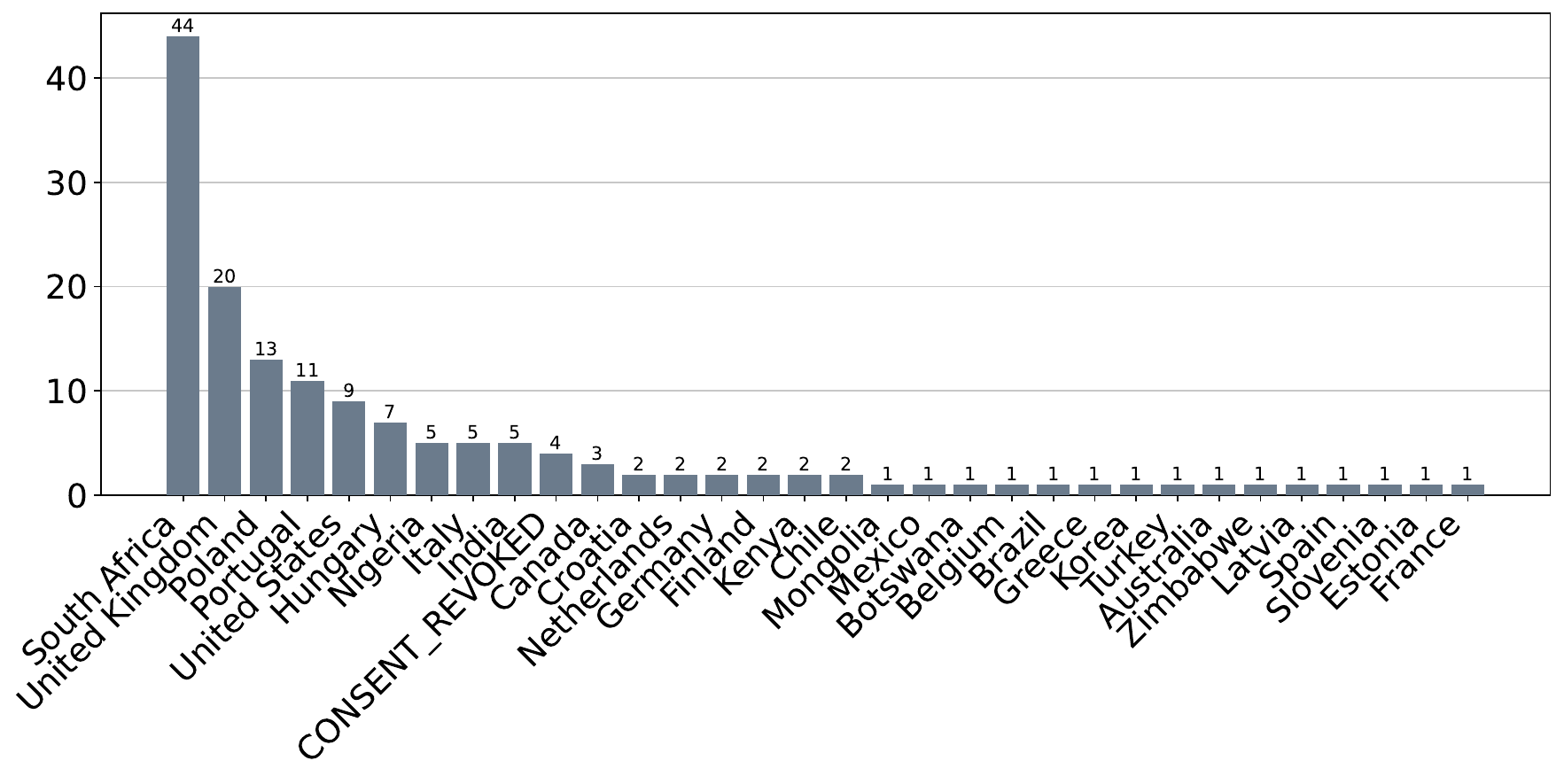}
    \caption{Country Distribution}
    \label{fig:country_distr}
\end{figure}

\clearpage
\onecolumn

\section{Results Summary}
\label{sec:results-app}

\begin{table*}[h]
\centering
\renewcommand{\arraystretch}{1.1}

\begin{tabular}{
l
l
S[table-format=2.0]
S[table-format=2.0]
S[table-format=2.0]
S[table-format=2.0]
@{\hspace{20pt}}
S[table-format=1.2]
@{\hspace{20pt}}
S[table-format=1.2]
}
\toprule
\multicolumn{2}{c}{\textbf{Task}} 
& \multicolumn{4}{c}{\textbf{Encoding (Performance \%)}} 
& \multicolumn{2}{c}{\textbf{Statistics}} \\
\cmidrule(lr){1-2}
\cmidrule(lr){3-6}
\cmidrule(lr){7-8}

\textbf{ID} 
& \textbf{Name}
& {\bic}
& {\oms}
& {\oma}
& {\barc}
& {$p$}
& {$\varepsilon^2$} \\
\midrule

T1 & Structural Adjacency
& 91 & \cellcolor{best}97 & 91 & 96 & 0.24 & 0.01 \\

T2 & Adjacency by Central Tendency
& 82 & \cellcolor{best}94 & 92 & 90 & 0.00** & 0.08 \\

T3 & Adjacency by Dispersion
& 68 & 91 & 80 & \cellcolor{best}95 & 0.00** & 0.27 \\

T4 & Adjacency by Attribute Combination
& 56 & 71 & 53 & \cellcolor{best}73 & 0.02* & 0.05 \\

T5 & Extreme Detection
& 55 & \cellcolor{best}80 & 67 & 70 & 0.00** & 0.09 \\

T6 & Classification by Dispersion
& 60 & 89 & 81 & \cellcolor{best}90 & 0.00** & 0.23 \\

T7 & Cluster Central Tendency Estimation
& 83 & \cellcolor{best}91 & 83 & 89 & 0.36 & 0.00 \\

T8 & Cluster Dispersion Estimation
& 70 & 92 & 84 & \cellcolor{best}95 & 0.00** & 0.09 \\

\bottomrule
\end{tabular}

\caption{
Performance across tasks (T1--T8).
Overall effects are assessed using the Kruskal--Wallis test;
$\varepsilon^2$ denotes effect size.
Highest-performing encoding per task is highlighted.
* $p < .05$, ** $p < .01$.
}
\label{tab:task-results}
\end{table*}

\begin{table*}[h]
\centering
\renewcommand{\arraystretch}{1.1}

\begin{tabular}{
l
l
l
S[table-format=1.2]
S[table-format=1.2]
S[table-format=1.2]
S[table-format=1.2]
@{\hspace{20pt}}
S[table-format=1.2]
@{\hspace{20pt}}
S[table-format=1.2]
}
\toprule
\multicolumn{3}{c}{\textbf{Construct}} 
& \multicolumn{4}{c}{\textbf{Encoding (Likert 1--7)}} 
& \multicolumn{2}{c}{\textbf{Statistics}} \\
\cmidrule(lr){1-3}
\cmidrule(lr){4-7}
\cmidrule(lr){8-9}

\textbf{Scale}
& \textbf{Measure}
& {}
& {\bic}
& {\oms}
& {\oma}
& {\barc}
& {$p$}
& {$\varepsilon^2$} \\
\midrule

PREVis & Reading Data
& {} & 4.56 & 5.69 & \cellcolor{best}5.90 & 4.99 & 0.00** & 0.13 \\

PREVis & Understanding
& {} & 4.99 & 5.83 & \cellcolor{best}5.94 & 5.43 & 0.00** & 0.09 \\

PREVis & Layout
& {} & 3.76 & \cellcolor{best}4.85 & 4.42 & 3.86 & 0.01** & 0.06 \\

PREVis & Reading Features
& {} & 4.95 & 5.54 & \cellcolor{best}5.61 & 4.86 & 0.01** & 0.06 \\

BeauVis & Aesthetic Pleasure
& {} & 4.65 & 5.62 & \cellcolor{best}5.96 & 4.49 & 0.00** & 0.17 \\

\bottomrule
\end{tabular}

\caption{
Mean Likert scores (1--7) for PREVis and BeauVis.
Overall effects are assessed using the Kruskal--Wallis test;
$\varepsilon^2$ denotes effect size.
Higher values indicate more positive ratings.
Highest mean per measure is highlighted.
* $p < .05$, ** $p < .01$.
}
\label{tab:perceived-results}
\end{table*}

\clearpage
\section{Tasks}
\label{sec:tasks-screenshots}
\begin{flushleft}
In this section, we provide screenshots of all tasks evaluated in our study, along with the correct answers, task rationale, associated hypotheses, scoring methods, and analysis results. The tasks includes eight analytic task types (T1-T8, see \autoref{tab:tasks}) as well as subjective ratings for aesthetics and readability. For each task, we present one of the four stimuli as an example and provide a link to access all four stimuli. 
\end{flushleft}
\vspace{1em}

\begin{figure*}[hbt]
\centering
\begin{minipage}{\textwidth}
{\fontsize{20}{22}\selectfont\textbf {T1: Structural Adjacency}} \label{sec:T1}
\\

\textbf{Task Rationale:} Assess whether different encodings influence the perception of basic connectivity. \\
\textbf{Hypothesis:} H1.1 \\
\textbf{Scoring:} F1-Score \\
\textbf{Correct Answer:} [
      "Florida",
      "Georgia",
      "Illinois",
      "Maryland",
      "Massachusetts",
      "Michigan",
      "New Jersey",
      "New York",
      "Ohio",
      "Pennsylvania",
      "Tennessee",
      "Texas",
      "Washington"
    ] \\
\textbf{Links:} \\
\inlinevis{bc} \url{https://jorgeacostaupm.github.io/revisit/matrices_barchart/d0pReDZ2RzkyTXZ4MFcrYzRqSWRJQT09}, \\
\inlinevis{b} \url{https://jorgeacostaupm.github.io/revisit/matrices_bivariate/d0pReDZ2RzkyTXZ4MFcrYzRqSWRJQT09}, \\
\inlinevis{oms} \url{https://jorgeacostaupm.github.io/revisit/matrices_size/d0pReDZ2RzkyTXZ4MFcrYzRqSWRJQT09}, \\
\inlinevis{oma} \url{https://jorgeacostaupm.github.io/revisit/matrices_angle/d0pReDZ2RzkyTXZ4MFcrYzRqSWRJQT09}
\end{minipage}
\vspace{1em}

\begin{subfigure}[b]{\linewidth}
    \includegraphics[width=\linewidth]{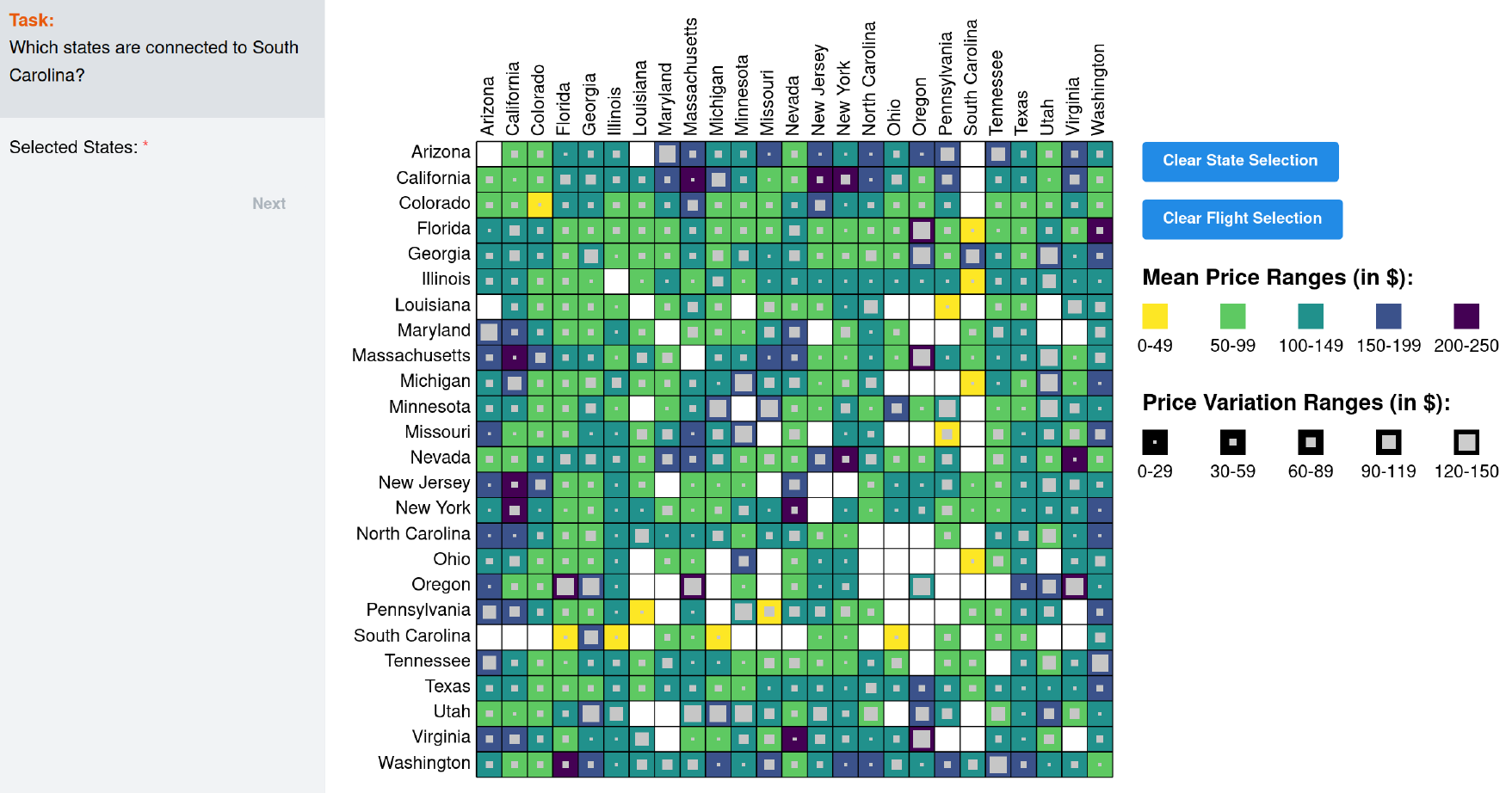}
    \phantomcaption
    \label{fig:T1_screenshot}
\end{subfigure}
\vspace{1em}

\begin{subfigure}[b]{0.23\textwidth}
    \centering
    \includegraphics[width=\linewidth]{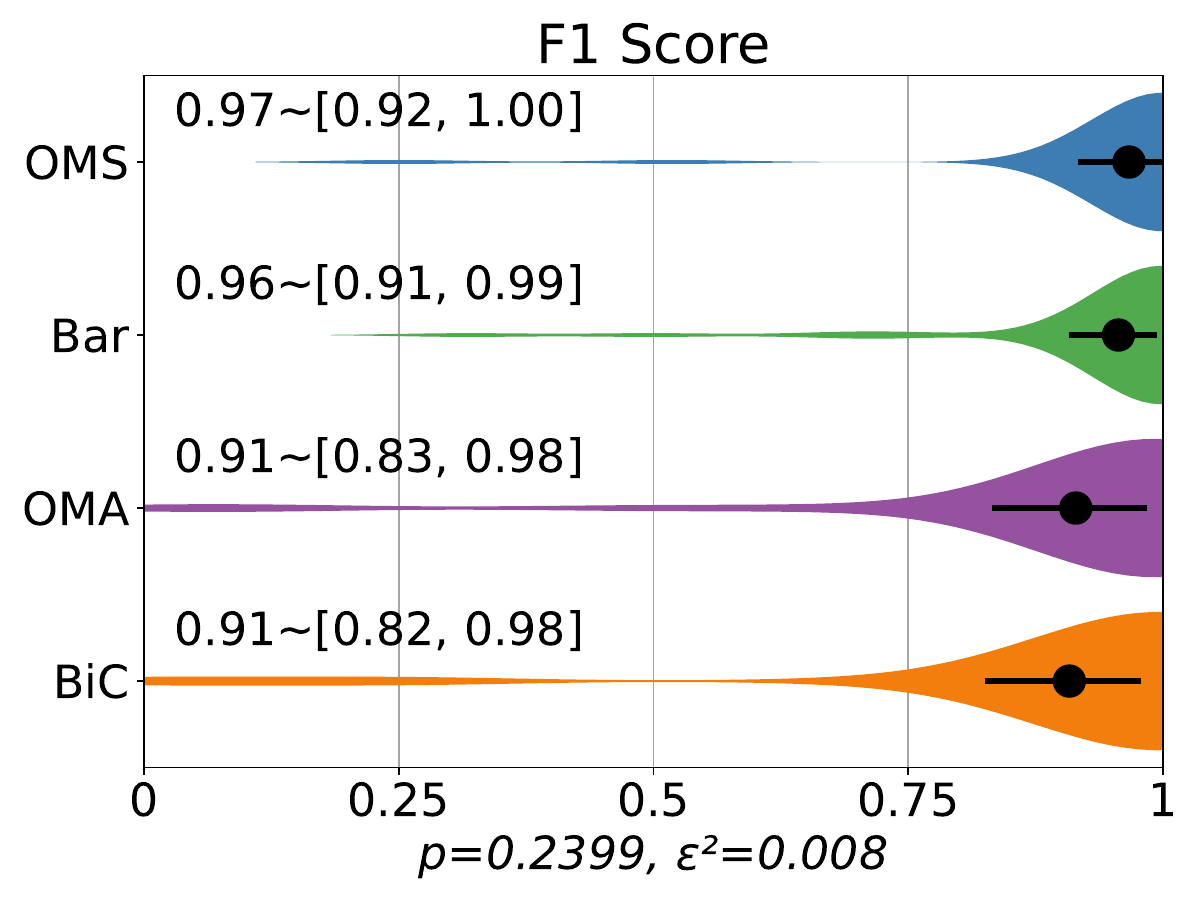}
\end{subfigure}
\hfill
\begin{subfigure}[b]{0.23\textwidth}
    \centering
    \includegraphics[width=\linewidth]{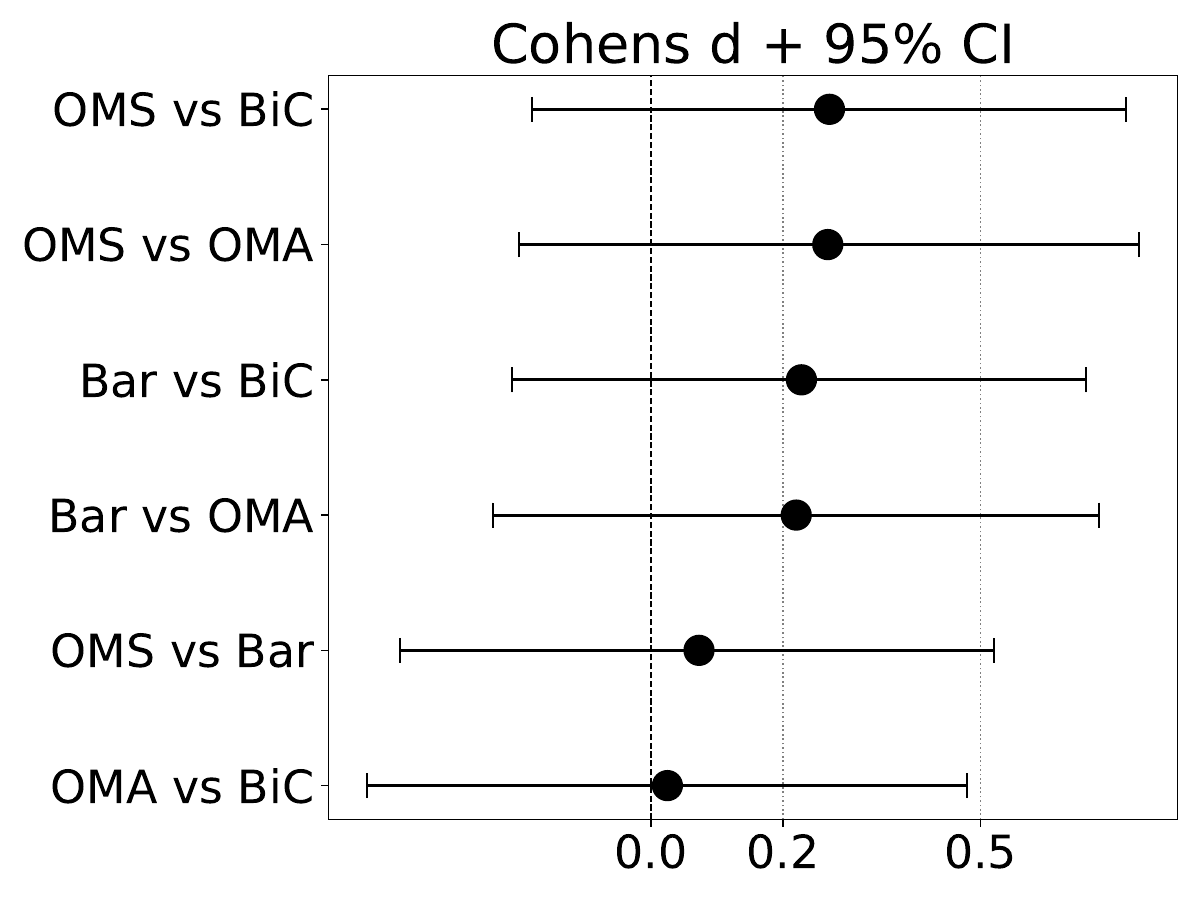}
\end{subfigure}
\hfill
\hspace{3em}
\begin{subfigure}[b]{0.23\textwidth}
    \centering
    \includegraphics[width=\linewidth]{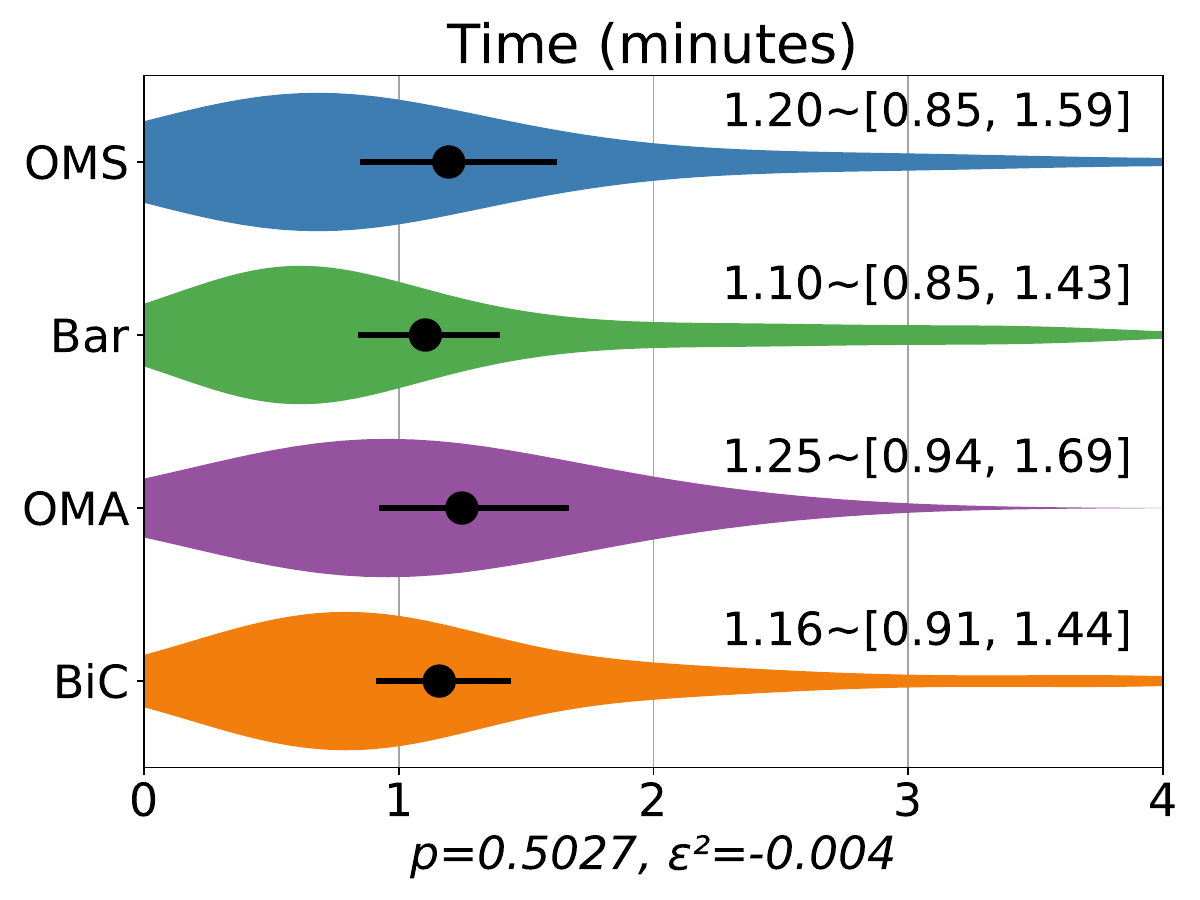}
\end{subfigure}
\hfill
\begin{subfigure}[b]{0.23\textwidth}
    \centering
    \includegraphics[width=\linewidth]{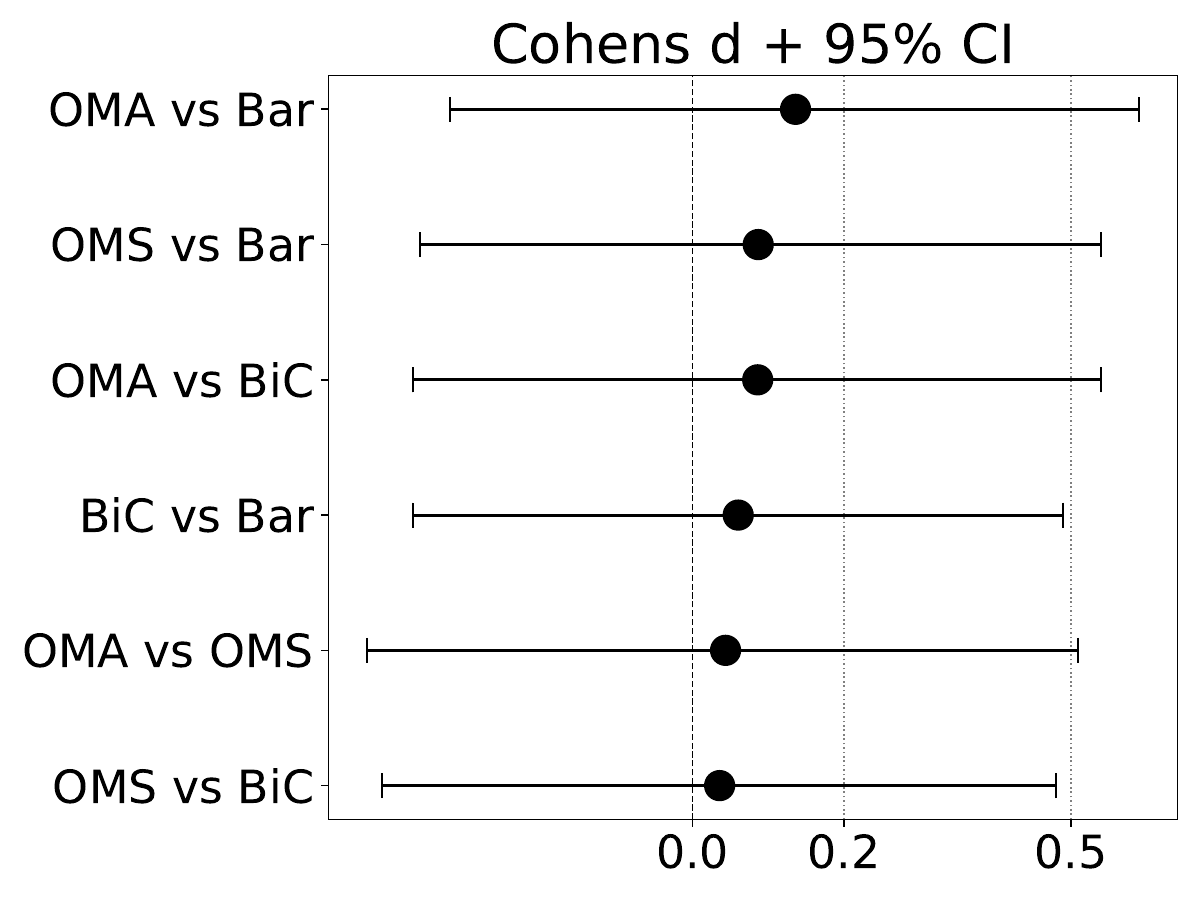}
\end{subfigure}

\caption{T1}
\label{fig:task1_app}
\end{figure*}

\begin{figure*}[hbt]
\centering

\begin{minipage}{\textwidth}
{\fontsize{20}{22}\selectfont\textbf {T2: Adjacency by Central Tendency - Task Prompt 1}} \label{sec:T2_1}\\

\textbf{Task Rationale:} Assess the effectiveness of the encodings to filter links when focusing on central tendency values. \\
\textbf{Hypothesis:} H1.2 and H1.3 \\
\textbf{Scoring:} F1-Score \\
\textbf{Correct Answer:} [
      "Maryland",
      "Massachusetts",
      "New Jersey",
      "New York",
      "Virginia"
    ] \\
\textbf{Links:} \\
\inlinevis{bc} \url{https://jorgeacostaupm.github.io/revisit/matrices_barchart/WndHOERsNlQrL3N5RWd2c1FUbnhCUT09}, \\
\inlinevis{b} \url{https://jorgeacostaupm.github.io/revisit/matrices_bivariate/WndHOERsNlQrL3N5RWd2c1FUbnhCUT09}, \\
\inlinevis{oms} \url{https://jorgeacostaupm.github.io/revisit/matrices_size/WndHOERsNlQrL3N5RWd2c1FUbnhCUT09}, \\
\inlinevis{oma} \url{https://jorgeacostaupm.github.io/revisit/matrices_angle/WndHOERsNlQrL3N5RWd2c1FUbnhCUT09},
\end{minipage}

\vspace{1em}
\begin{subfigure}[b]{\linewidth}
    \includegraphics[width=\linewidth]{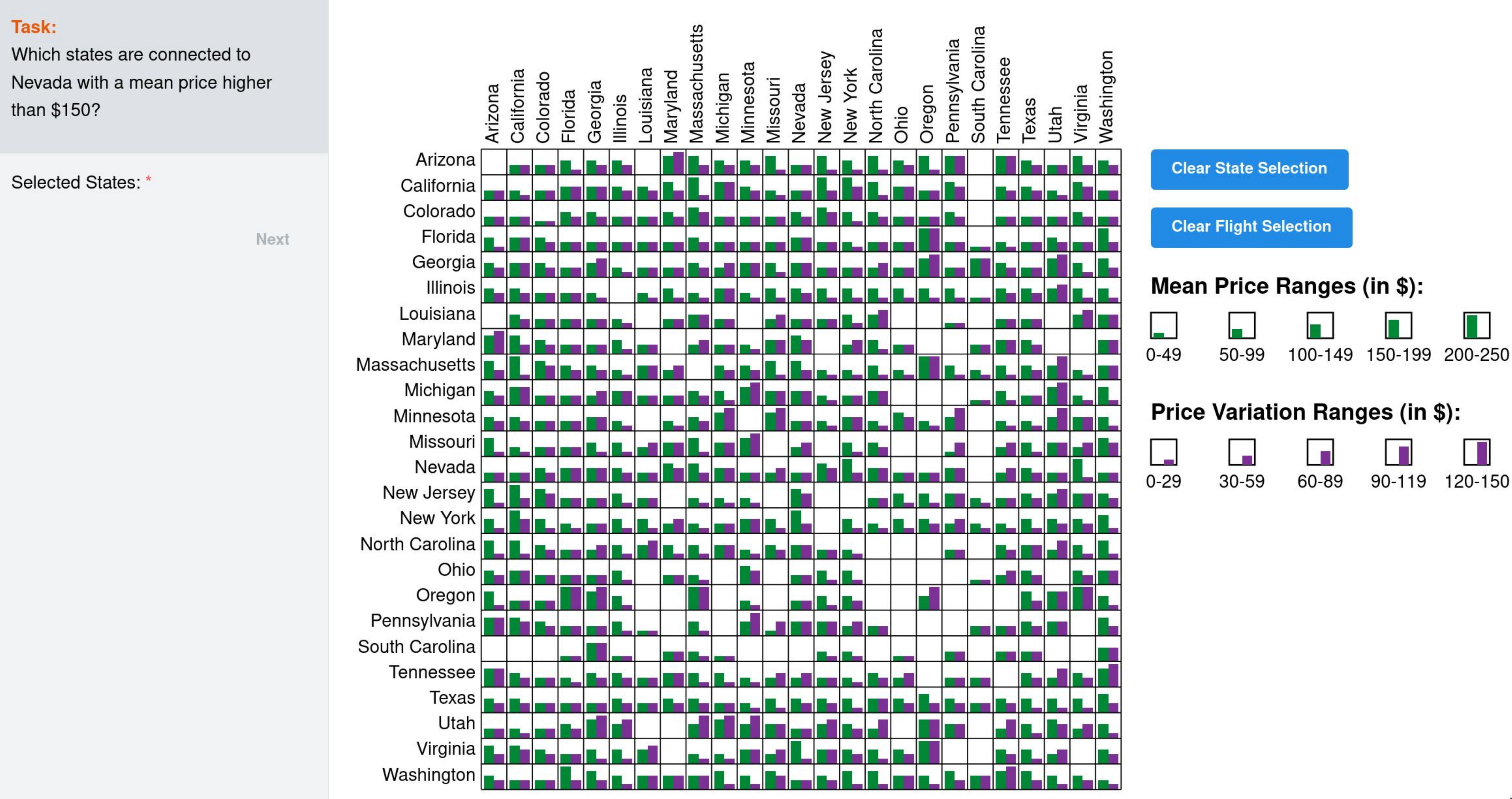}
\end{subfigure}
\vspace{1em}

\begin{subfigure}[b]{0.23\textwidth}
    \centering
    \includegraphics[width=\linewidth]{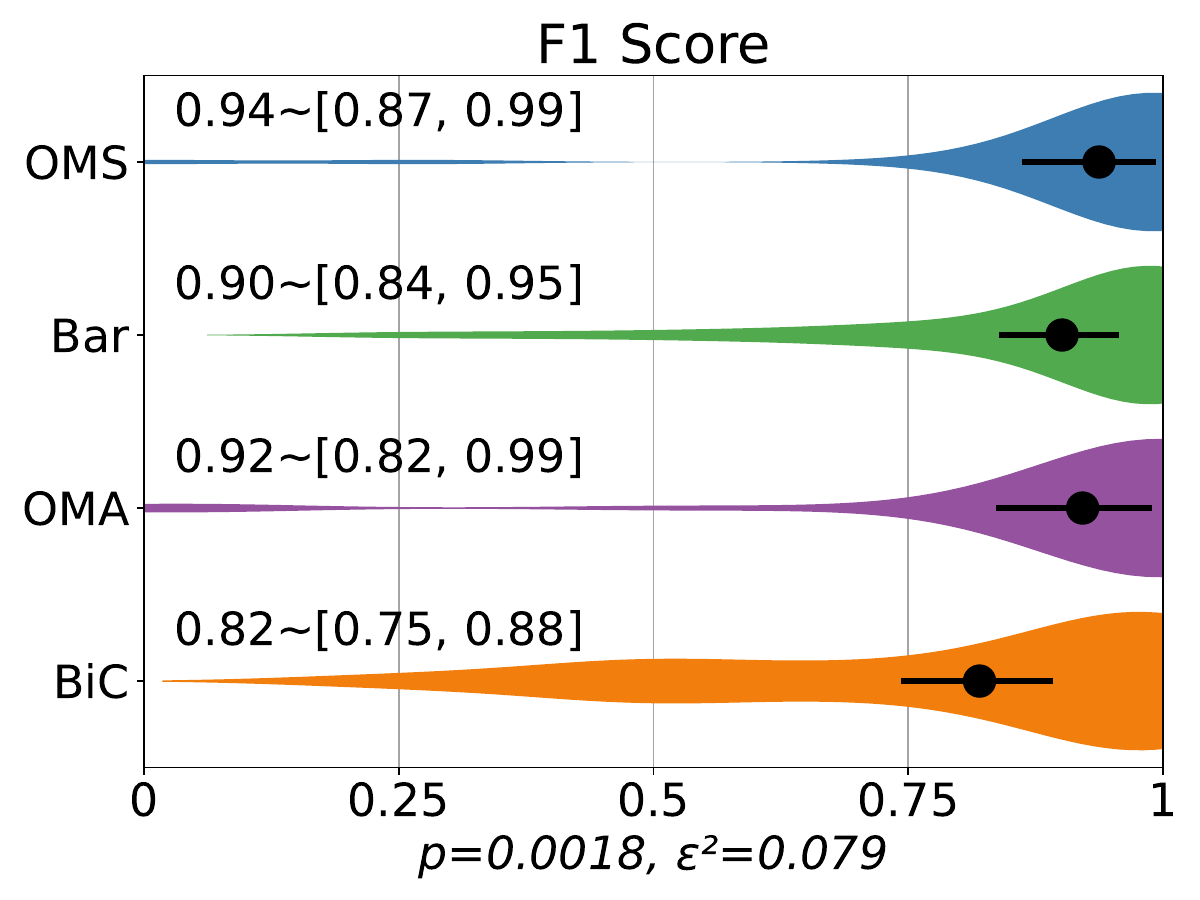}
\end{subfigure}
\hfill
\begin{subfigure}[b]{0.23\textwidth}
    \centering
    \includegraphics[width=\linewidth]{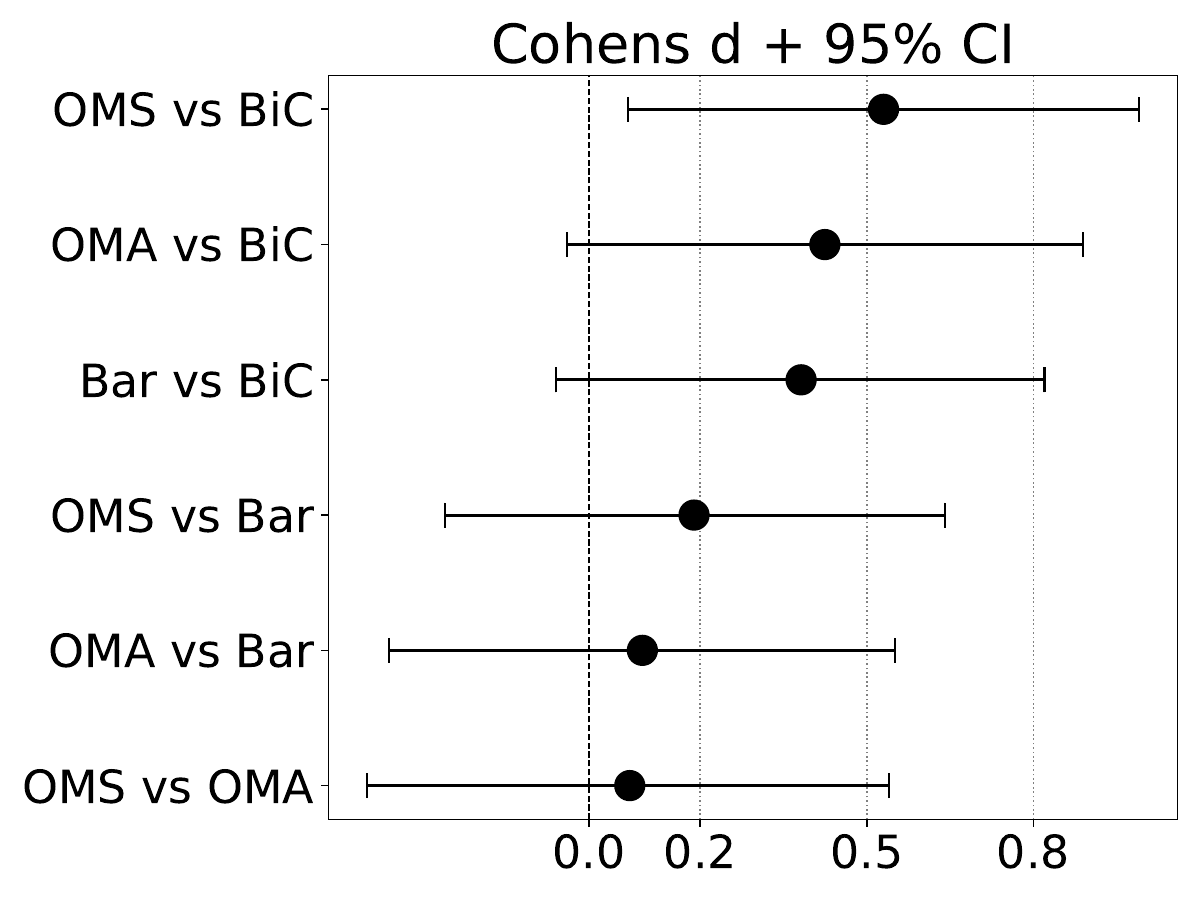}
\end{subfigure}
\hfill
\hspace{3em}
\begin{subfigure}[b]{0.23\textwidth}
    \centering
    \includegraphics[width=\linewidth]{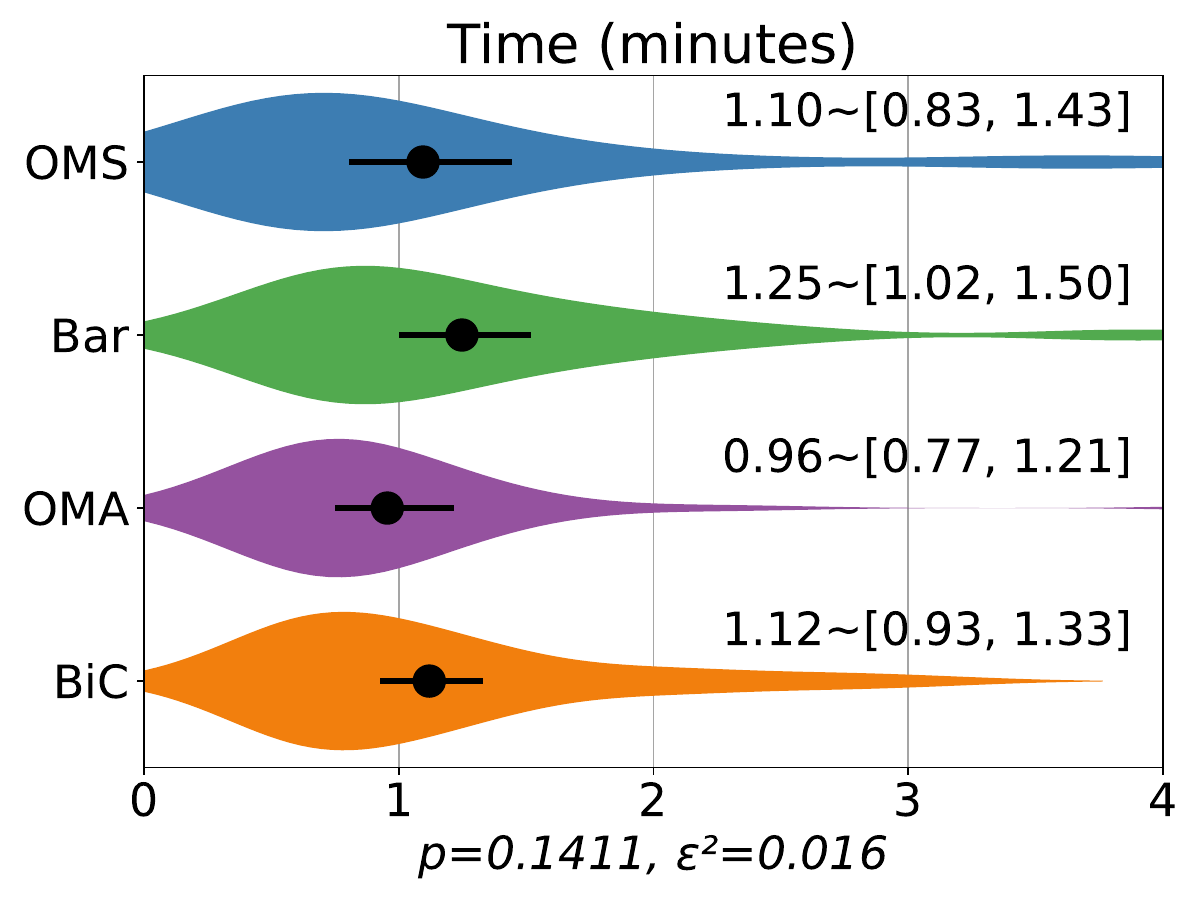}
\end{subfigure}
\hfill
\begin{subfigure}[b]{0.23\textwidth}
    \centering
    \includegraphics[width=\linewidth]{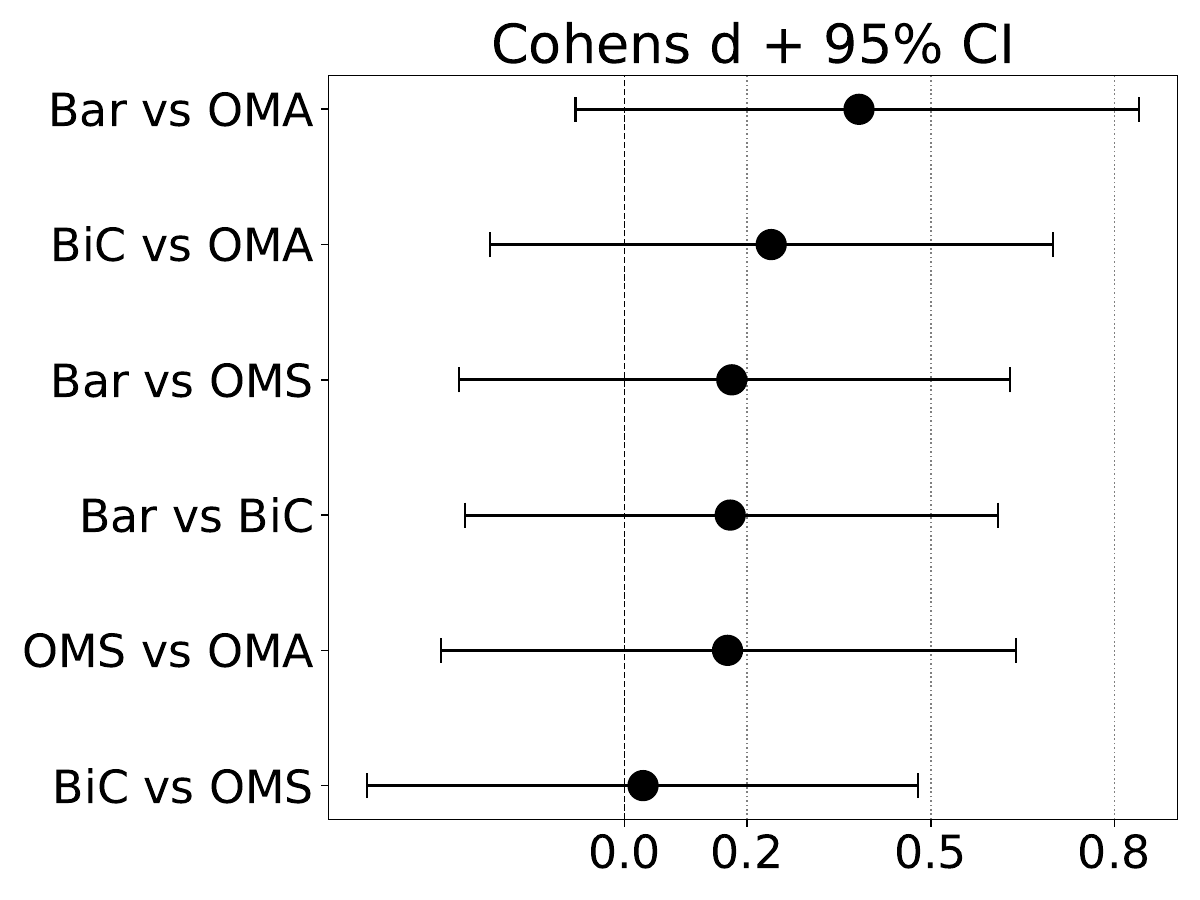}
\end{subfigure}

\caption{T2 - Task Prompt 1}
\label{fig:task2_1_app}
\end{figure*}

\begin{figure*}[hbt]
\centering

\begin{minipage}{\textwidth}
{\fontsize{20}{22}\selectfont\textbf {T2: Adjacency by Central Tendency - Task Prompt 2}} \label{sec:T2_2}\\

\textbf{Task Rationale:} Assess the effectiveness of the encodings to filter links when focusing on central tendency values. \\
\textbf{Hypothesis:} H1.2 and H1.3 \\
\textbf{Scoring:} F1-Score \\
\textbf{Correct Answer:} [
      "California",
      "Colorado",
      "Nevada",
      "Utah"
    ] \\
\textbf{Links:} \\
\inlinevis{bc} \url{https://jorgeacostaupm.github.io/revisit/matrices_barchart/dFpXVWhwWUNtSXRTRkRmcHFIb0NNZz09}, \\
\inlinevis{b} \url{https://jorgeacostaupm.github.io/revisit/matrices_bivariate/dFpXVWhwWUNtSXRTRkRmcHFIb0NNZz09}, \\
\inlinevis{oms} \url{https://jorgeacostaupm.github.io/revisit/matrices_size/dFpXVWhwWUNtSXRTRkRmcHFIb0NNZz09}, \\
\inlinevis{oma} \url{https://jorgeacostaupm.github.io/revisit/matrices_angle/dFpXVWhwWUNtSXRTRkRmcHFIb0NNZz09}
\end{minipage}

\vspace{1em}
\begin{subfigure}[b]{\linewidth}
    \includegraphics[width=\linewidth]{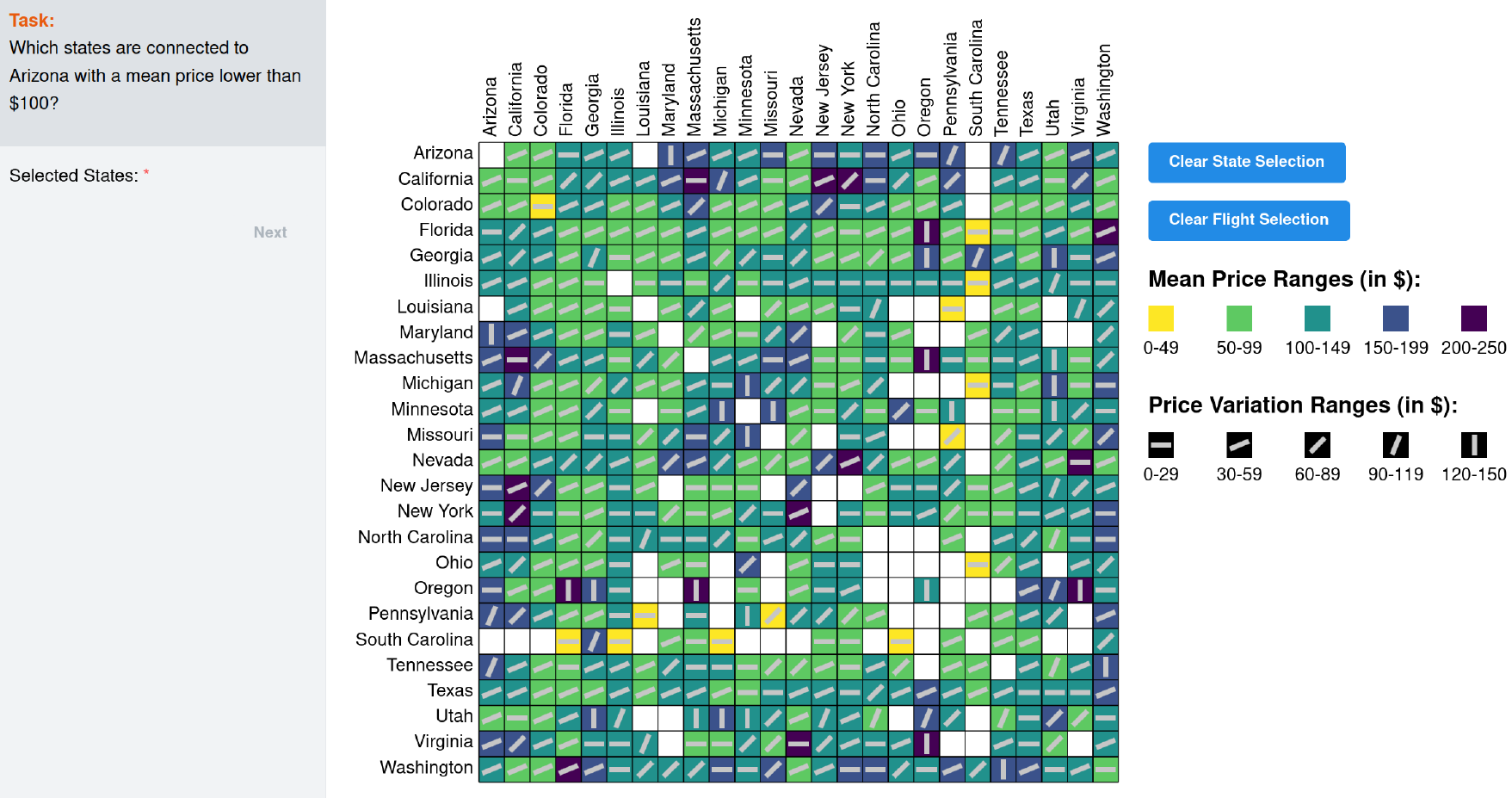}
\end{subfigure}
\vspace{1em}

\begin{subfigure}[b]{0.23\textwidth}
    \centering
    \includegraphics[width=\linewidth]{figs/violinplots/violin_f1_adj_mean.pdf}
\end{subfigure}
\hfill
\begin{subfigure}[b]{0.23\textwidth}
    \centering
    \includegraphics[width=\linewidth]{figs/forestplots/effects_f1_adj_mean.pdf}
\end{subfigure}
\hfill
\hspace{3em}
\begin{subfigure}[b]{0.23\textwidth}
    \centering
    \includegraphics[width=\linewidth]{figs/violinplots/violin_time_adj_mean.pdf}
\end{subfigure}
\hfill
\begin{subfigure}[b]{0.23\textwidth}
    \centering
    \includegraphics[width=\linewidth]{figs/forestplots/effects_time_adj_mean.pdf}
\end{subfigure}

\caption{T2 - Task Prompt 2}
\label{fig:task2_2_app}
\end{figure*}

\begin{figure*}[hbt]
\centering

\begin{minipage}{\textwidth}
{\fontsize{20}{22}\selectfont\textbf {T3: Adjacency by Dispersion - Task Prompt 1}} \label{sec:T3_1}\\

\textbf{Task Rationale:} Assess the effectiveness of the encodings to filter links when focusing on dispersion values. \\
\textbf{Hypothesis:} H1.2 and H1.3 \\
\textbf{Scoring:} F1-Score \\
\textbf{Correct Answer:} [
      "Georgia",
      "Illinois",
      "Michigan",
      "Massachusetts",
      "Minnesota",
      "New Jersey",
      "North Carolina",
      "Tennessee",
      "Oregon"
    ]\\
\textbf{Links:} \\
\inlinevis{bc} \url{https://jorgeacostaupm.github.io/revisit/matrices_barchart/UVltQ2dCbjF3RmNKTnlpZ0hmVXlBZz09}, \\
\inlinevis{b} \url{https://jorgeacostaupm.github.io/revisit/matrices_bivariate/UVltQ2dCbjF3RmNKTnlpZ0hmVXlBZz09}, \\
\inlinevis{oms} \url{https://jorgeacostaupm.github.io/revisit/matrices_size/UVltQ2dCbjF3RmNKTnlpZ0hmVXlBZz09}, \\
\inlinevis{oma} \url{https://jorgeacostaupm.github.io/revisit/matrices_angle/UVltQ2dCbjF3RmNKTnlpZ0hmVXlBZz09}

\vspace{1em}
\end{minipage}
\begin{subfigure}[b]{\linewidth}
    \includegraphics[width=\linewidth]{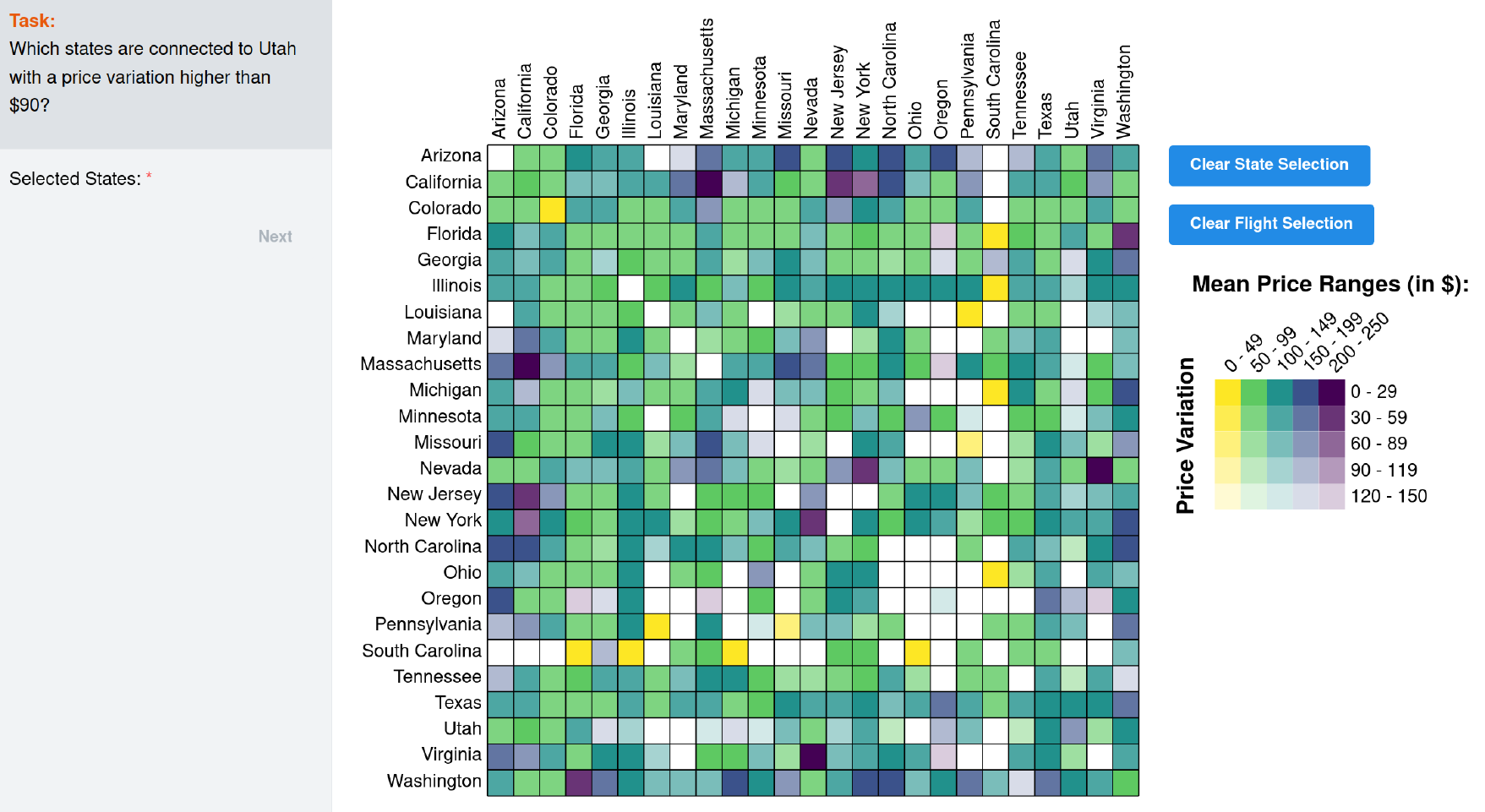}
\end{subfigure}
\vspace{1em}

\begin{subfigure}[b]{0.23\textwidth}
    \centering
    \includegraphics[width=\linewidth]{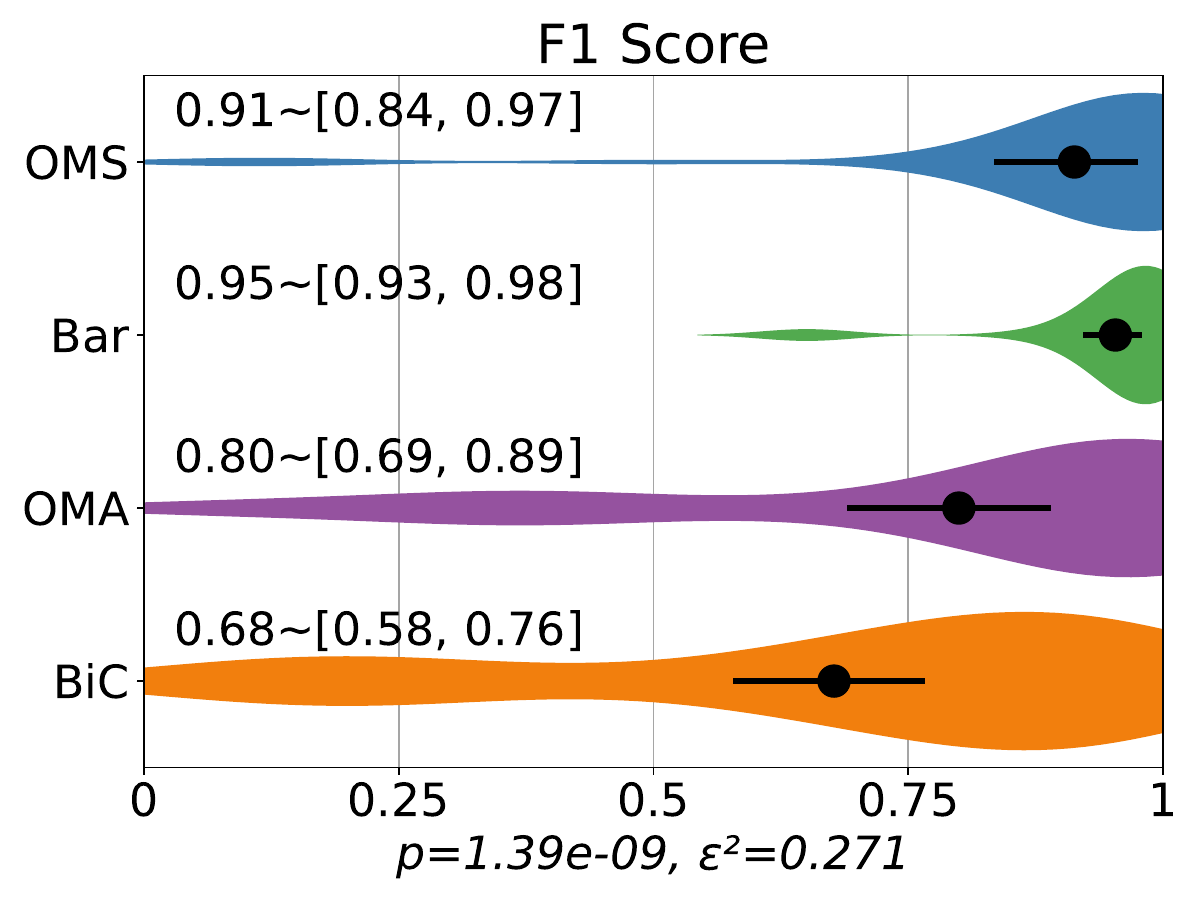}
\end{subfigure}
\hfill
\begin{subfigure}[b]{0.23\textwidth}
    \centering
    \includegraphics[width=\linewidth]{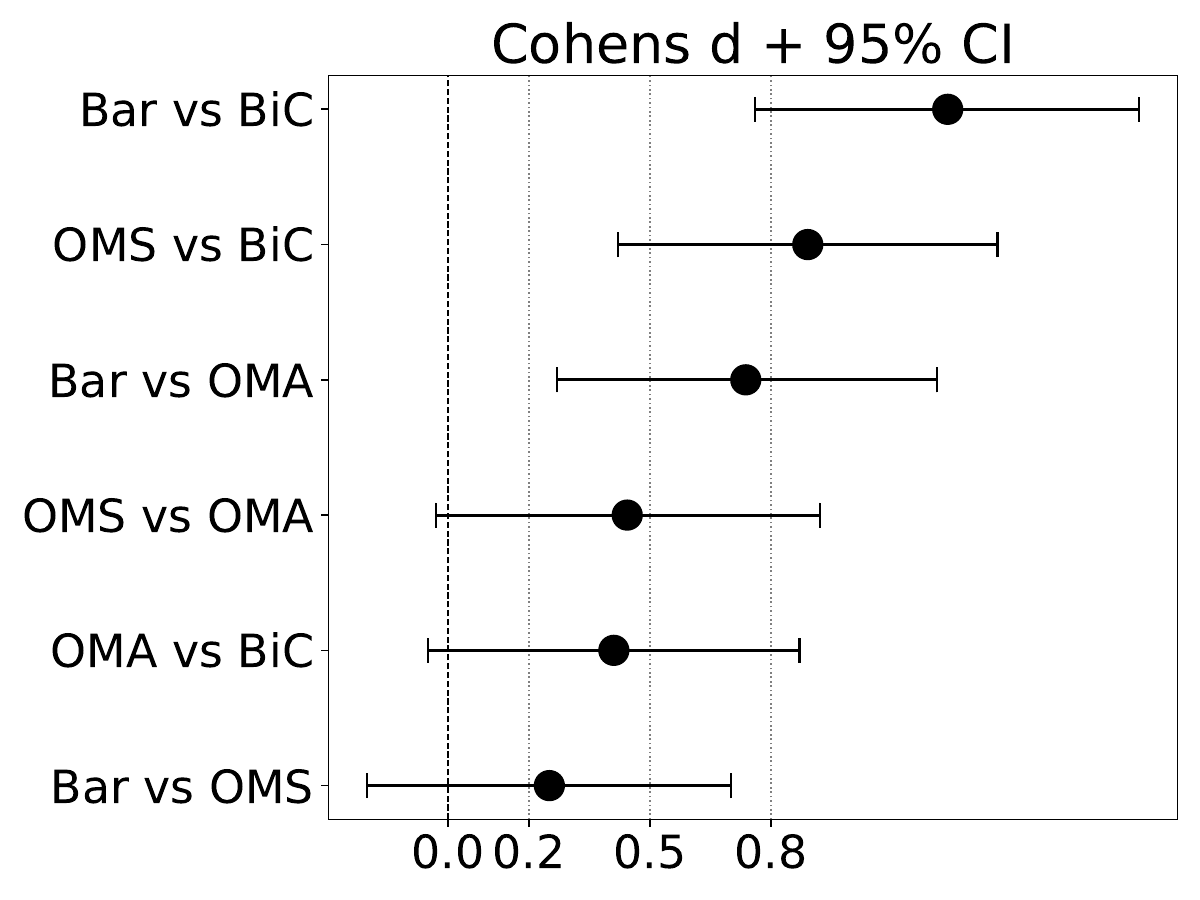}
\end{subfigure}
\hfill
\hspace{3em}
\begin{subfigure}[b]{0.23\textwidth}
    \centering
    \includegraphics[width=\linewidth]{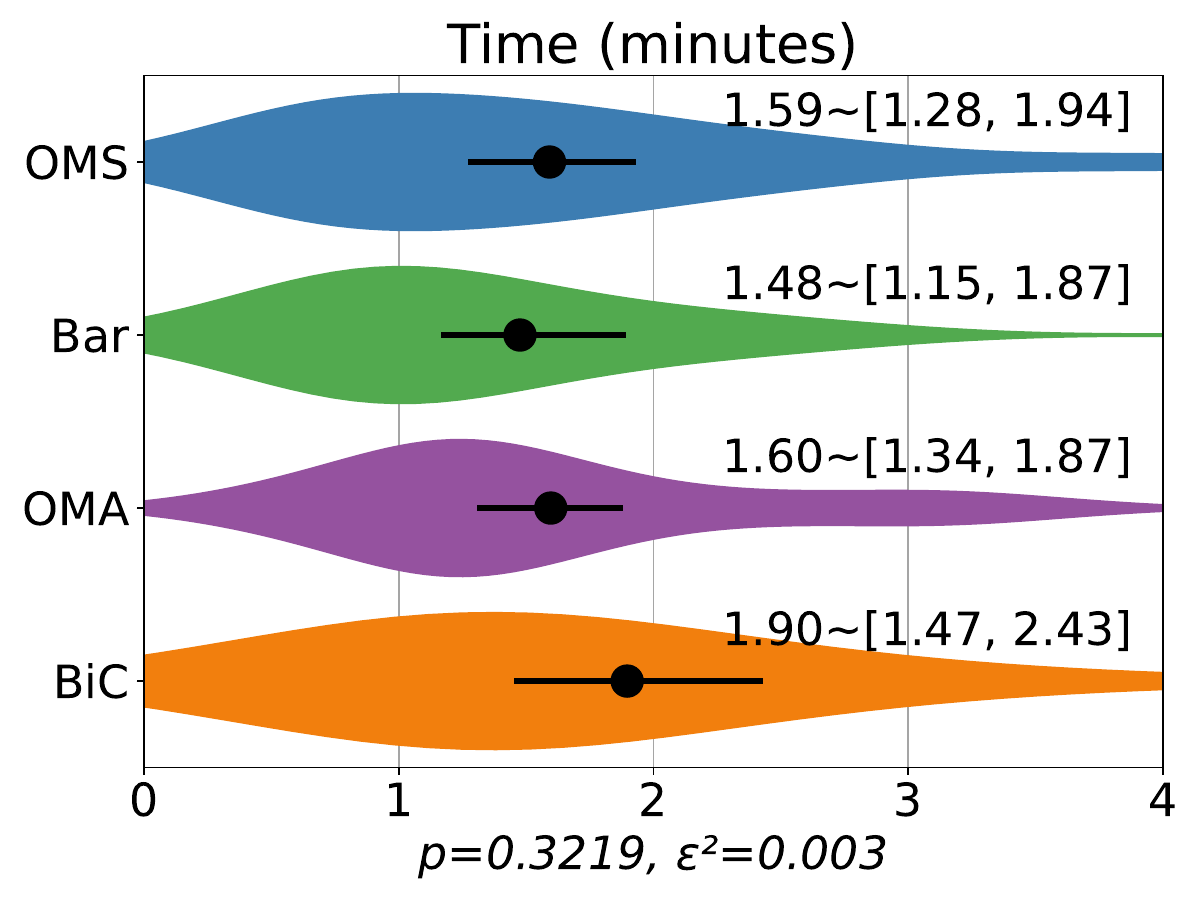}
\end{subfigure}
\hfill
\begin{subfigure}[b]{0.23\textwidth}
    \centering
    \includegraphics[width=\linewidth]{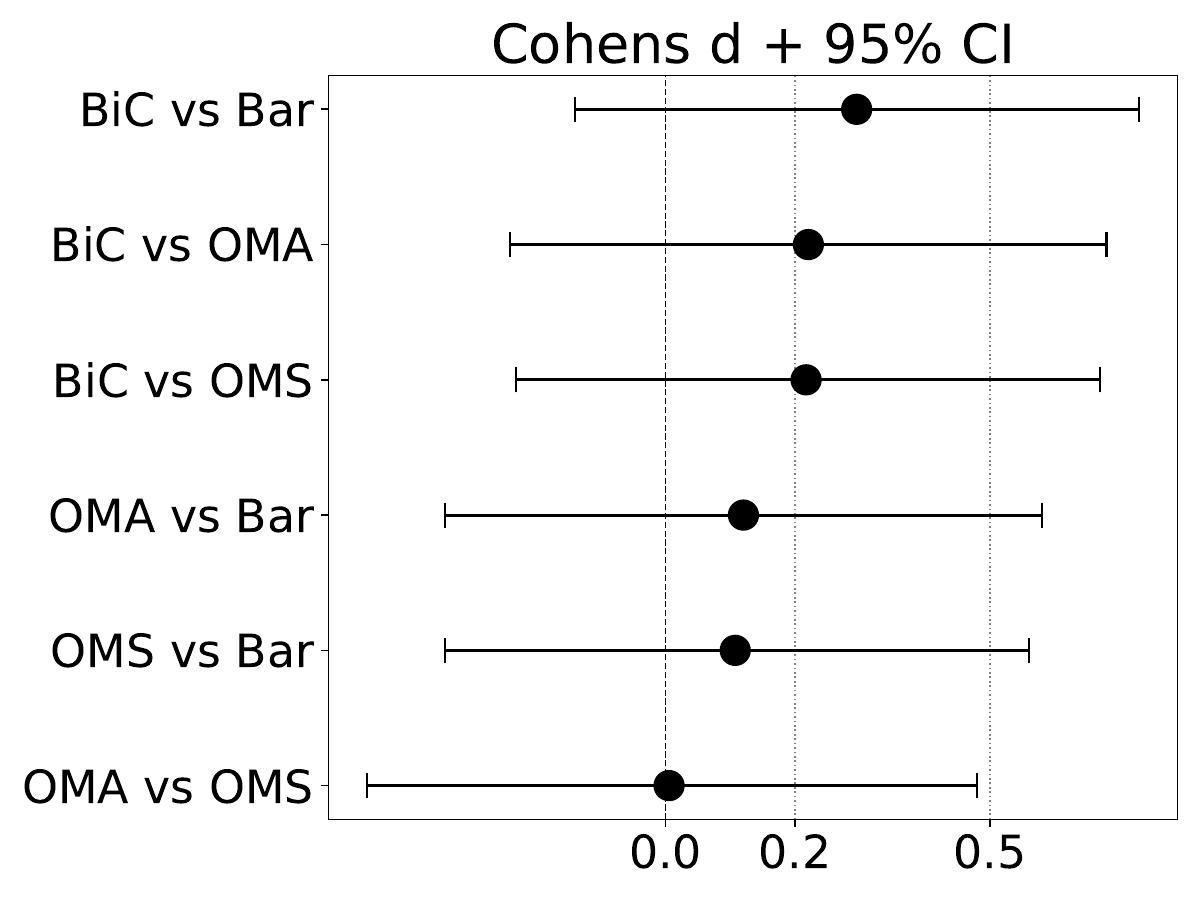}
\end{subfigure}

\caption{T3 - Task Prompt 1}
\label{fig:task3_1_app}
\end{figure*}

\begin{figure*}[hbt]
\centering

\begin{minipage}{\textwidth}
{\fontsize{20}{22}\selectfont\textbf {T3: Adjacency by Dispersion - Task Prompt 2}} \label{sec:T3_2}\\

\textbf{Task Rationale:} Assess the effectiveness of the encodings to filter links when focusing on dispersion values. \\
\textbf{Hypothesis:} H1.2 and H1.3 \\
\textbf{Scoring:} F1-Score \\
\textbf{Correct Answer:} [
      "Washington",
      "Colorado",
      "Florida",
      "Illinois",
      "Georgia",
      "Louisiana",
      "Massachusetts",
      "North Carolina",
      "South Carolina",
      "Tennessee",
      "Texas"
    ]\\
\textbf{Links:} \\
\inlinevis{bc} \url{https://jorgeacostaupm.github.io/revisit/matrices_barchart/MTU1S3V4ZDZiR0VVdVBENVArTStjUT09}, \\
\inlinevis{b} \url{https://jorgeacostaupm.github.io/revisit/matrices_bivariate/MTU1S3V4ZDZiR0VVdVBENVArTStjUT09}, \\
\inlinevis{oms} \url{https://jorgeacostaupm.github.io/revisit/matrices_size/MTU1S3V4ZDZiR0VVdVBENVArTStjUT09}, \\
\inlinevis{oma} \url{https://jorgeacostaupm.github.io/revisit/matrices_angle/MTU1S3V4ZDZiR0VVdVBENVArTStjUT09}
\end{minipage}

\vspace{1em}
\begin{subfigure}[b]{\linewidth}
    \includegraphics[width=\linewidth]{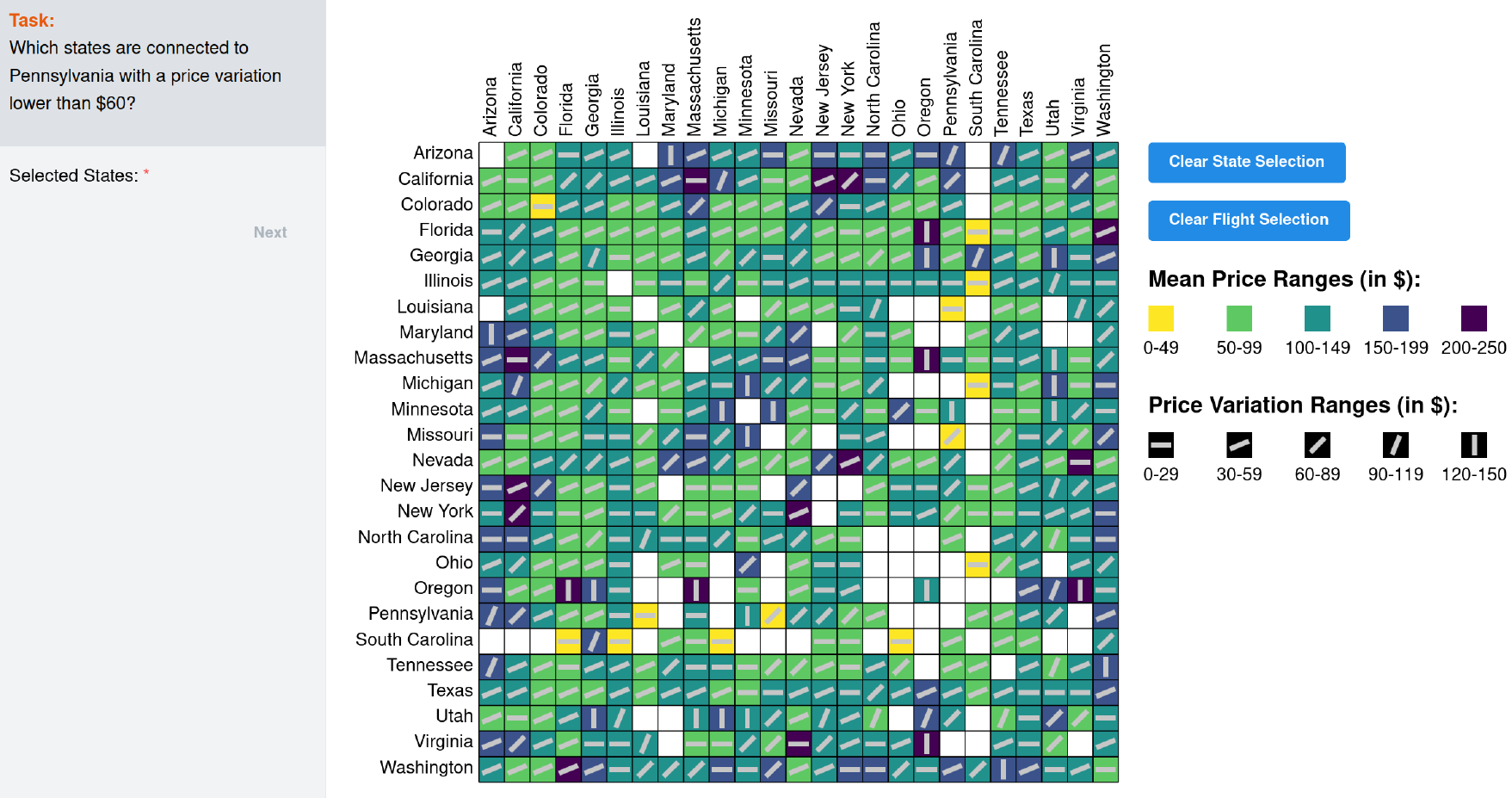}
\end{subfigure}
\vspace{1em}

\begin{subfigure}[b]{0.23\textwidth}
    \centering
    \includegraphics[width=\linewidth]{figs/violinplots/violin_f1_adj_var.pdf}
\end{subfigure}
\hfill
\begin{subfigure}[b]{0.23\textwidth}
    \centering
    \includegraphics[width=\linewidth]{figs/forestplots/effects_f1_adj_var.pdf}
\end{subfigure}
\hfill
\hspace{3em}
\begin{subfigure}[b]{0.23\textwidth}
    \centering
    \includegraphics[width=\linewidth]{figs/violinplots/violin_time_adj_var.pdf}
\end{subfigure}
\hfill
\begin{subfigure}[b]{0.23\textwidth}
    \centering
    \includegraphics[width=\linewidth]{figs/forestplots/effects_time_adj_var.pdf}
\end{subfigure}

\caption{Task 3 - Task Prompt 2}
\label{fig:task3_2_app}
\end{figure*}

\begin{figure*}[hbt]
\centering

\begin{minipage}{\textwidth}
{\fontsize{20}{22}\selectfont\textbf {T4: Adjacency by Attribute Combination - Task Prompt 1}} \label{sec:T4_1}\\

\textbf{Task Rationale:} Assess the effectiveness of the encodings when combining attributes to filter links. \\
\textbf{Hypothesis:} H1.2 and H1.3 \\
\textbf{Scoring:} F1-Score \\
\textbf{Correct Answer:} \\
\textbf{Links:} \\
\inlinevis{bc} \url{https://jorgeacostaupm.github.io/revisit/matrices_barchart/WXo1cnRDTHVZaElFbTdmMGtjU252QT09}, \\
\inlinevis{b} \url{https://jorgeacostaupm.github.io/revisit/matrices_bivariate/WXo1cnRDTHVZaElFbTdmMGtjU252QT09}, \\
\inlinevis{oms} \url{https://jorgeacostaupm.github.io/revisit/matrices_size/WXo1cnRDTHVZaElFbTdmMGtjU252QT09}, \\
\inlinevis{oma} \url{https://jorgeacostaupm.github.io/revisit/matrices_angle/WXo1cnRDTHVZaElFbTdmMGtjU252QT09}

\vspace{1em}
\end{minipage}
\begin{subfigure}[b]{\linewidth}
    \includegraphics[width=\linewidth]{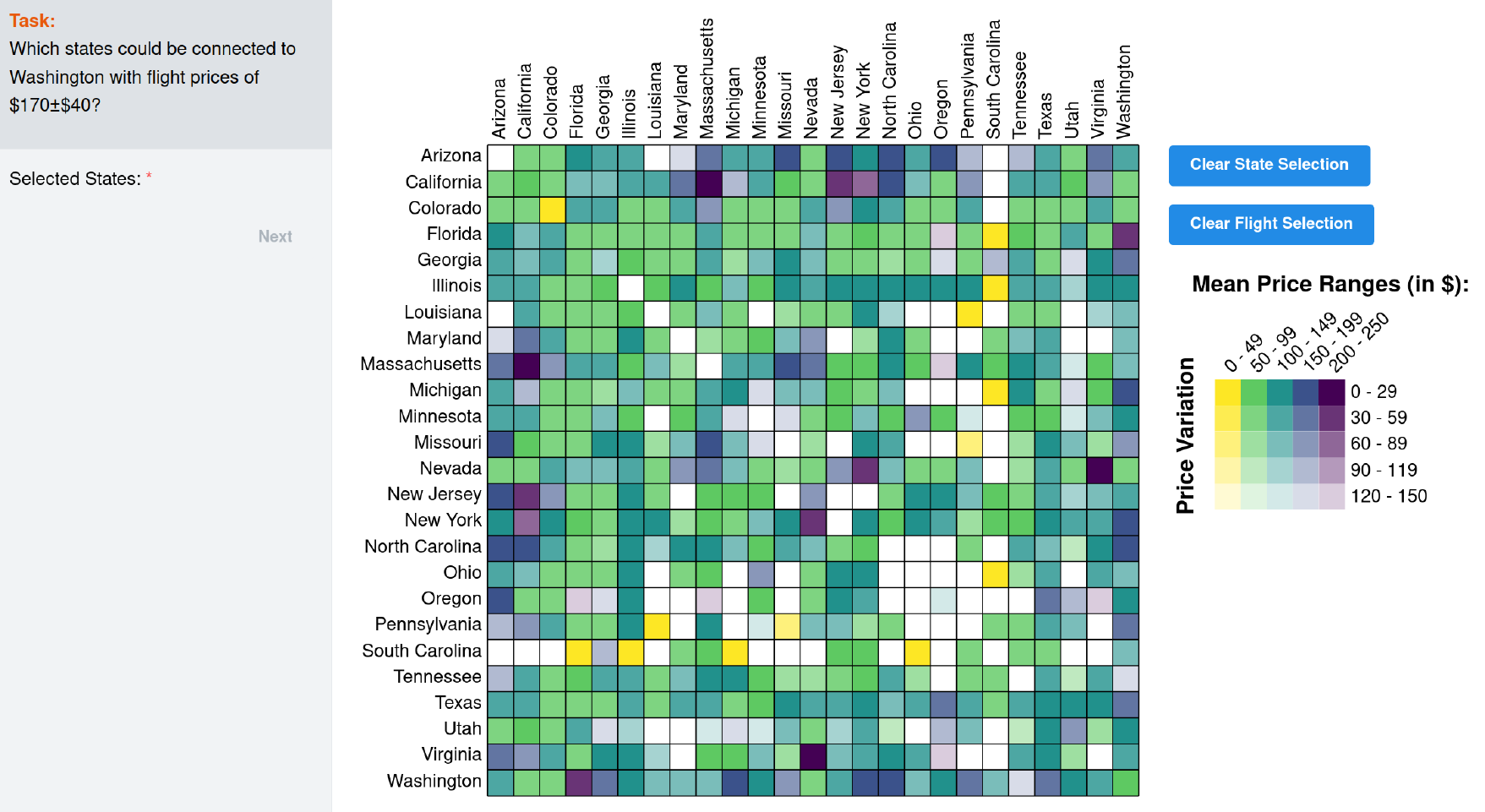}
\end{subfigure}
\vspace{1em}

\begin{subfigure}[b]{0.23\textwidth}
    \centering
    \includegraphics[width=\linewidth]{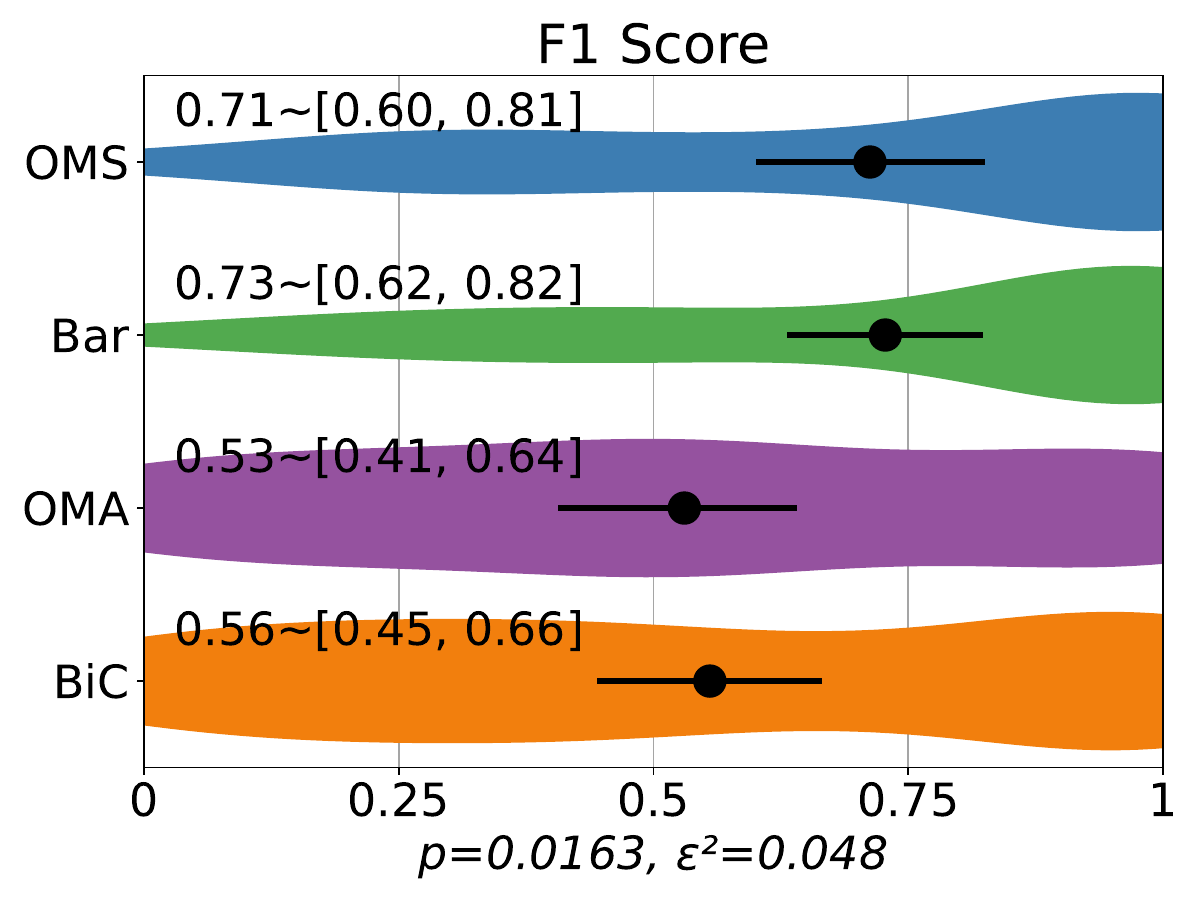}
\end{subfigure}
\hfill
\begin{subfigure}[b]{0.23\textwidth}
    \centering
    \includegraphics[width=\linewidth]{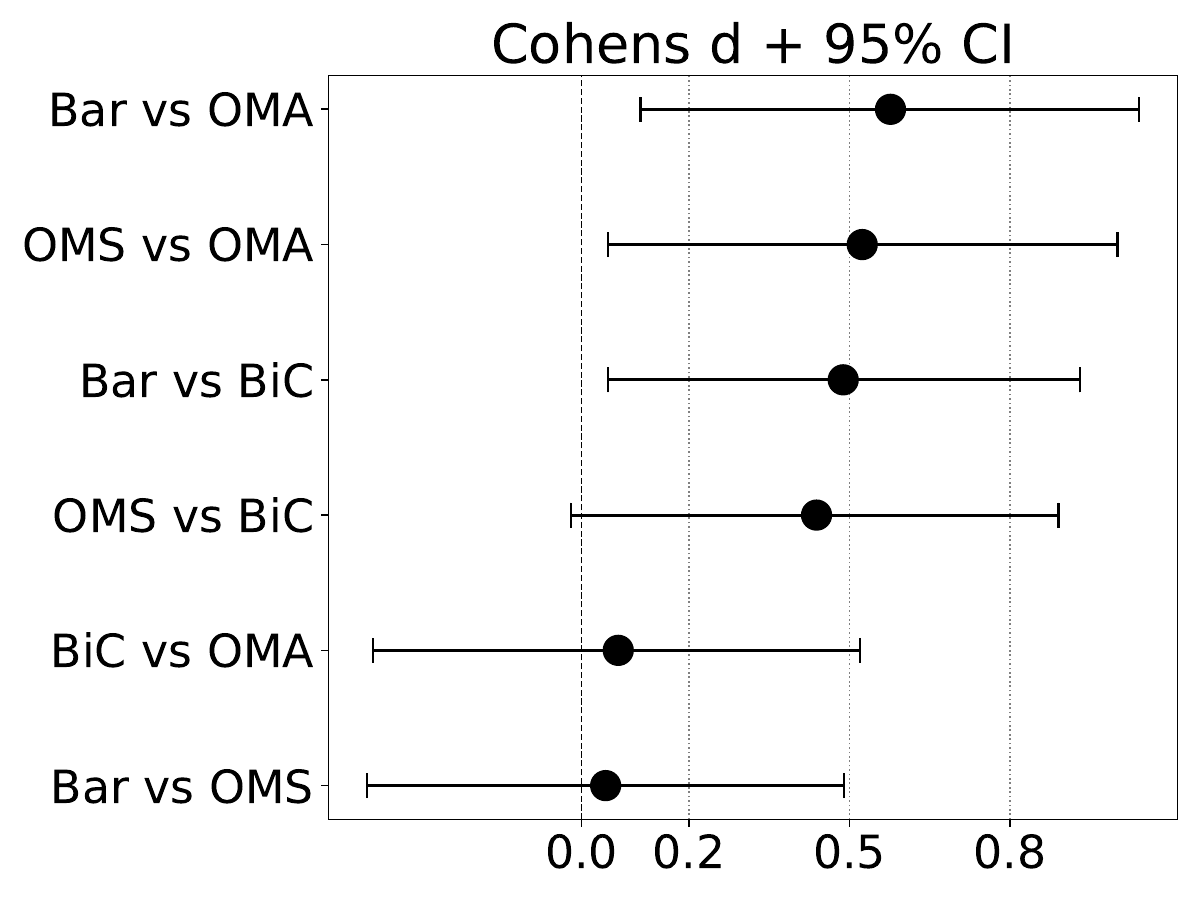}
\end{subfigure}
\hfill
\hspace{3em}
\begin{subfigure}[b]{0.23\textwidth}
    \centering
    \includegraphics[width=\linewidth]{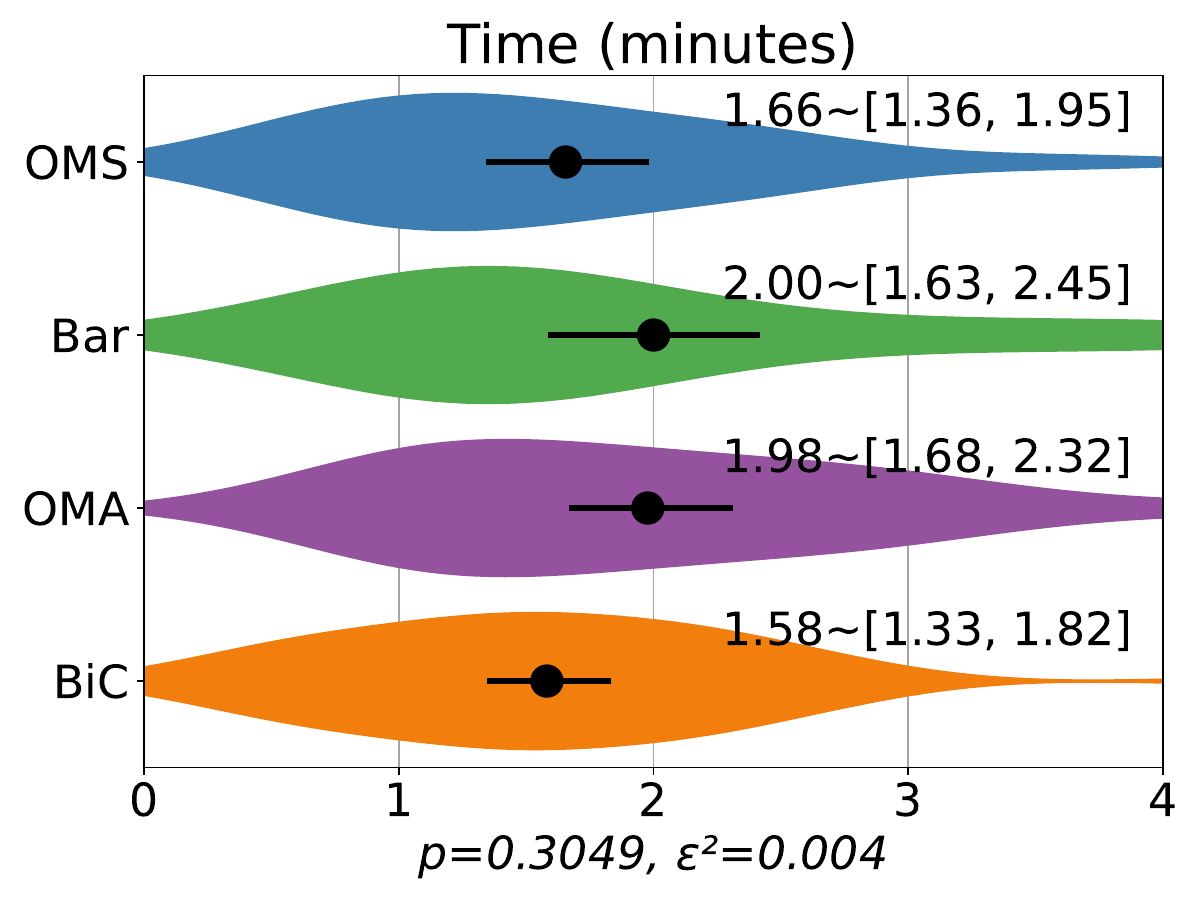}
\end{subfigure}
\hfill
\begin{subfigure}[b]{0.23\textwidth}
    \centering
    \includegraphics[width=\linewidth]{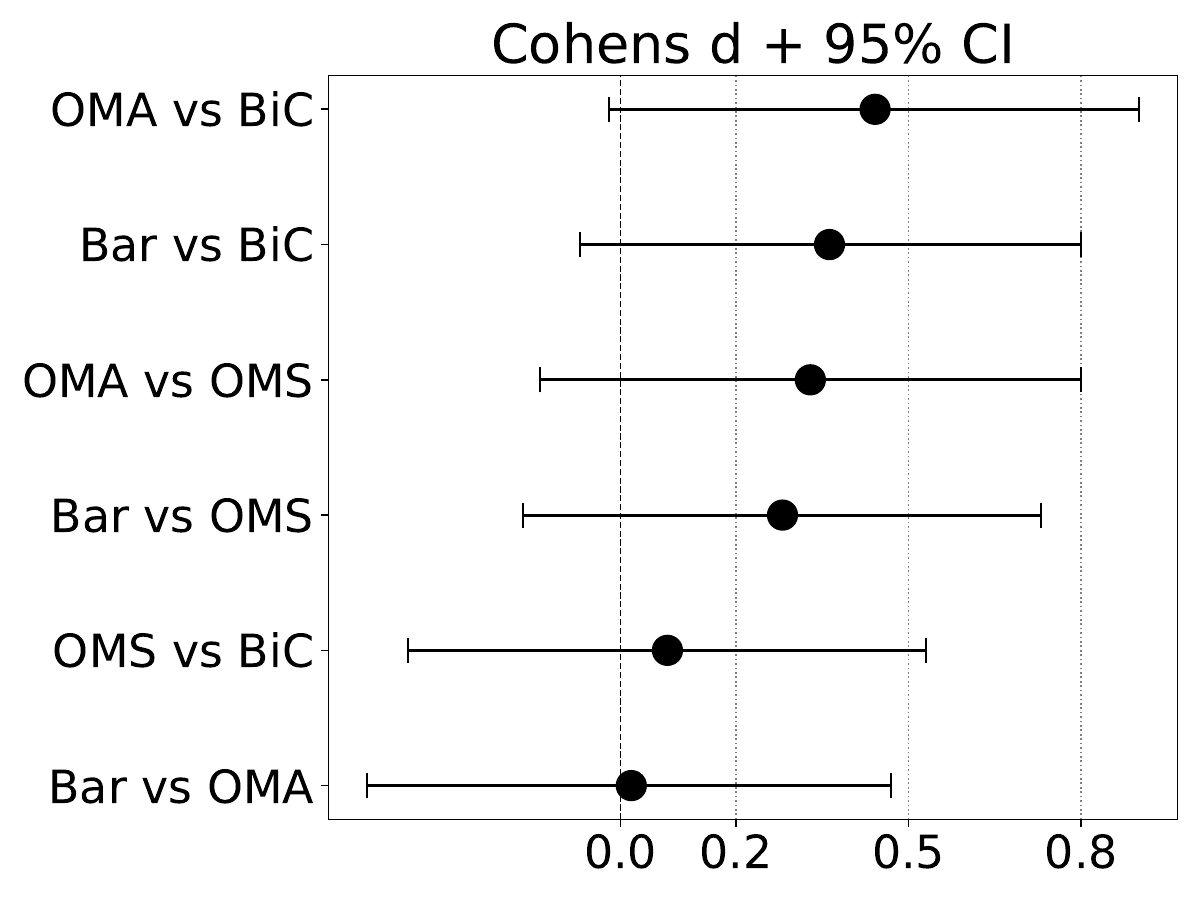}
\end{subfigure}

\caption{T4 - Task Prompt 1}
\label{fig:task4_1_app}
\end{figure*}

\begin{figure*}[hbt]
\centering

\begin{minipage}{\textwidth}
{\fontsize{20}{22}\selectfont\textbf {T4: Adjacency by Attribute Combination - Task Prompt 2}} \label{sec:T4_2}\\

\textbf{Task Rationale:} Assess the effectiveness of the encodings when combining attributes to filter links. \\
\textbf{Hypothesis:} H1.2 and H1.3 \\
\textbf{Scoring:} F1-Score \\
\textbf{Correct Answer:} [
      "Tennessee",
      "Washington",
      "Missouri"
    ]\\
\textbf{Links:} \\
\inlinevis{bc} \url{https://jorgeacostaupm.github.io/revisit/matrices_barchart/b2tnVmUzTVZXMGhxZThBTW1DblNGZz09}, \\
\inlinevis{b} \url{https://jorgeacostaupm.github.io/revisit/matrices_bivariate/b2tnVmUzTVZXMGhxZThBTW1DblNGZz09}, \\
\inlinevis{oms} \url{https://jorgeacostaupm.github.io/revisit/matrices_size/b2tnVmUzTVZXMGhxZThBTW1DblNGZz09}, \\
\inlinevis{oma} \url{https://jorgeacostaupm.github.io/revisit/matrices_angle/b2tnVmUzTVZXMGhxZThBTW1DblNGZz09}

\vspace{1em}
\end{minipage}
\begin{subfigure}[b]{\linewidth}
    \includegraphics[width=\linewidth]{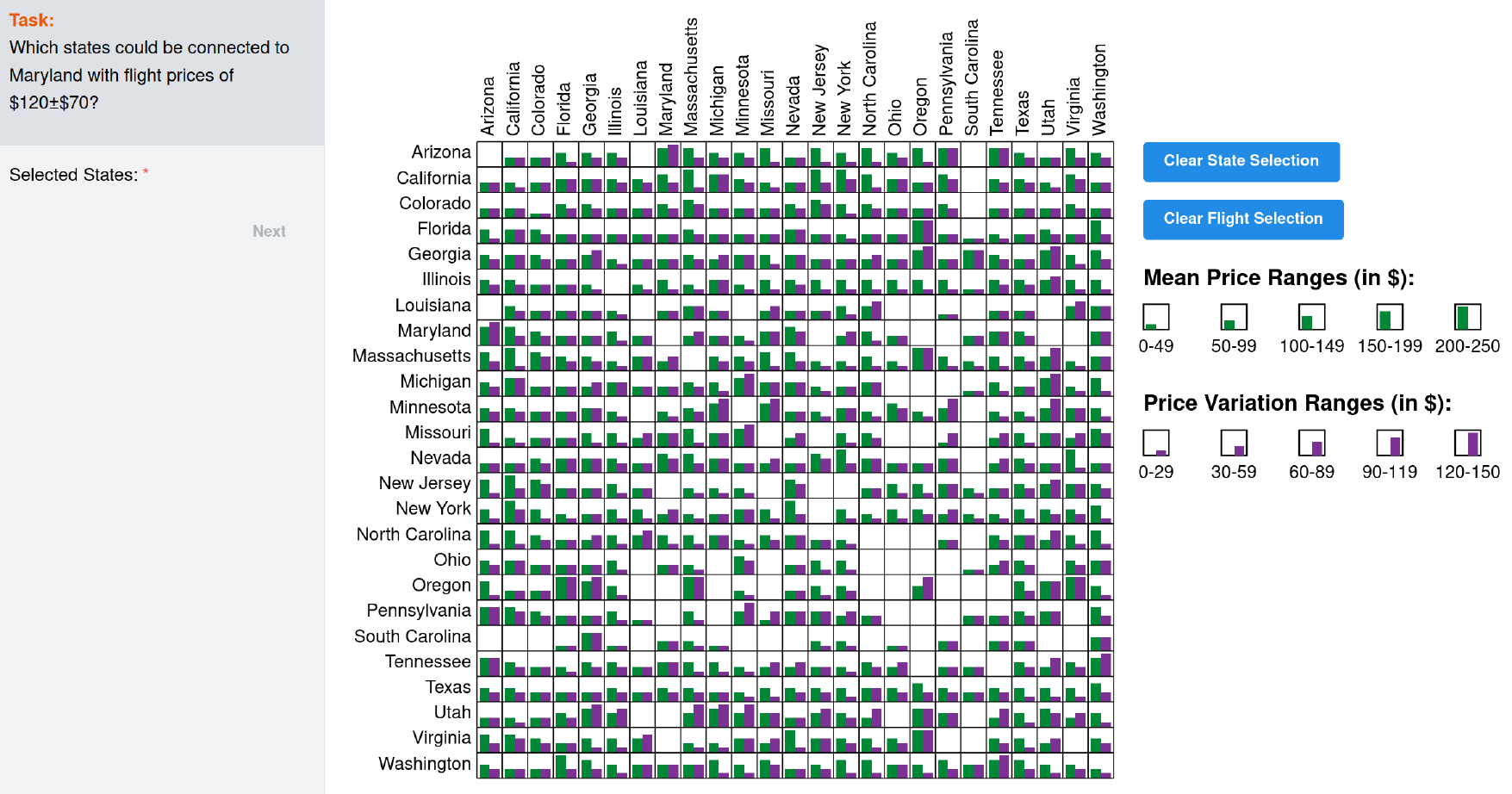}
\end{subfigure}
\vspace{1em}

\begin{subfigure}[b]{0.23\textwidth}
    \centering
    \includegraphics[width=\linewidth]{figs/violinplots/violin_f1_attr_comb.pdf}
\end{subfigure}
\hfill
\begin{subfigure}[b]{0.23\textwidth}
    \centering
    \includegraphics[width=\linewidth]{figs/forestplots/effects_f1_attr_comb.pdf}
\end{subfigure}
\hfill
\hspace{3em}
\begin{subfigure}[b]{0.23\textwidth}
    \centering
    \includegraphics[width=\linewidth]{figs/violinplots/violin_time_attr_comb.pdf}
\end{subfigure}
\hfill
\begin{subfigure}[b]{0.23\textwidth}
    \centering
    \includegraphics[width=\linewidth]{figs/forestplots/effects_time_attr_comb.pdf}
\end{subfigure}

\caption{T4 - Task Prompt 2}
\label{fig:task4_2_app}
\end{figure*}

\begin{figure*}[hbt]
\centering

\begin{minipage}{\textwidth}
{\fontsize{20}{22}\selectfont\textbf {T5: Extreme Detection - Task Prompt 1}} \label{sec:T5_1}\\

\textbf{Task Rationale:} Assess the effectiveness of the encodings to detect extreme attribute values.\\
\textbf{Hypothesis:} H1.2 and H1.3 \\
\textbf{Scoring:} F1-Score \\
\textbf{Correct Answer:} [
      "Minnesota",
      "Pennsylvania"
    ]\\
\textbf{Links:} \\
\inlinevis{bc} \url{https://jorgeacostaupm.github.io/revisit/matrices_barchart/SEVFVUNGaWFCVEhiMU55ajRqczliUT09}, \\
\inlinevis{b} \url{https://jorgeacostaupm.github.io/revisit/matrices_bivariate/SEVFVUNGaWFCVEhiMU55ajRqczliUT09}, \\
\inlinevis{oms} \url{https://jorgeacostaupm.github.io/revisit/matrices_size/SEVFVUNGaWFCVEhiMU55ajRqczliUT09}, \\
\inlinevis{oma} \url{https://jorgeacostaupm.github.io/revisit/matrices_angle/SEVFVUNGaWFCVEhiMU55ajRqczliUT09}
\end{minipage}

\vspace{1em}
\begin{subfigure}[b]{\linewidth}
    \includegraphics[width=\linewidth]{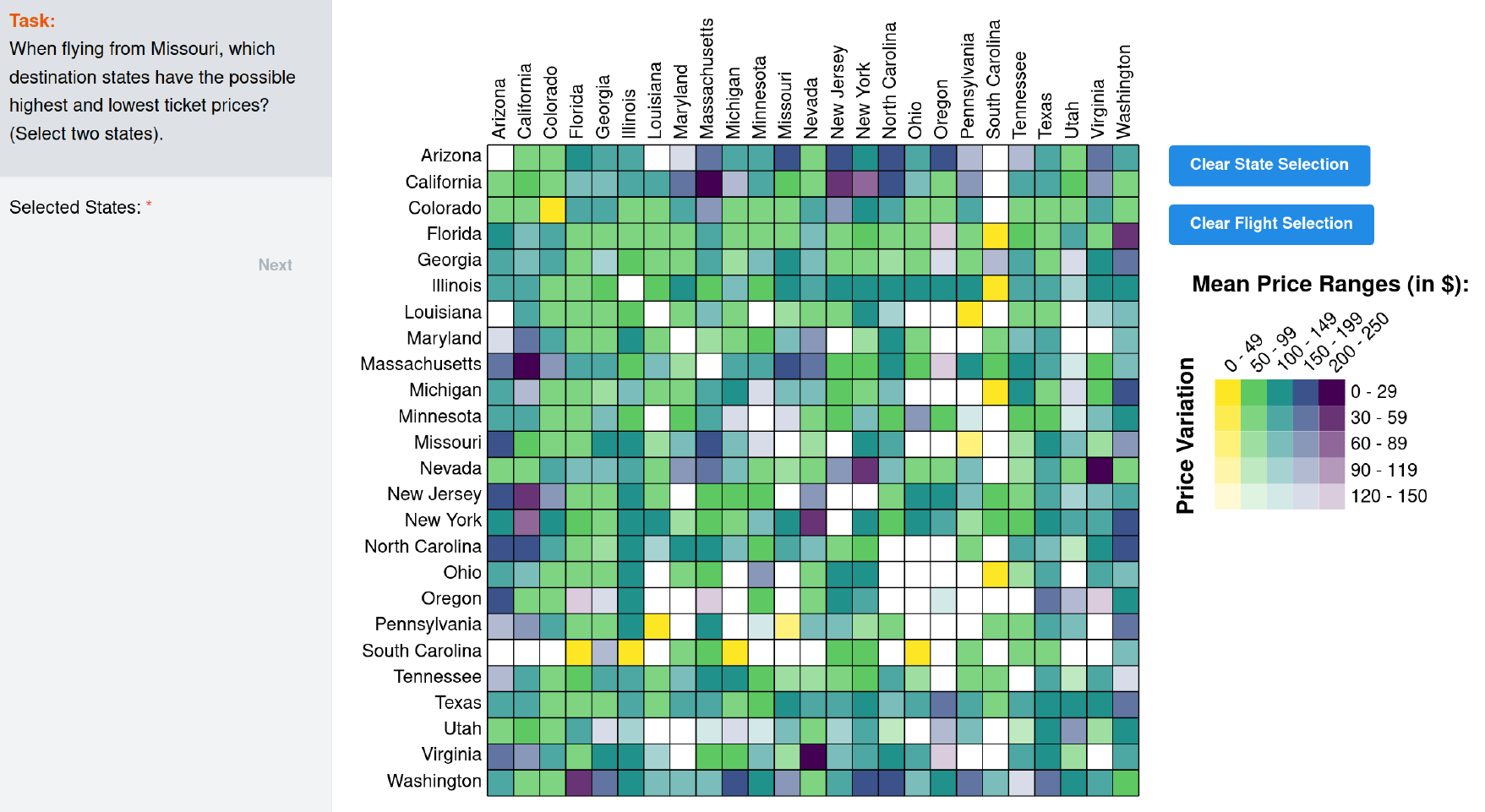}
\end{subfigure}
\vspace{1em}

\begin{subfigure}[b]{0.23\textwidth}
    \centering
    \includegraphics[width=\linewidth]{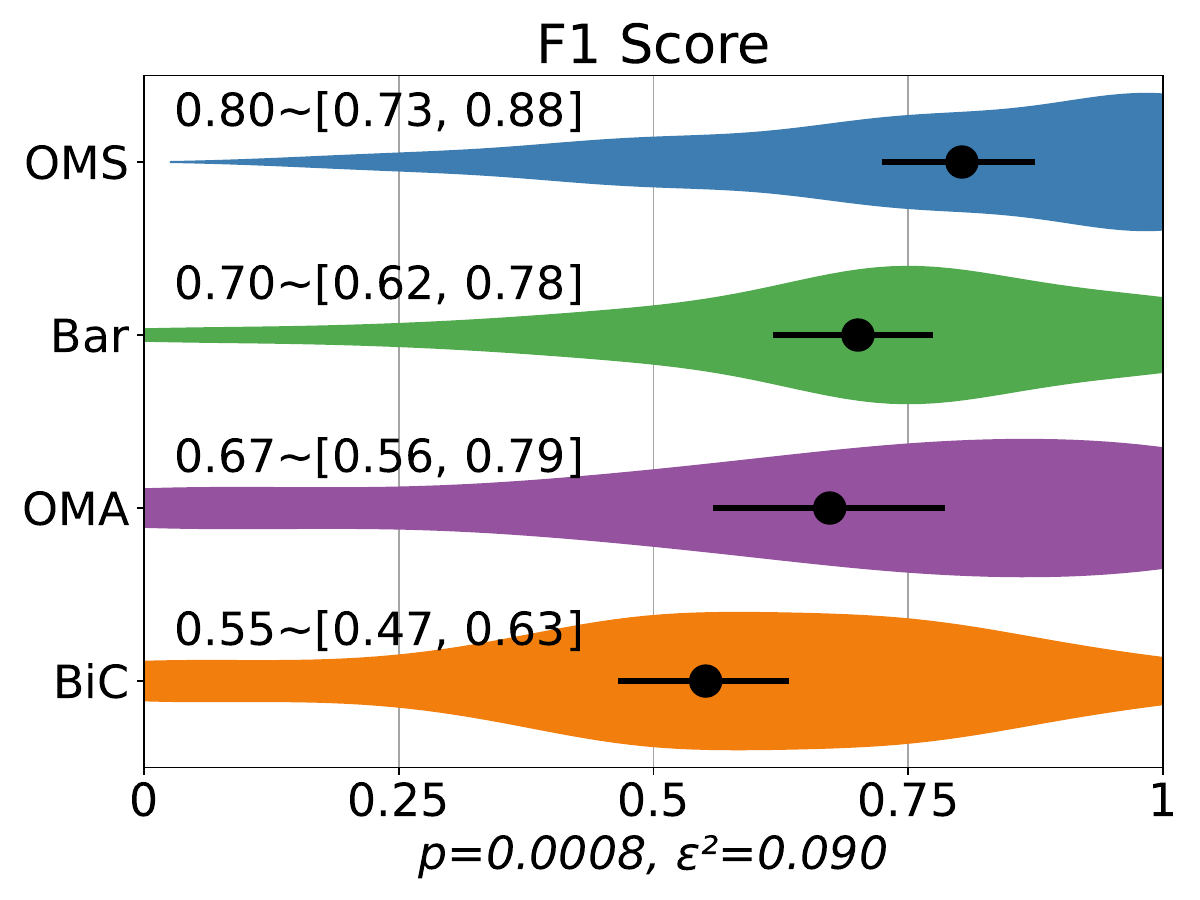}
\end{subfigure}
\hfill
\begin{subfigure}[b]{0.23\textwidth}
    \centering
    \includegraphics[width=\linewidth]{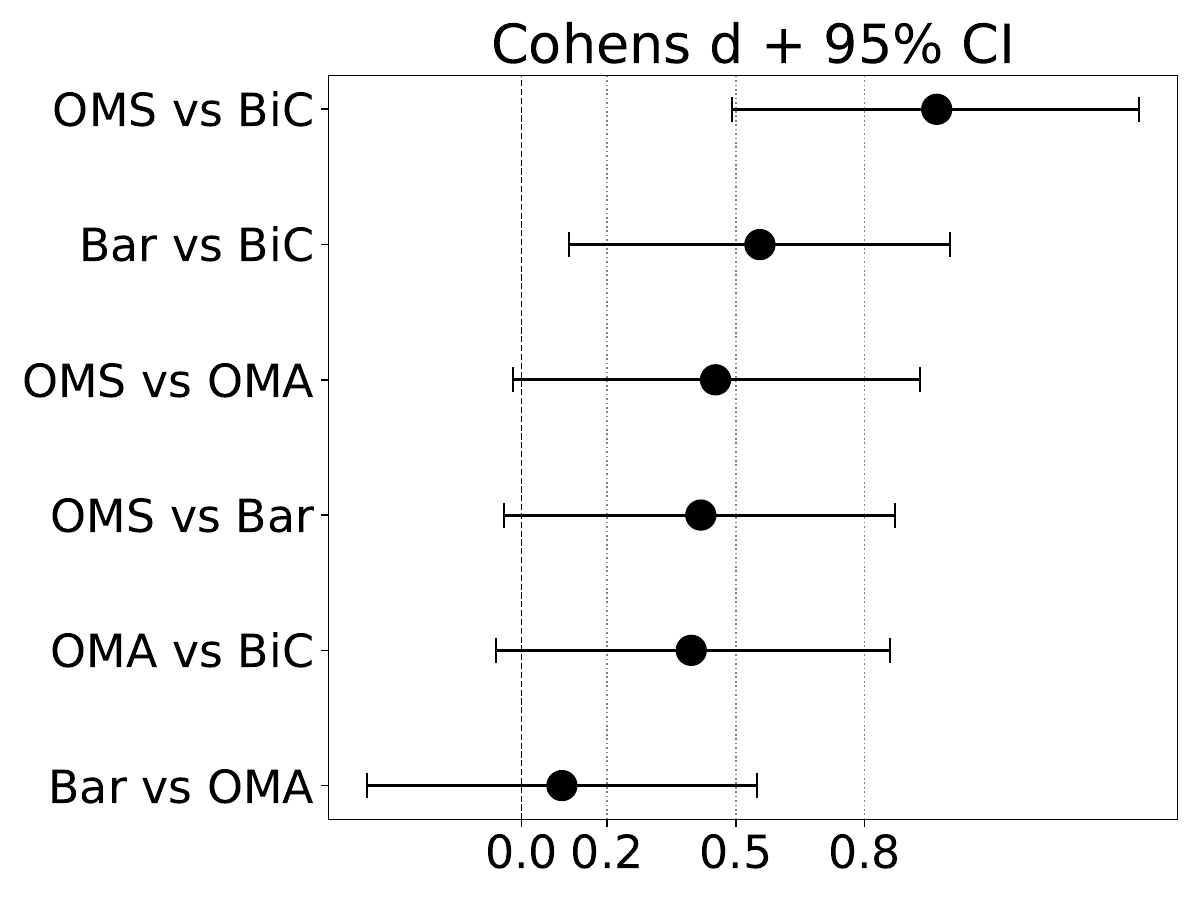}
\end{subfigure}
\hfill
\hspace{3em}
\begin{subfigure}[b]{0.23\textwidth}
    \centering
    \includegraphics[width=\linewidth]{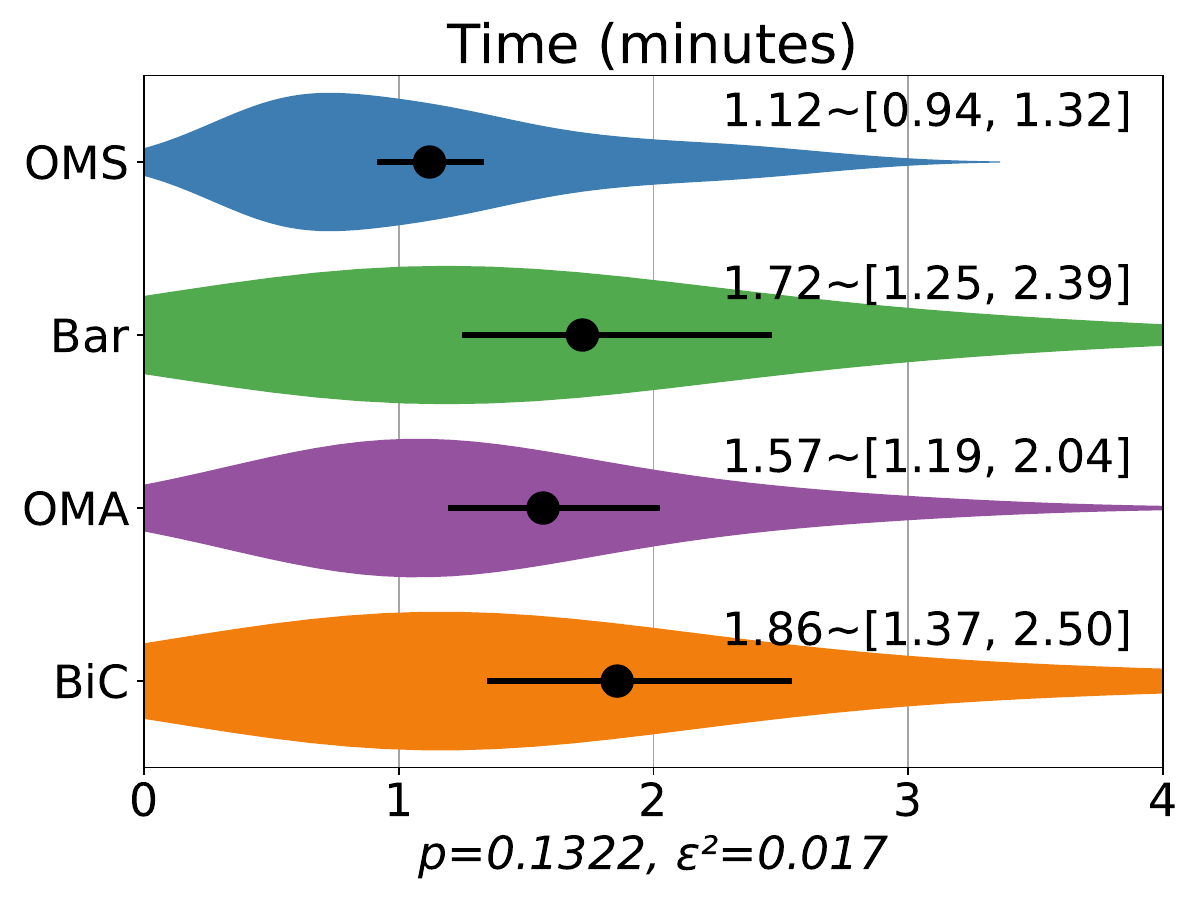}
\end{subfigure}
\hfill
\begin{subfigure}[b]{0.23\textwidth}
    \centering
    \includegraphics[width=\linewidth]{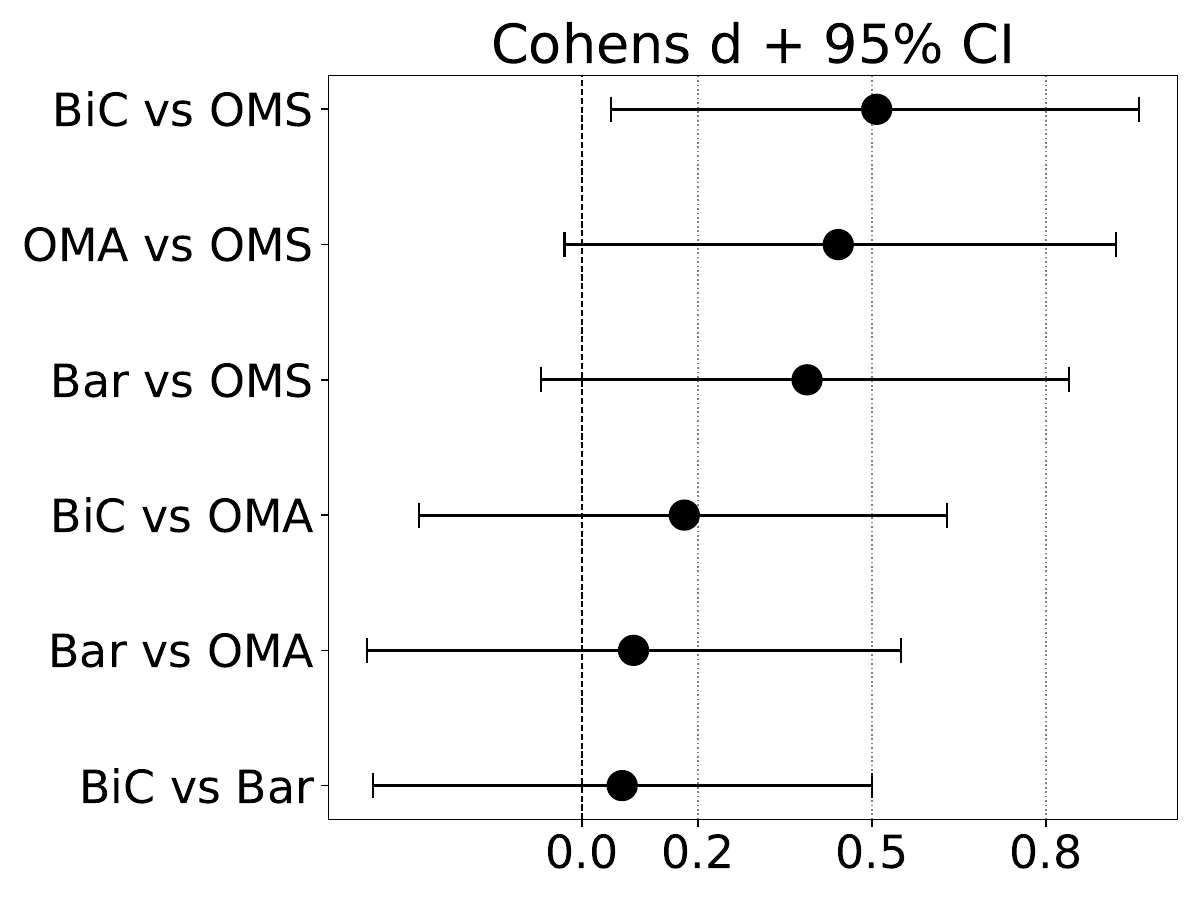}
\end{subfigure}

\caption{T5 - Task Prompt 1}
\label{fig:task5_1_app}
\end{figure*}

\begin{figure*}[hbt]
\centering

\begin{minipage}{\textwidth}
{\fontsize{20}{22}\selectfont\textbf {T5: Extreme Detection - Task Prompt 2}} \label{sec:T5_2}\\

\textbf{Task Rationale:} Assess the effectiveness of the encodings to detect extreme attribute values.\\
\textbf{Hypothesis:} H1.2 and H1.3 \\
\textbf{Scoring:} F1-Score \\
\textbf{Correct Answer:} [
      "South Carolina",
      "Utah"
    ]\\
\textbf{Links:} \\
\inlinevis{bc} \url{https://jorgeacostaupm.github.io/revisit/matrices_barchart/SkhFOWpNT0FCL2syb3pSS3JZYW9jdz09}, \\
\inlinevis{b} \url{https://jorgeacostaupm.github.io/revisit/matrices_bivariate/SkhFOWpNT0FCL2syb3pSS3JZYW9jdz09}, \\
\inlinevis{oms} \url{https://jorgeacostaupm.github.io/revisit/matrices_size/SkhFOWpNT0FCL2syb3pSS3JZYW9jdz09}, \\
\inlinevis{oma} \url{https://jorgeacostaupm.github.io/revisit/matrices_angle/SkhFOWpNT0FCL2syb3pSS3JZYW9jdz09}
\end{minipage}

\vspace{1em}
\begin{subfigure}[b]{\linewidth}
    \includegraphics[width=\linewidth]{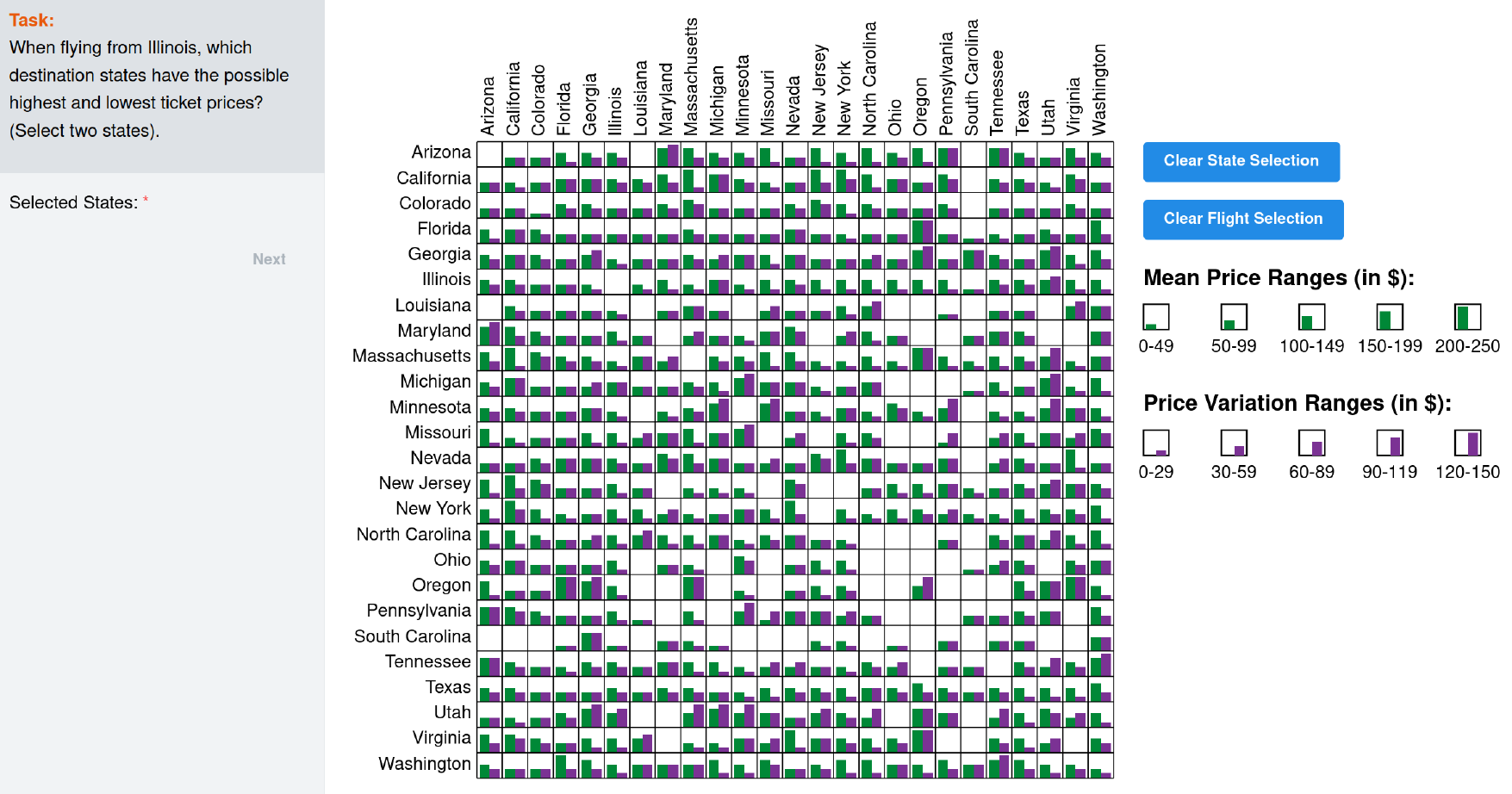}
\end{subfigure}
\vspace{1em}

\begin{subfigure}[b]{0.23\textwidth}
    \centering
    \includegraphics[width=\linewidth]{figs/violinplots/violin_f1_attr_extremes.pdf}
\end{subfigure}
\hfill
\begin{subfigure}[b]{0.23\textwidth}
    \centering
    \includegraphics[width=\linewidth]{figs/forestplots/effects_f1_attr_extremes.pdf}
\end{subfigure}
\hfill
\hspace{3em}
\begin{subfigure}[b]{0.23\textwidth}
    \centering
    \includegraphics[width=\linewidth]{figs/violinplots/violin_time_attr_extremes.pdf}
\end{subfigure}
\hfill
\begin{subfigure}[b]{0.23\textwidth}
    \centering
    \includegraphics[width=\linewidth]{figs/forestplots/effects_time_attr_extremes.pdf}
\end{subfigure}

\caption{T5 - Task Prompt 2}
\label{fig:task5_2_app}
\end{figure*}

\begin{figure*}[hbt]
\centering

\begin{minipage}{\textwidth}
{\fontsize{20}{22}\selectfont\textbf {T6: Classification by Dispersion - Task Prompt 1}} \label{sec:T6_1}\\

\textbf{Task Rationale:} Assess the effectiveness of encodings for estimating and comparing overall node link attribute values between nodes. \\
\textbf{Hypothesis:} H1.2 and H1.3 \\
\textbf{Scoring:} F1-Score \\
\textbf{Correct Answer:} \{
      "Utah": "High Prices Variation",
      "California": "Medium Prices Variation",
      "Illinois": "Low Prices Variation"
    \}\\
\textbf{Links:} \\
\inlinevis{bc} \url{https://jorgeacostaupm.github.io/revisit/matrices_barchart/U2dDOFVybzJsUHpobWVFak8rZHRsZz09}, \\
\inlinevis{b} \url{https://jorgeacostaupm.github.io/revisit/matrices_bivariate/U2dDOFVybzJsUHpobWVFak8rZHRsZz09}, \\
\inlinevis{oms} \url{https://jorgeacostaupm.github.io/revisit/matrices_size/U2dDOFVybzJsUHpobWVFak8rZHRsZz09}, \\
\inlinevis{oma} \url{https://jorgeacostaupm.github.io/revisit/matrices_angle/U2dDOFVybzJsUHpobWVFak8rZHRsZz09}
\end{minipage}

\vspace{1em}
\begin{subfigure}[b]{\linewidth}
    \includegraphics[width=\linewidth]{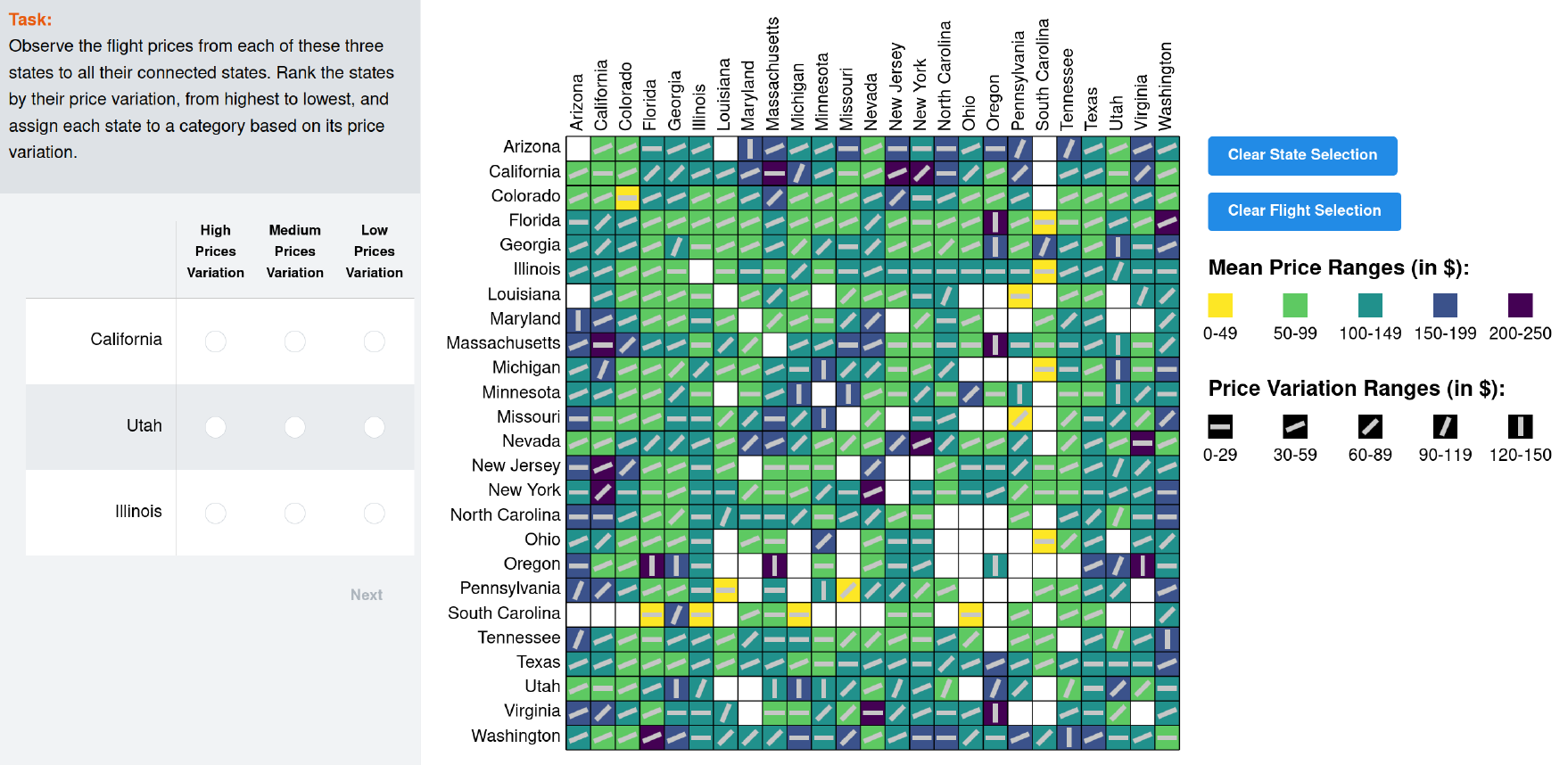} 
\end{subfigure}
\vspace{1em}

\begin{subfigure}[b]{0.23\textwidth}
    \centering
    \includegraphics[width=\linewidth]{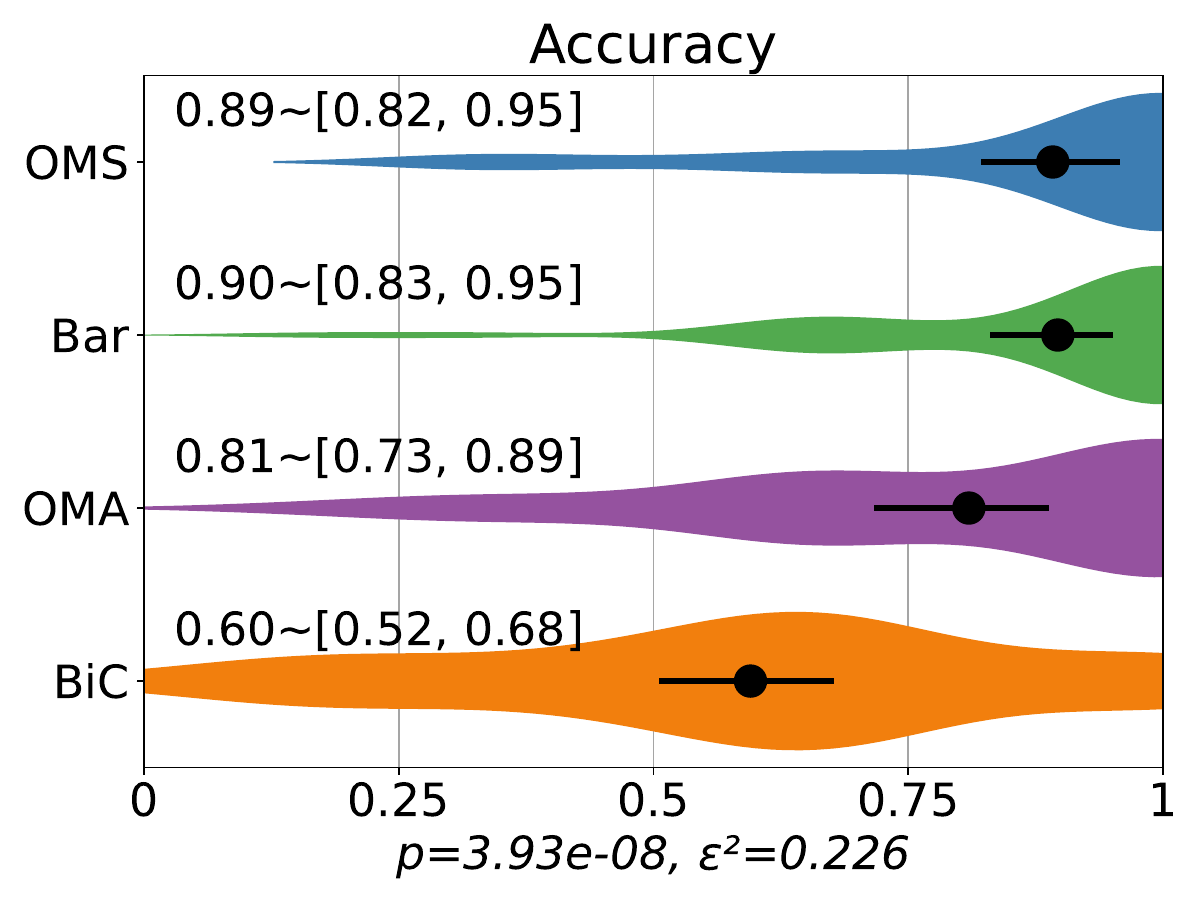}
\end{subfigure}
\hfill
\begin{subfigure}[b]{0.23\textwidth}
    \centering
    \includegraphics[width=\linewidth]{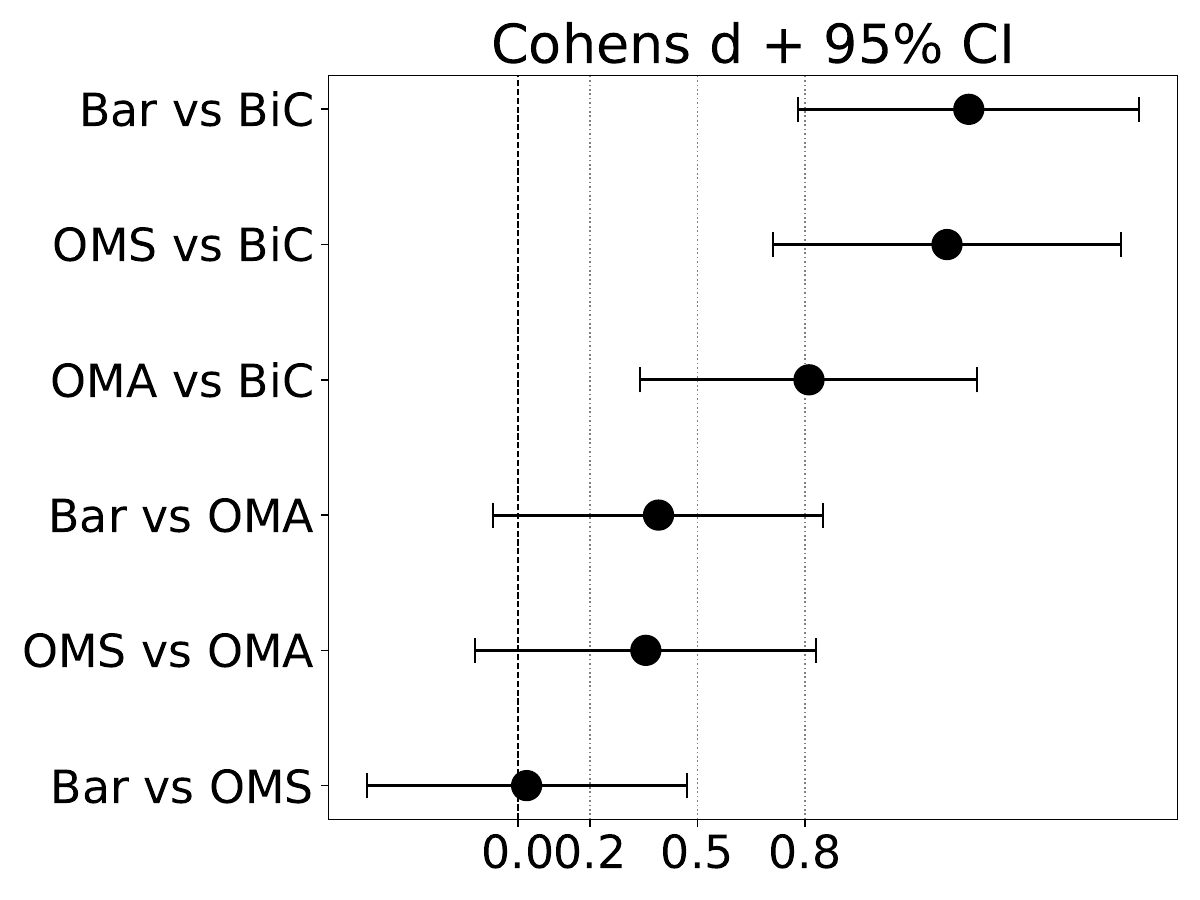}
\end{subfigure}
\hfill
\hspace{3em}
\begin{subfigure}[b]{0.23\textwidth}
    \centering
    \includegraphics[width=\linewidth]{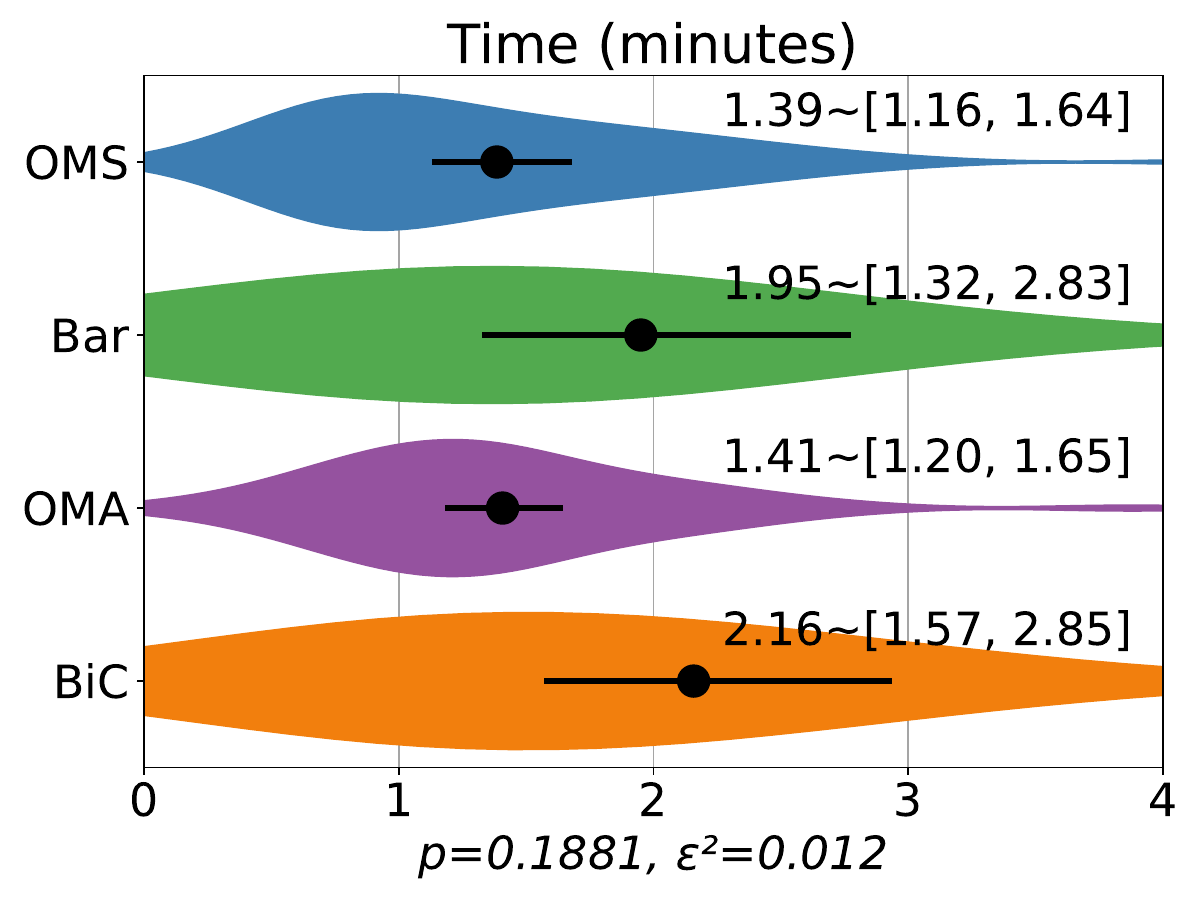}
\end{subfigure}
\hfill
\begin{subfigure}[b]{0.23\textwidth}
    \centering
    \includegraphics[width=\linewidth]{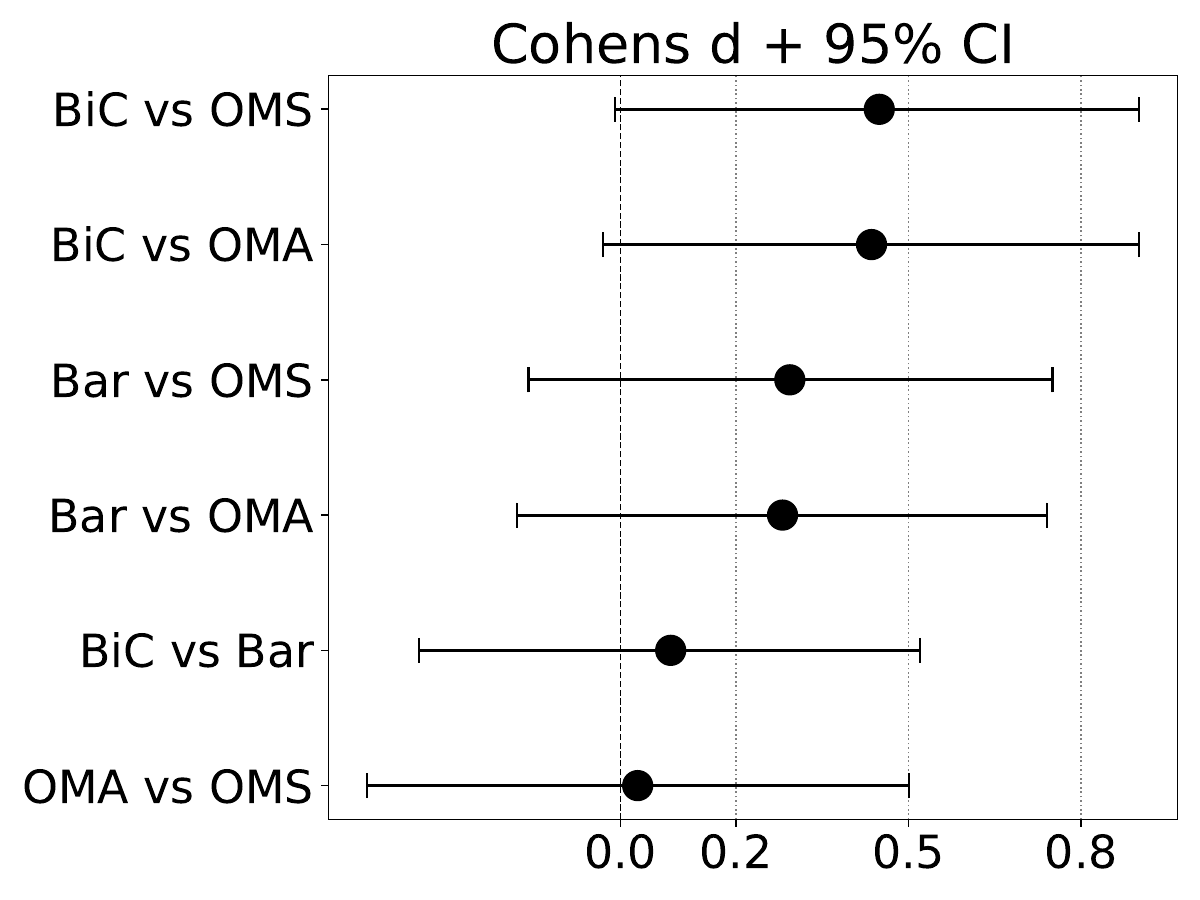}
\end{subfigure}

\caption{T6 - Task Prompt 1}
\label{fig:task6_1_app}
\end{figure*}

\begin{figure*}[hbt]
\centering

\begin{minipage}{\textwidth}
{\fontsize{20}{22}\selectfont\textbf {T6: Classification by Dispersion - Task Prompt 2}} \label{sec:T6_2}\\

\textbf{Task Rationale:} Assess the effectiveness of encodings for estimating and comparing overall node link attribute values between nodes. \\
\textbf{Hypothesis:} H1.2 and H1.3 \\
\textbf{Scoring:} F1-Score \\
\textbf{Correct Answer:} \{
      "Oregon": "High Prices Variation",
      "Pennsylvania": "Medium Prices Variation",
      "Texas": "Low Prices Variation"
    \}\\
\textbf{Links:} \\
\inlinevis{bc} \url{https://jorgeacostaupm.github.io/revisit/matrices_barchart/VHVnMXlTRk5oaUtrZ1NuT0RldGZVQT09}, \\
\inlinevis{b} \url{https://jorgeacostaupm.github.io/revisit/matrices_bivariate/VHVnMXlTRk5oaUtrZ1NuT0RldGZVQT09}, \\
\inlinevis{oms} \url{https://jorgeacostaupm.github.io/revisit/matrices_size/VHVnMXlTRk5oaUtrZ1NuT0RldGZVQT09}, \\
\inlinevis{oma} \url{https://jorgeacostaupm.github.io/revisit/matrices_angle/VHVnMXlTRk5oaUtrZ1NuT0RldGZVQT09}
\end{minipage}

\vspace{1em}
\begin{subfigure}[b]{\linewidth}
    \includegraphics[width=\linewidth]{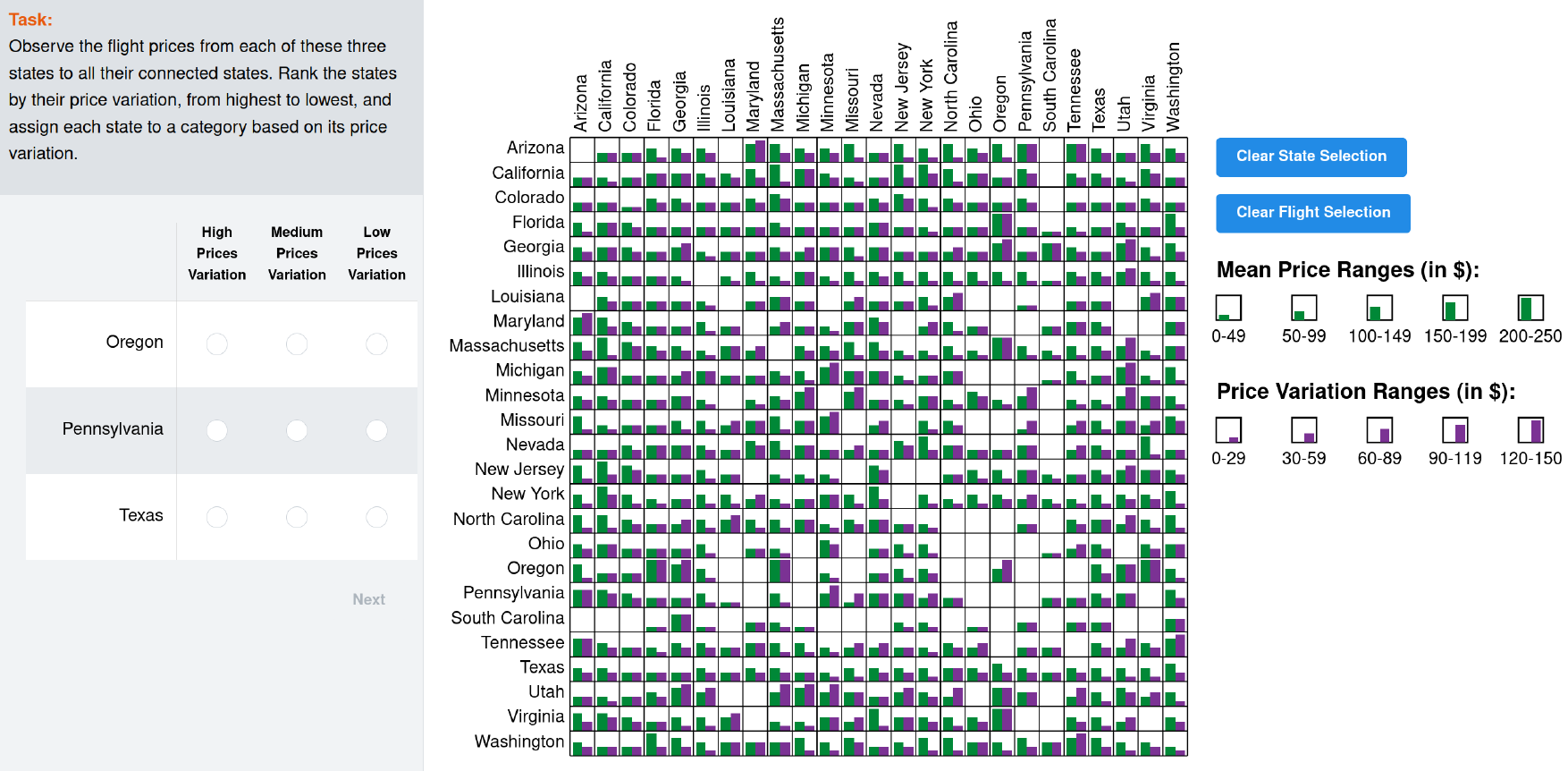} 
\end{subfigure}
\vspace{1em}

\begin{subfigure}[b]{0.23\textwidth}
    \centering
    \includegraphics[width=\linewidth]{figs/violinplots/violin_accuracy_classification.pdf}
\end{subfigure}
\hfill
\begin{subfigure}[b]{0.23\textwidth}
    \centering
    \includegraphics[width=\linewidth]{figs/forestplots/effects_accuracy_classification.pdf}
\end{subfigure}
\hfill
\hspace{3em}
\begin{subfigure}[b]{0.23\textwidth}
    \centering
    \includegraphics[width=\linewidth]{figs/violinplots/violin_time_classification.pdf}
\end{subfigure}
\hfill
\begin{subfigure}[b]{0.23\textwidth}
    \centering
    \includegraphics[width=\linewidth]{figs/forestplots/effects_time_classification.pdf}
\end{subfigure}

\caption{T6 - Task Prompt 2}
\label{fig:task6_2_app}
\end{figure*}

\begin{figure*}[hbt]
\centering

\begin{minipage}{\textwidth}
{\fontsize{20}{22}\selectfont\textbf {T7: Cluster by Central Tendency - Task Prompt 1}} \label{sec:T7_1}\\

\textbf{Task Rationale:} Assess whether different encodings influence the perception of average cluster central tendency. \\
\textbf{Hypothesis:} H1.2 and H1.3 \\
\textbf{Scoring:} Accuracy \\
\textbf{Correct Answer:} C\\
\textbf{Links:} \\
\inlinevis{bc} \url{https://jorgeacostaupm.github.io/revisit/matrices_barchart/S3hVamdOYkNZYU00dlMwTHVRZmFSUT09}, \\
\inlinevis{b} \url{https://jorgeacostaupm.github.io/revisit/matrices_bivariate/S3hVamdOYkNZYU00dlMwTHVRZmFSUT09}, \\
\inlinevis{oms} \url{https://jorgeacostaupm.github.io/revisit/matrices_size/S3hVamdOYkNZYU00dlMwTHVRZmFSUT09}, \\
\inlinevis{oma} \url{https://jorgeacostaupm.github.io/revisit/matrices_angle/S3hVamdOYkNZYU00dlMwTHVRZmFSUT09}
\end{minipage}

\vspace{1em}
\begin{subfigure}[b]{\linewidth}
    \includegraphics[width=\linewidth]{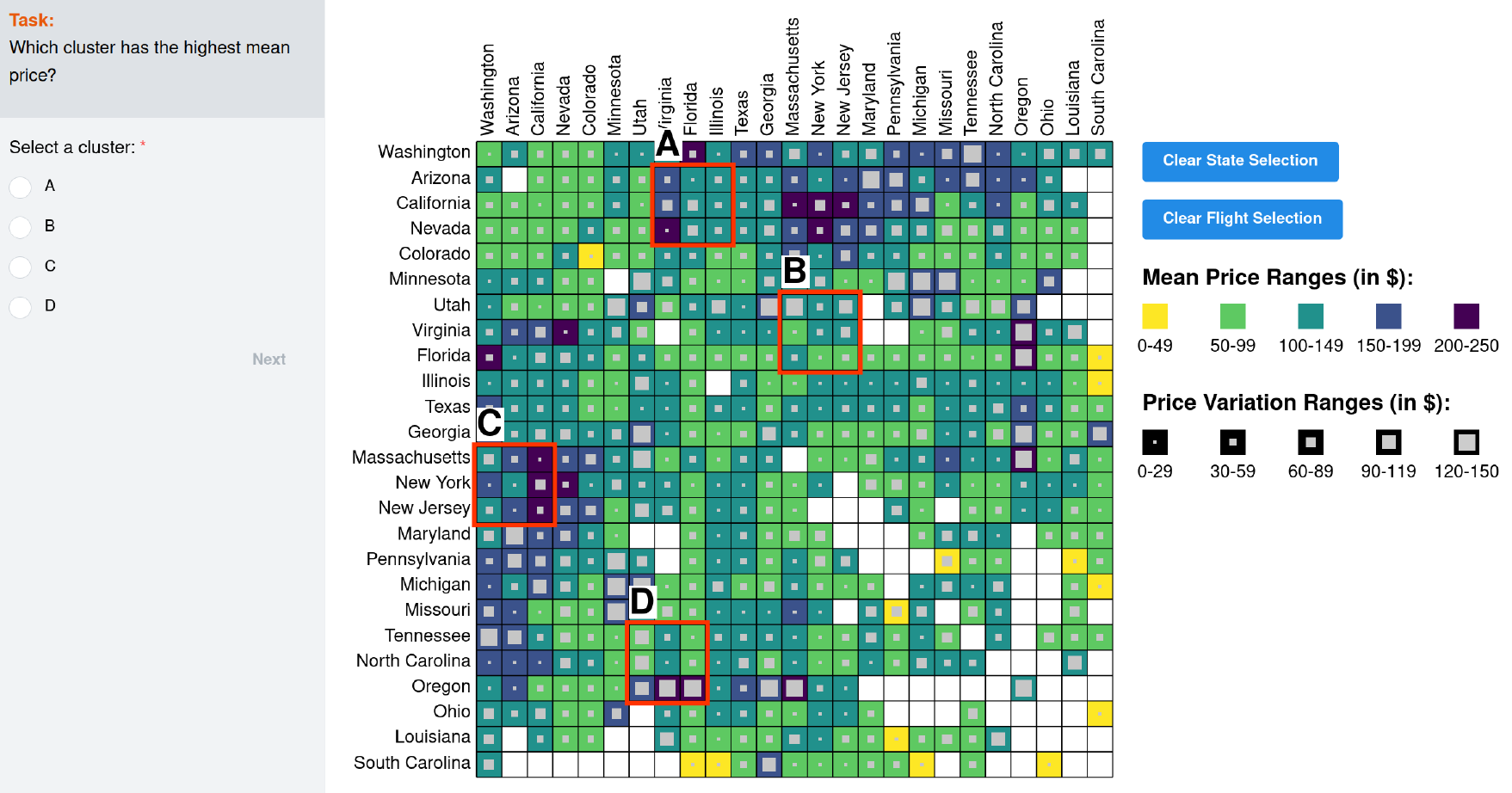}
\end{subfigure}
\vspace{1em}

\begin{subfigure}[b]{0.23\textwidth}
    \centering
    \includegraphics[width=\linewidth]{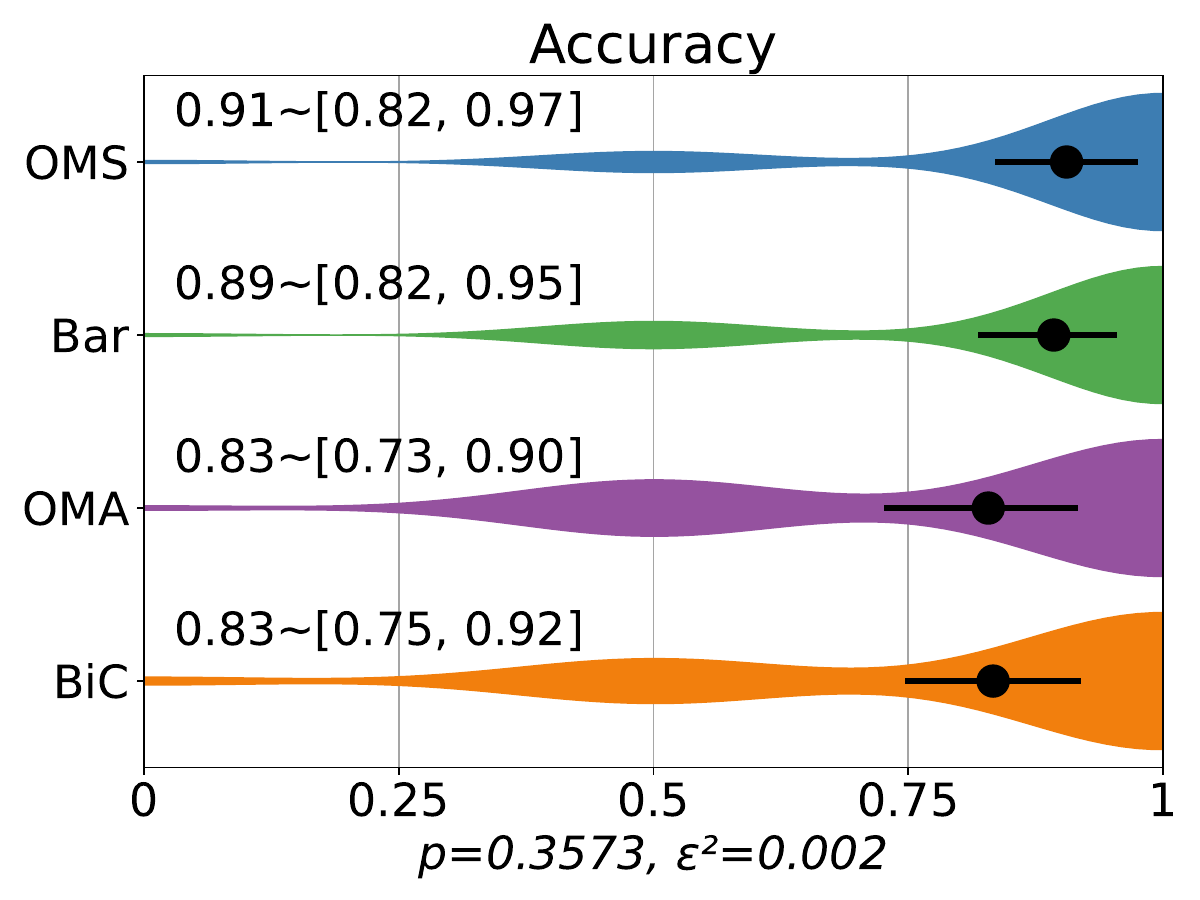}
\end{subfigure}
\hfill
\begin{subfigure}[b]{0.23\textwidth}
    \centering
    \includegraphics[width=\linewidth]{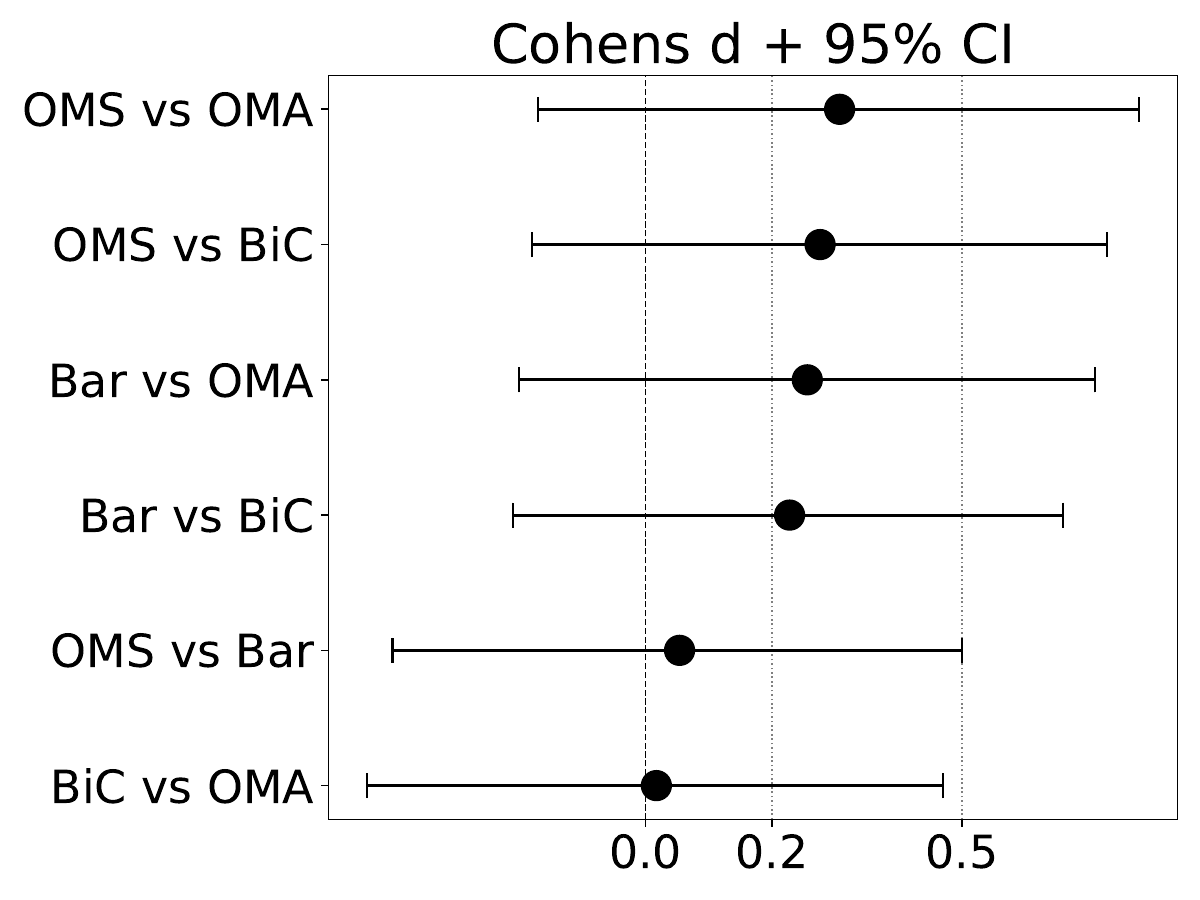}
\end{subfigure}
\hfill
\hspace{3em}
\begin{subfigure}[b]{0.23\textwidth}
    \centering
    \includegraphics[width=\linewidth]{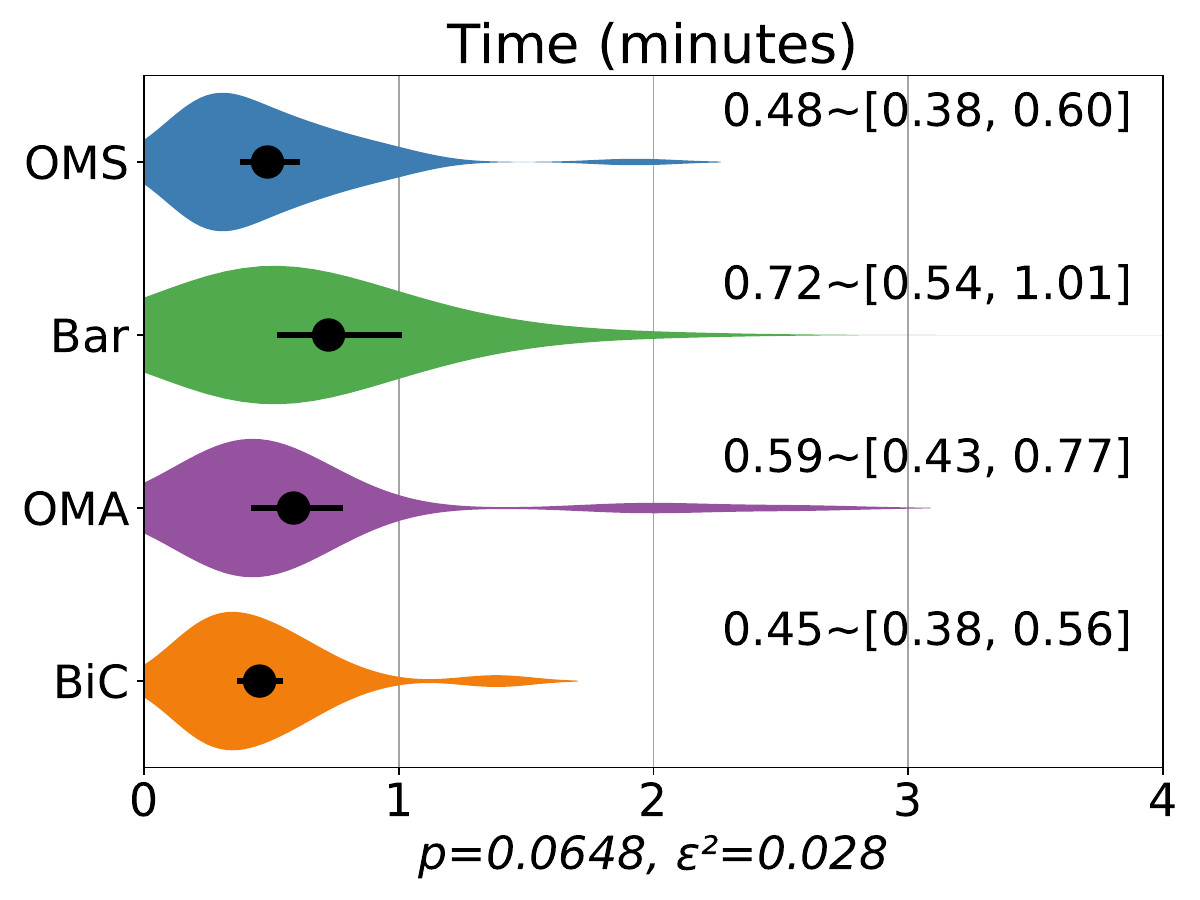}
\end{subfigure}
\hfill
\begin{subfigure}[b]{0.23\textwidth}
    \centering
    \includegraphics[width=\linewidth]{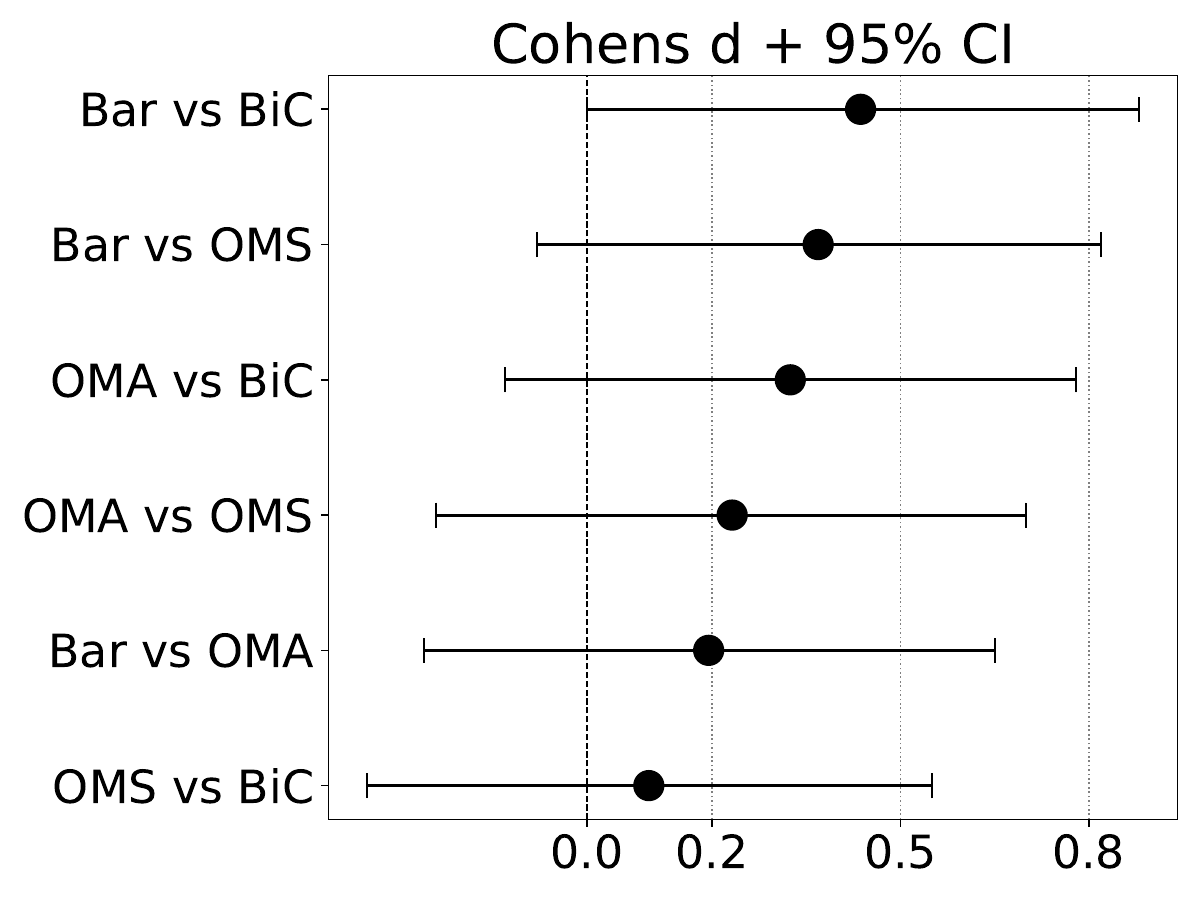}
\end{subfigure}

\caption{T7 - Task Prompt 1}
\label{fig:task7_1_app}
\end{figure*}

\begin{figure*}[hbt]
\centering

\begin{minipage}{\textwidth}
{\fontsize{20}{22}\selectfont\textbf {T7: Cluster by Central Tendency - Task Prompt 2}} \label{sec:T7_2}\\

\textbf{Task Rationale:} Assess whether different encodings influence the perception of average cluster central tendency. \\
\textbf{Hypothesis:} H1.2 and H1.3 \\
\textbf{Scoring:} Accuracy \\
\textbf{Correct Answer:} A\\
\textbf{Links:} \\
\inlinevis{bc} \url{https://jorgeacostaupm.github.io/revisit/matrices_barchart/d290SWVGV1NQZFM0Z2UrUDRaVkxRQT09}, \\
\inlinevis{b} \url{https://jorgeacostaupm.github.io/revisit/matrices_bivariate/d290SWVGV1NQZFM0Z2UrUDRaVkxRQT09}, \\
\inlinevis{oms} \url{https://jorgeacostaupm.github.io/revisit/matrices_size/d290SWVGV1NQZFM0Z2UrUDRaVkxRQT09}, \\
\inlinevis{oma} \url{https://jorgeacostaupm.github.io/revisit/matrices_angle/d290SWVGV1NQZFM0Z2UrUDRaVkxRQT09}
\end{minipage}

\vspace{1em}
\begin{subfigure}[b]{\linewidth}
    \includegraphics[width=\linewidth]{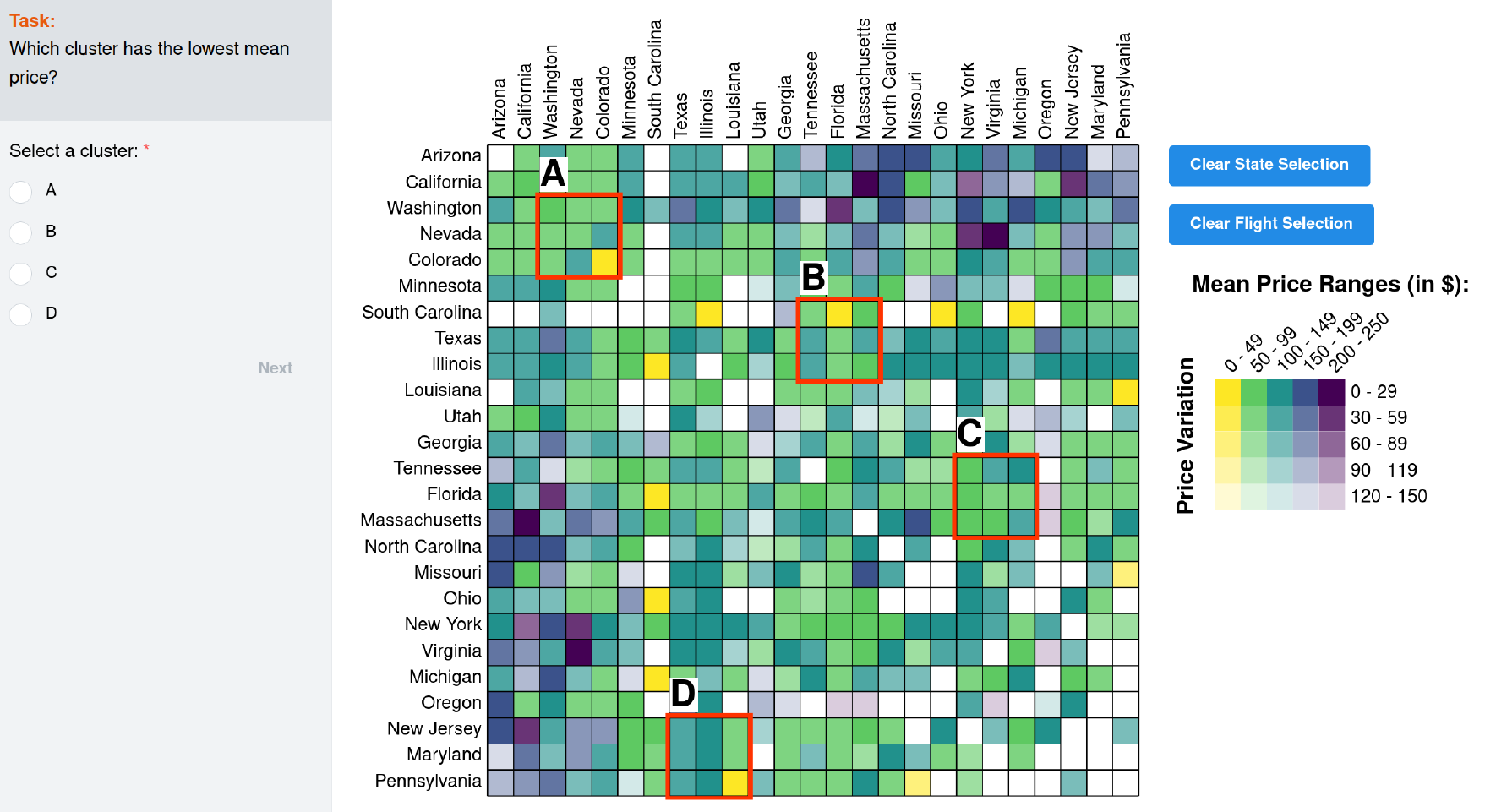}
\end{subfigure}

\vspace{1em}

\begin{subfigure}[b]{0.23\textwidth}
    \centering
    \includegraphics[width=\linewidth]{figs/violinplots/violin_accuracy_cluster_mean.pdf}
\end{subfigure}
\hfill
\begin{subfigure}[b]{0.23\textwidth}
    \centering
    \includegraphics[width=\linewidth]{figs/forestplots/effects_accuracy_cluster_mean.pdf}
\end{subfigure}
\hfill
\hspace{3em}
\begin{subfigure}[b]{0.23\textwidth}
    \centering
    \includegraphics[width=\linewidth]{figs/violinplots/violin_time_cluster_mean.pdf}
\end{subfigure}
\hfill
\begin{subfigure}[b]{0.23\textwidth}
    \centering
    \includegraphics[width=\linewidth]{figs/forestplots/effects_time_cluster_mean.pdf}
\end{subfigure}

\caption{T7 - Task Prompt 2}
\label{fig:task7_2_app}
\end{figure*}

\begin{figure*}[hbt]
\centering

\begin{minipage}{\textwidth}
{\fontsize{20}{22}\selectfont\textbf {T8: Cluster by Dispersion - Task Prompt 1}} \label{sec:T8_1}\\

\textbf{Task Rationale:} Assess whether different encodings influence the perception of average cluster dispersion. \\
\textbf{Hypothesis:} H1.2 and H1.3 \\
\textbf{Scoring:} Accuracy \\
\textbf{Correct Answer:} C\\
\textbf{Links:} \\
\inlinevis{bc} \url{https://jorgeacostaupm.github.io/revisit/matrices_barchart/UHBsS0E3VlVtUWRsLzVGK0kzRHRKZz09}, \\
\inlinevis{b} \url{https://jorgeacostaupm.github.io/revisit/matrices_bivariate/UHBsS0E3VlVtUWRsLzVGK0kzRHRKZz09}, \\
\inlinevis{oms} \url{https://jorgeacostaupm.github.io/revisit/matrices_size/UHBsS0E3VlVtUWRsLzVGK0kzRHRKZz09}, \\
\inlinevis{oma} \url{https://jorgeacostaupm.github.io/revisit/matrices_angle/UHBsS0E3VlVtUWRsLzVGK0kzRHRKZz09}
\end{minipage}

\vspace{1em}
\begin{subfigure}[b]{\linewidth}
    \includegraphics[width=\linewidth]{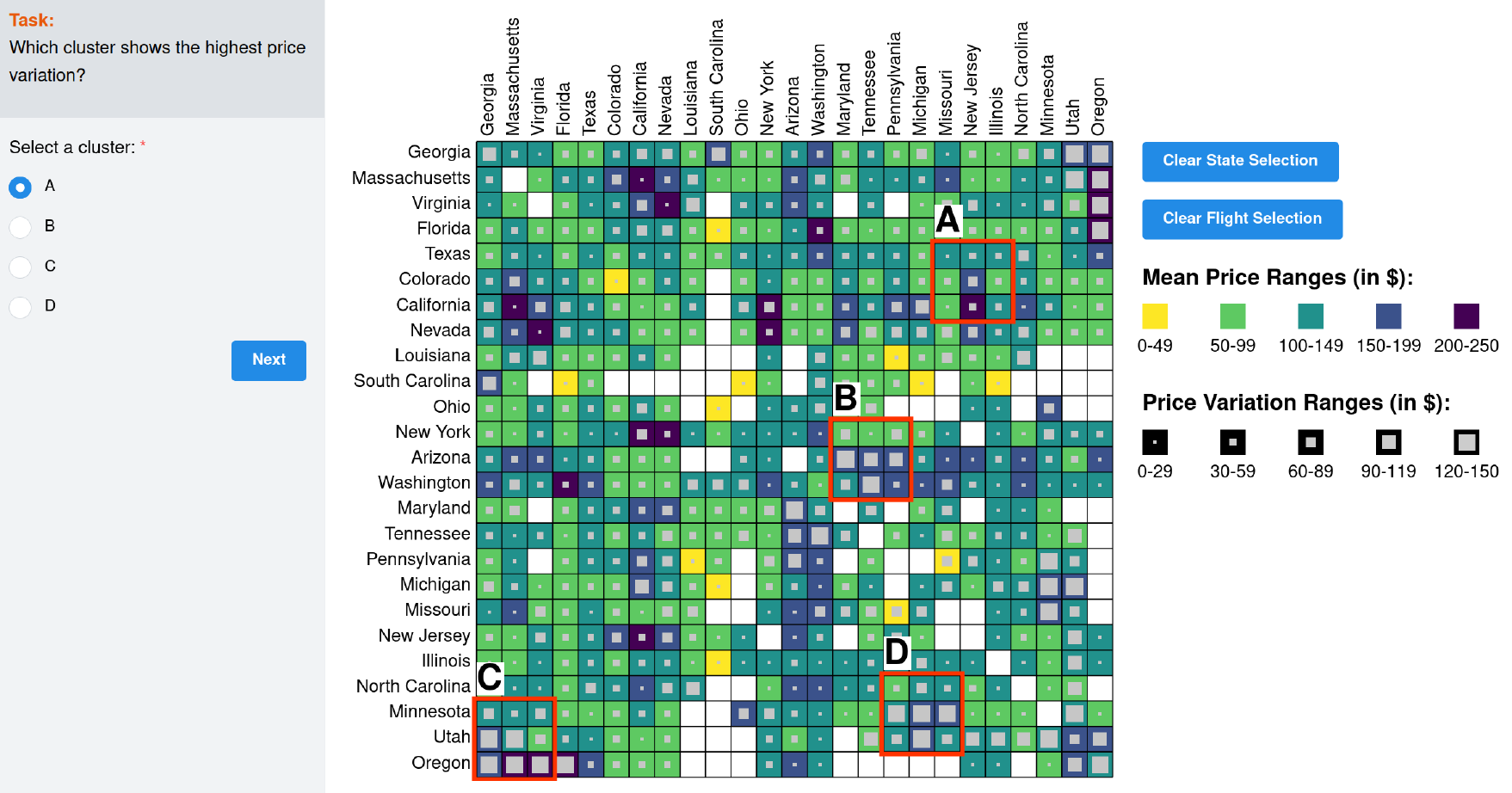}
\end{subfigure}
\vspace{1em}

\begin{subfigure}[b]{0.23\textwidth}
    \centering
    \includegraphics[width=\linewidth]{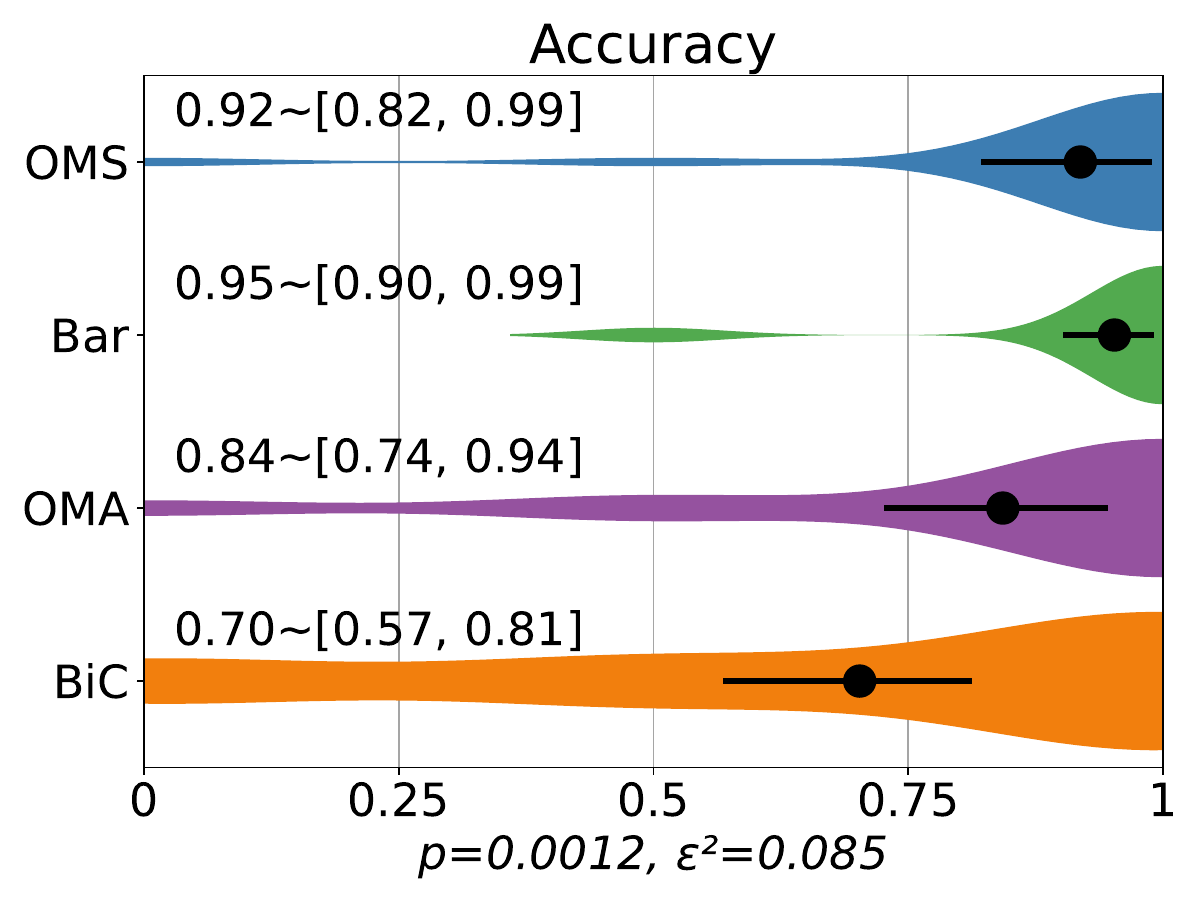}
\end{subfigure}
\hfill
\begin{subfigure}[b]{0.23\textwidth}
    \centering
    \includegraphics[width=\linewidth]{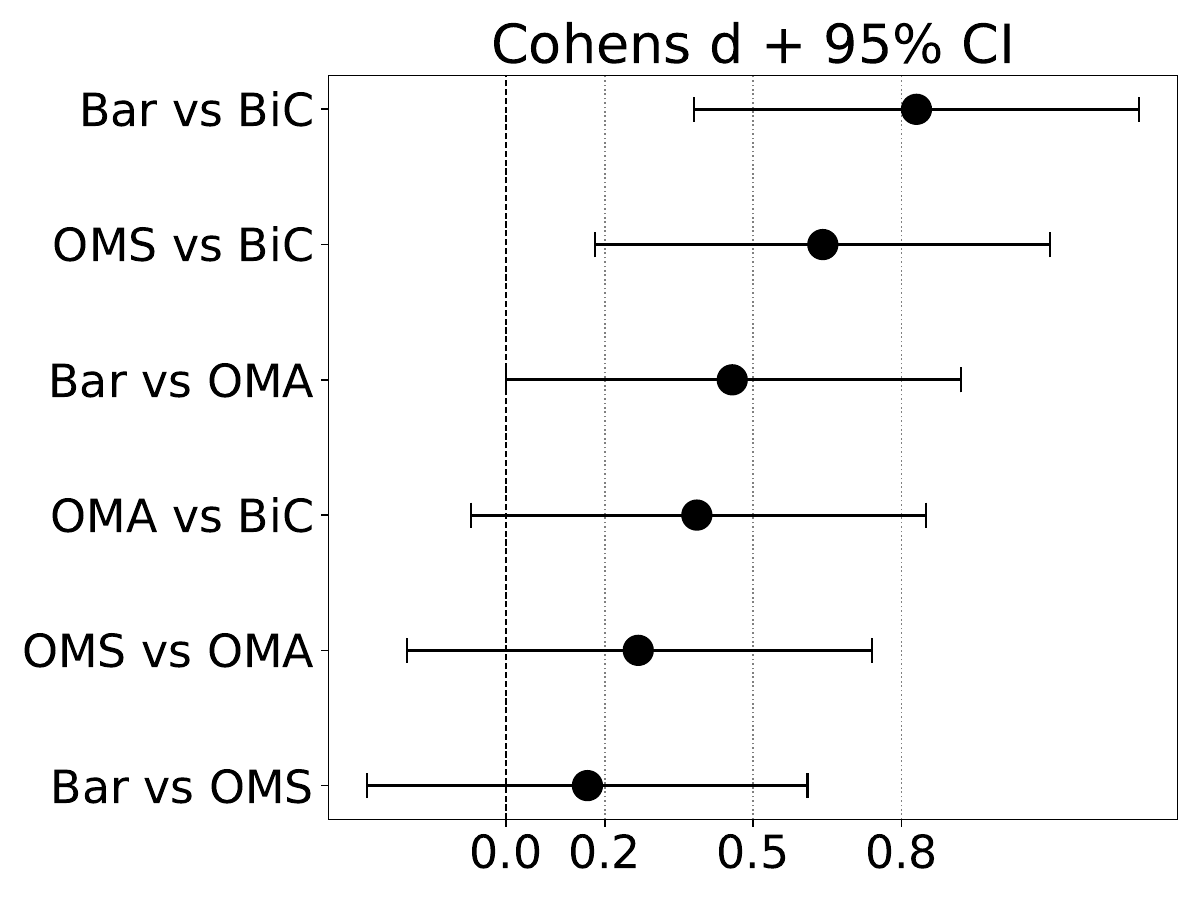}
\end{subfigure}
\hfill
\hspace{3em}
\begin{subfigure}[b]{0.23\textwidth}
    \centering
    \includegraphics[width=\linewidth]{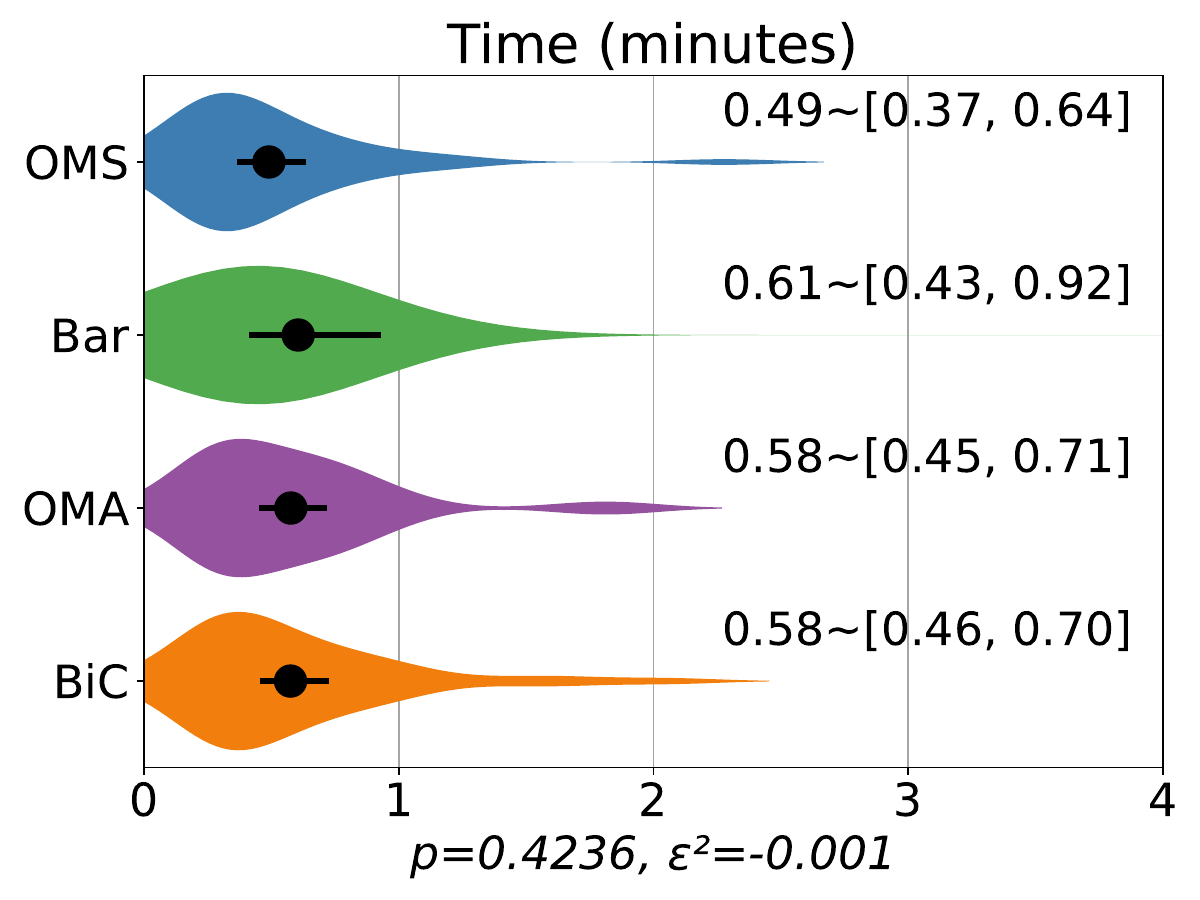}
\end{subfigure}
\hfill
\begin{subfigure}[b]{0.23\textwidth}
    \centering
    \includegraphics[width=\linewidth]{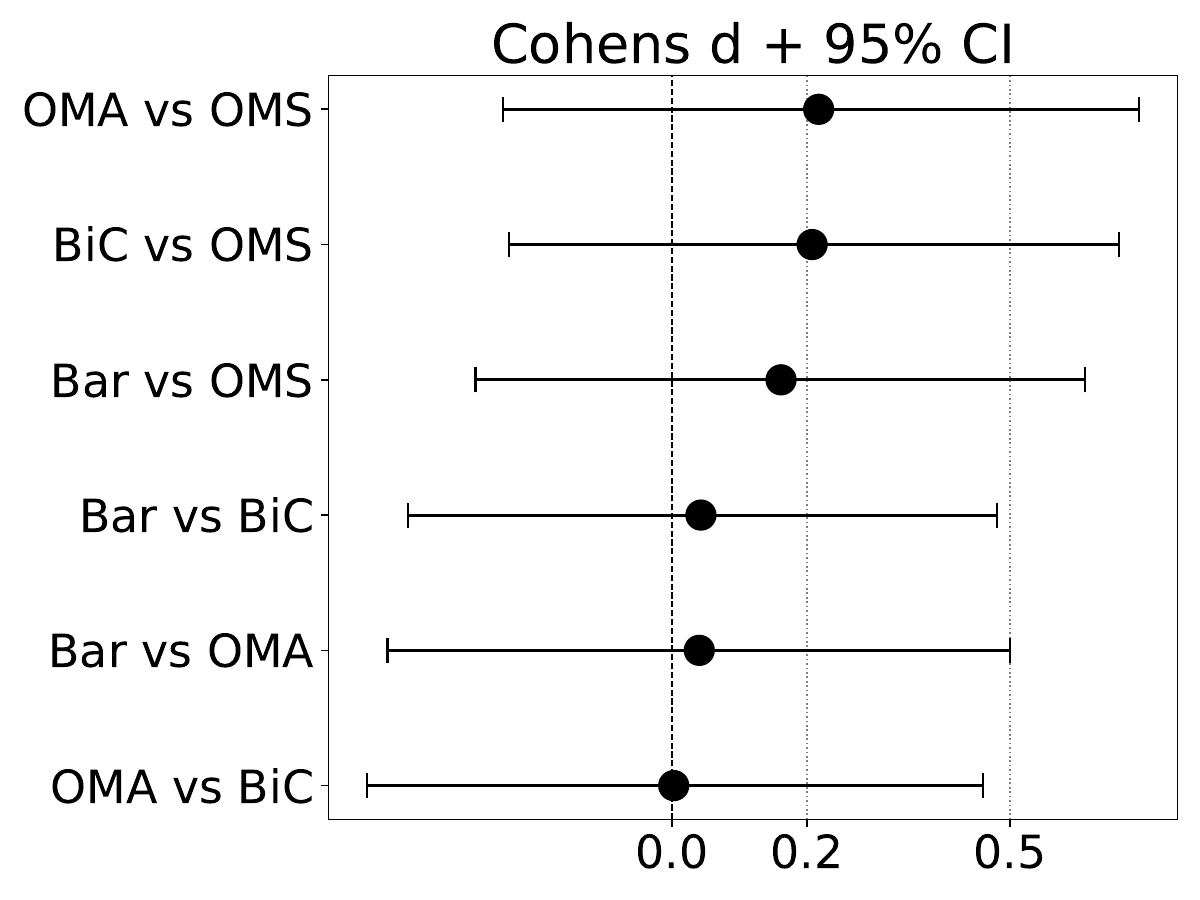}
\end{subfigure}

\caption{Task 8 - Task Prompt 1}
\label{fig:task8_1_app}
\end{figure*}

\begin{figure*}[hbt]
\centering

\begin{minipage}{\textwidth}
{\fontsize{20}{22}\selectfont\textbf {T8: Cluster by Dispersion - Task Prompt 2}} \label{sec:T8_2}\\

\textbf{Task Rationale:} Assess whether different encodings influence the perception of average cluster dispersion. \\
\textbf{Hypothesis:} H1.2 and H1.3 \\
\textbf{Scoring:} Accuracy \\
\textbf{Correct Answer:} A\\
\textbf{Links:} \\
\inlinevis{bc} \url{https://jorgeacostaupm.github.io/revisit/matrices_barchart/YTVMWWI3NE1oc0NEblpnL2tYSUJpZz09}, \\
\inlinevis{b} \url{https://jorgeacostaupm.github.io/revisit/matrices_bivariate/YTVMWWI3NE1oc0NEblpnL2tYSUJpZz09}, \\
\inlinevis{oms} \url{https://jorgeacostaupm.github.io/revisit/matrices_size/YTVMWWI3NE1oc0NEblpnL2tYSUJpZz09}, \\
\inlinevis{oma} \url{https://jorgeacostaupm.github.io/revisit/matrices_size/YTVMWWI3NE1oc0NEblpnL2tYSUJpZz09}
\end{minipage}

\vspace{1em}
\begin{subfigure}[b]{\linewidth}
    \includegraphics[width=\linewidth]{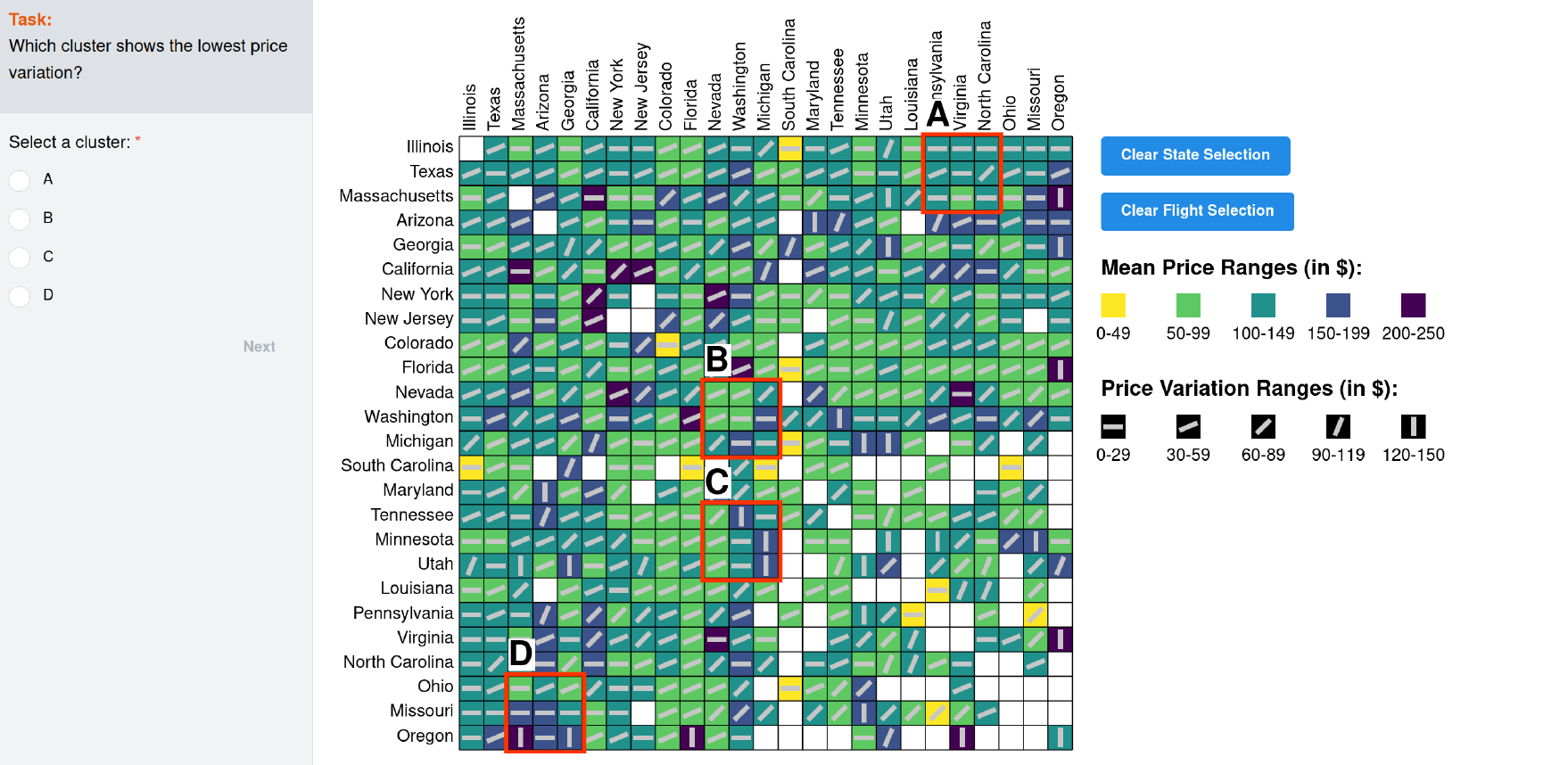}
\end{subfigure}
\vspace{1em}

\begin{subfigure}[b]{0.23\textwidth}
    \centering
    \includegraphics[width=\linewidth]{figs/violinplots/violin_accuracy_cluster_var.pdf}
\end{subfigure}
\hfill
\begin{subfigure}[b]{0.23\textwidth}
    \centering
    \includegraphics[width=\linewidth]{figs/forestplots/effects_accuracy_cluster_var.pdf}
\end{subfigure}
\hfill
\hspace{3em}
\begin{subfigure}[b]{0.23\textwidth}
    \centering
    \includegraphics[width=\linewidth]{figs/violinplots/violin_time_cluster_var.pdf}
\end{subfigure}
\hfill
\begin{subfigure}[b]{0.23\textwidth}
    \centering
    \includegraphics[width=\linewidth]{figs/forestplots/effects_time_cluster_var.pdf}
\end{subfigure}

\caption{T8 - Task Prompt 2}
\label{fig:task8_2_app}
\end{figure*}

\begin{figure*}[hbt]
\centering

\begin{minipage}{\textwidth}
{\fontsize{20}{22}\selectfont\textbf {PREVis - Reading Data}} \label{sec:pre_data}\\

\textbf{Rationale:} This item measures perception of high-level patterns — trends, outliers, extremes.\\
\textbf{Hypothesis:} H2 \\
\textbf{Scoring:} Likert Scale (1-7) \\
\textbf{Links:} \\
\inlinevis{bc} \url{https://jorgeacostaupm.github.io/revisit/matrices_barchart/STVxUHlHWGhoUFlKWDNyaS81OFNQQT09}, \\
\inlinevis{b} \url{https://jorgeacostaupm.github.io/revisit/matrices_bivariate/STVxUHlHWGhoUFlKWDNyaS81OFNQQT09}, \\
\inlinevis{oms} \url{https://jorgeacostaupm.github.io/revisit/matrices_size/STVxUHlHWGhoUFlKWDNyaS81OFNQQT09}, \\
\inlinevis{oma} \url{https://jorgeacostaupm.github.io/revisit/matrices_angle/STVxUHlHWGhoUFlKWDNyaS81OFNQQT09}
\end{minipage}

\vspace{1em}
\begin{subfigure}[b]{\linewidth}
    \includegraphics[width=\linewidth]{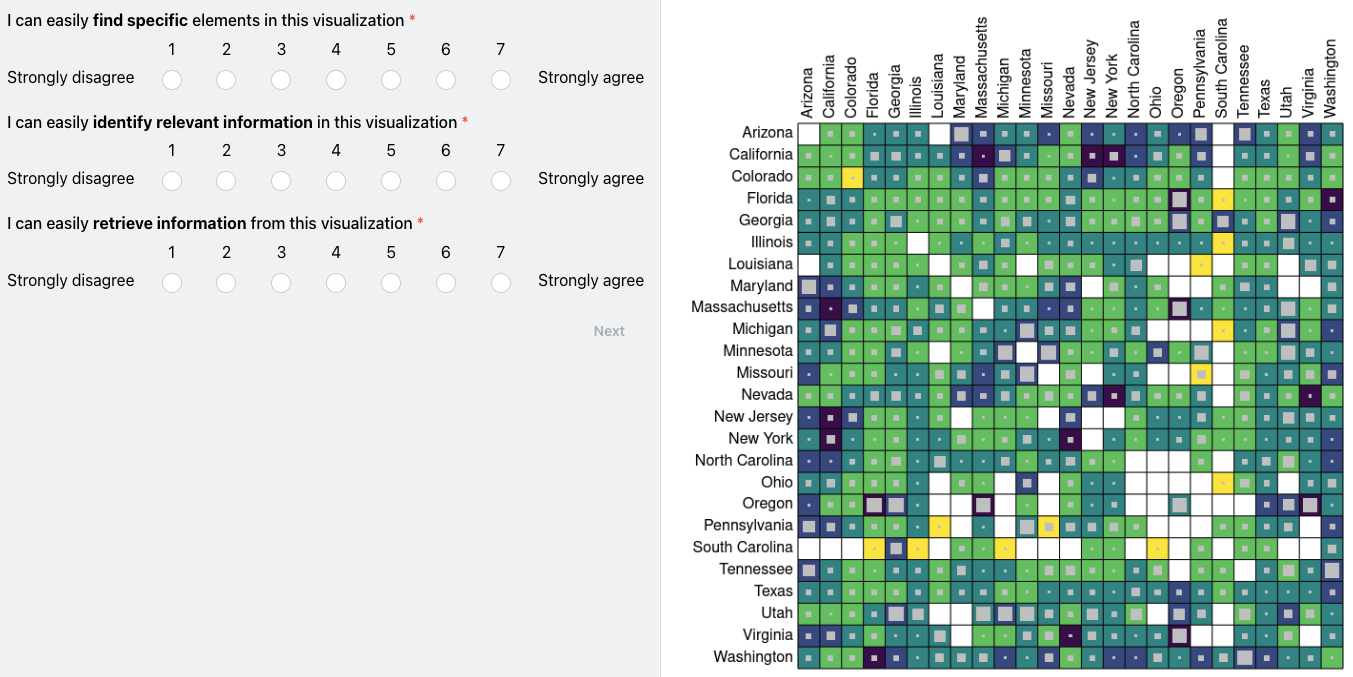} 
\end{subfigure}
\vspace{1em}

\begin{subfigure}[b]{0.48\textwidth}
    \centering
        \includegraphics[width=0.48\textwidth]{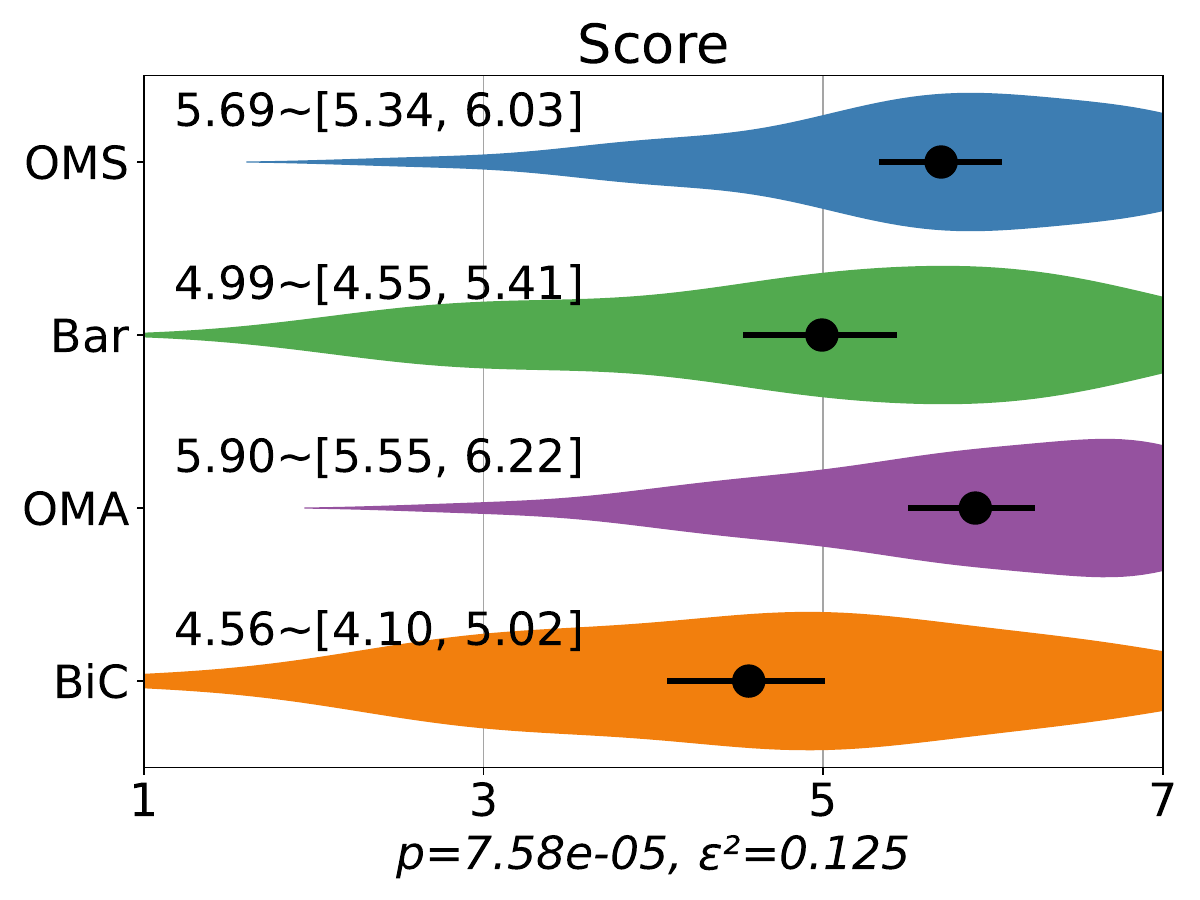}
    \hfill
    \includegraphics[width=0.48\textwidth]{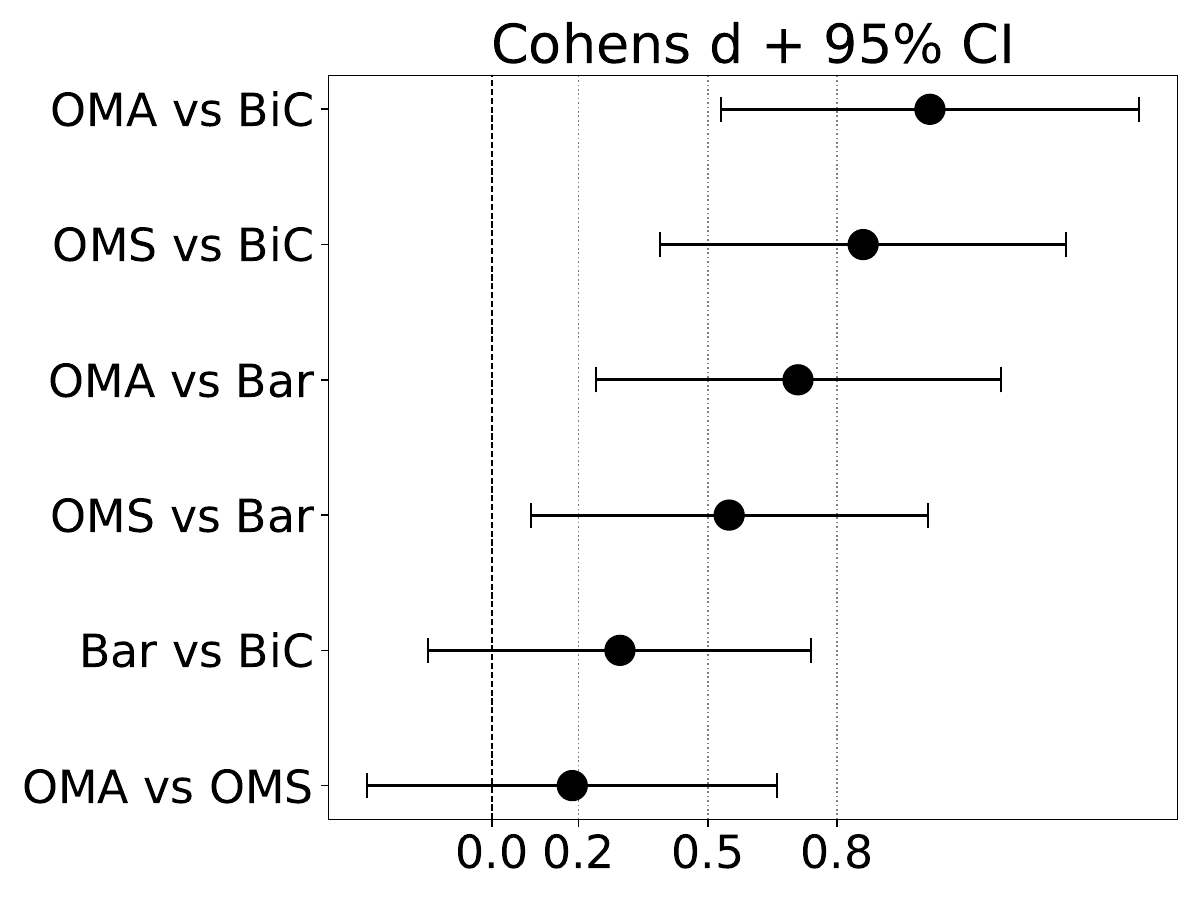}
    \label{fig:readability_data}
\end{subfigure}

\caption{Comparative evaluation of the Reading Data Item of the PREVis Scale across the four tested encoding techniques.}
\label{fig:readability_app}
\end{figure*}

\begin{figure*}[hbt]
\centering

\begin{minipage}{\textwidth}
{\fontsize{20}{22}\selectfont\textbf {PREVis - Understand}} \label{sec:pre_under}\\

\textbf{Rationale:} This item measures overall conceptual clarity — how easily the user understands what the visualization is and how to read it. \\
\textbf{Hypothesis:} H2 \\
\textbf{Scoring:} Likert Scale (1-7) \\
\textbf{Links:} \\
\inlinevis{bc} \url{https://jorgeacostaupm.github.io/revisit/matrices_barchart/L2VVUUdBQWZXcTFhbnVvREE0MDdCQT09}, \\
\inlinevis{b} \url{https://jorgeacostaupm.github.io/revisit/matrices_bivariate/L2VVUUdBQWZXcTFhbnVvREE0MDdCQT09}, \\
\inlinevis{oms} \url{https://jorgeacostaupm.github.io/revisit/matrices_size/L2VVUUdBQWZXcTFhbnVvREE0MDdCQT09}, \\
\inlinevis{oma} \url{https://jorgeacostaupm.github.io/revisit/matrices_angle/L2VVUUdBQWZXcTFhbnVvREE0MDdCQT09}
\end{minipage}

\vspace{1em}
\begin{subfigure}[b]{\linewidth}
    \includegraphics[width=\linewidth]{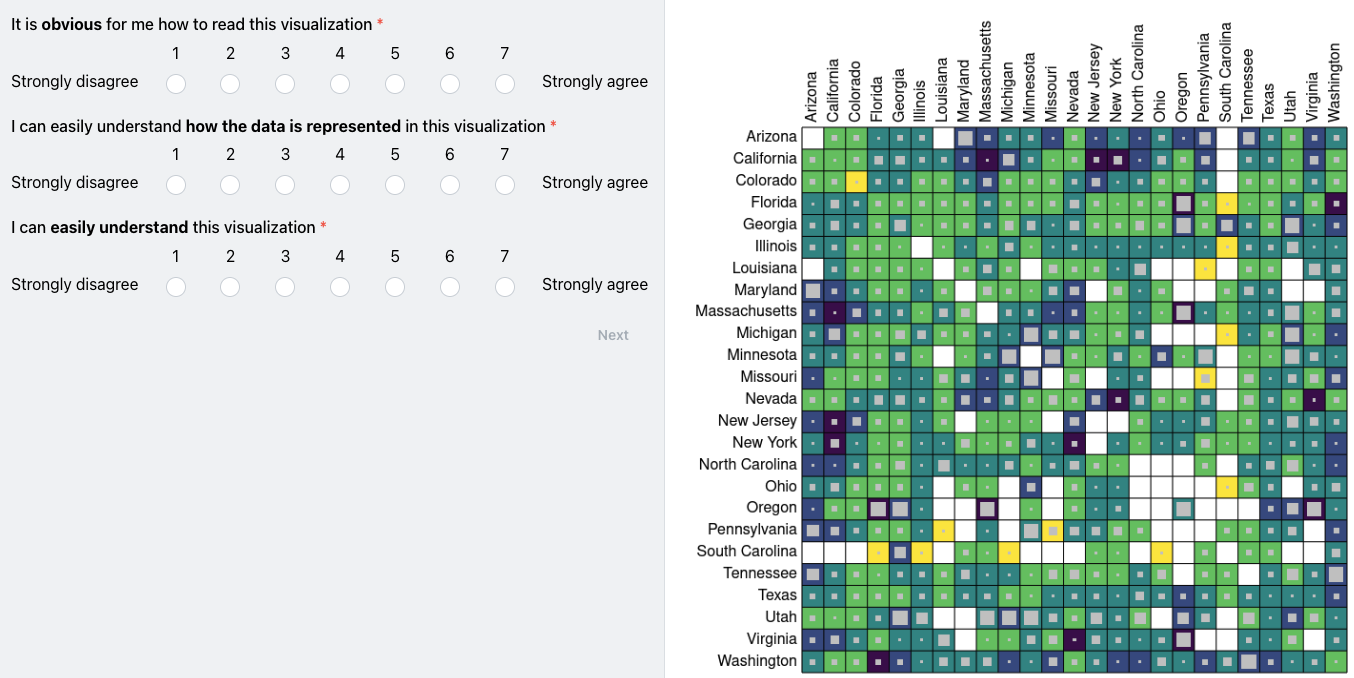}
\end{subfigure}
\vspace{1em}

\begin{subfigure}[b]{0.48\textwidth}
    \centering
        \includegraphics[width=0.48\textwidth]{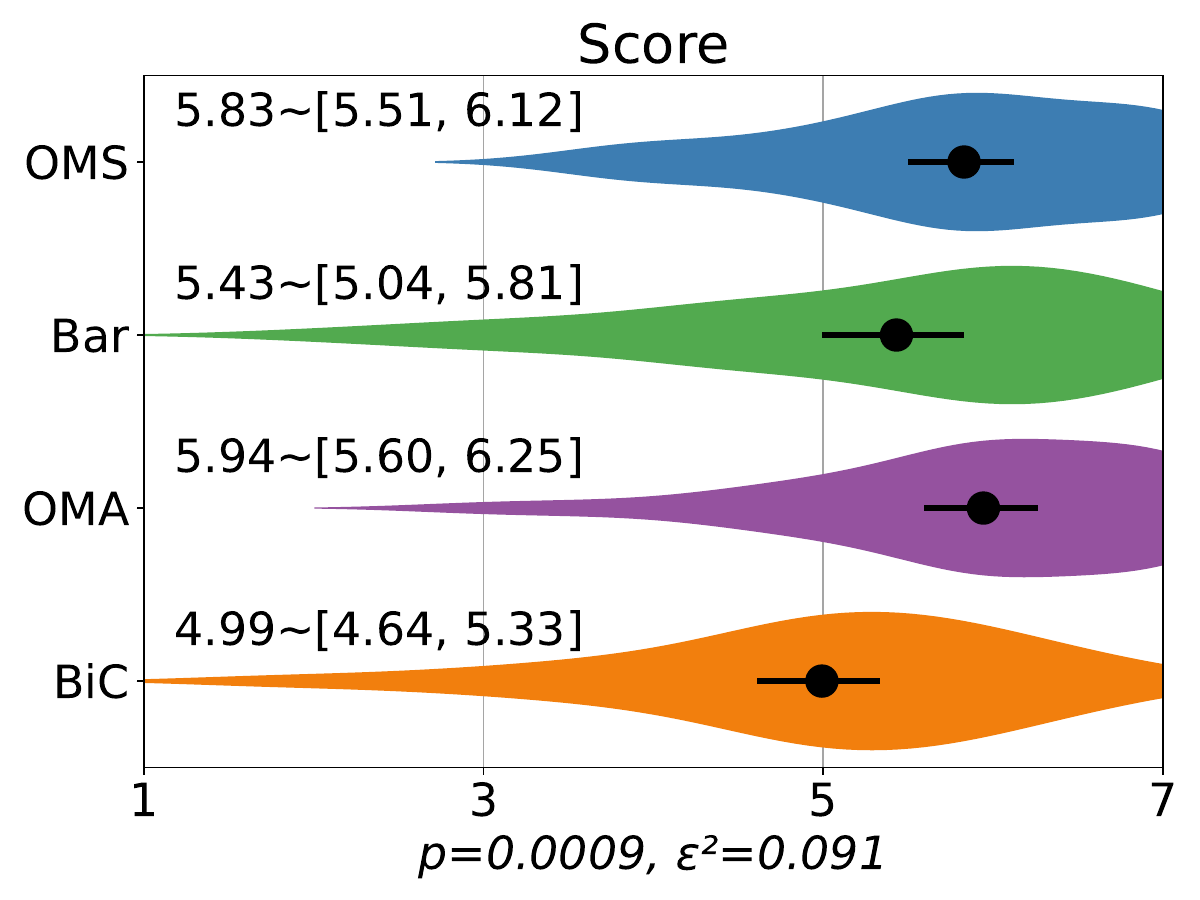}
    \hfill
    \includegraphics[width=0.48\textwidth]{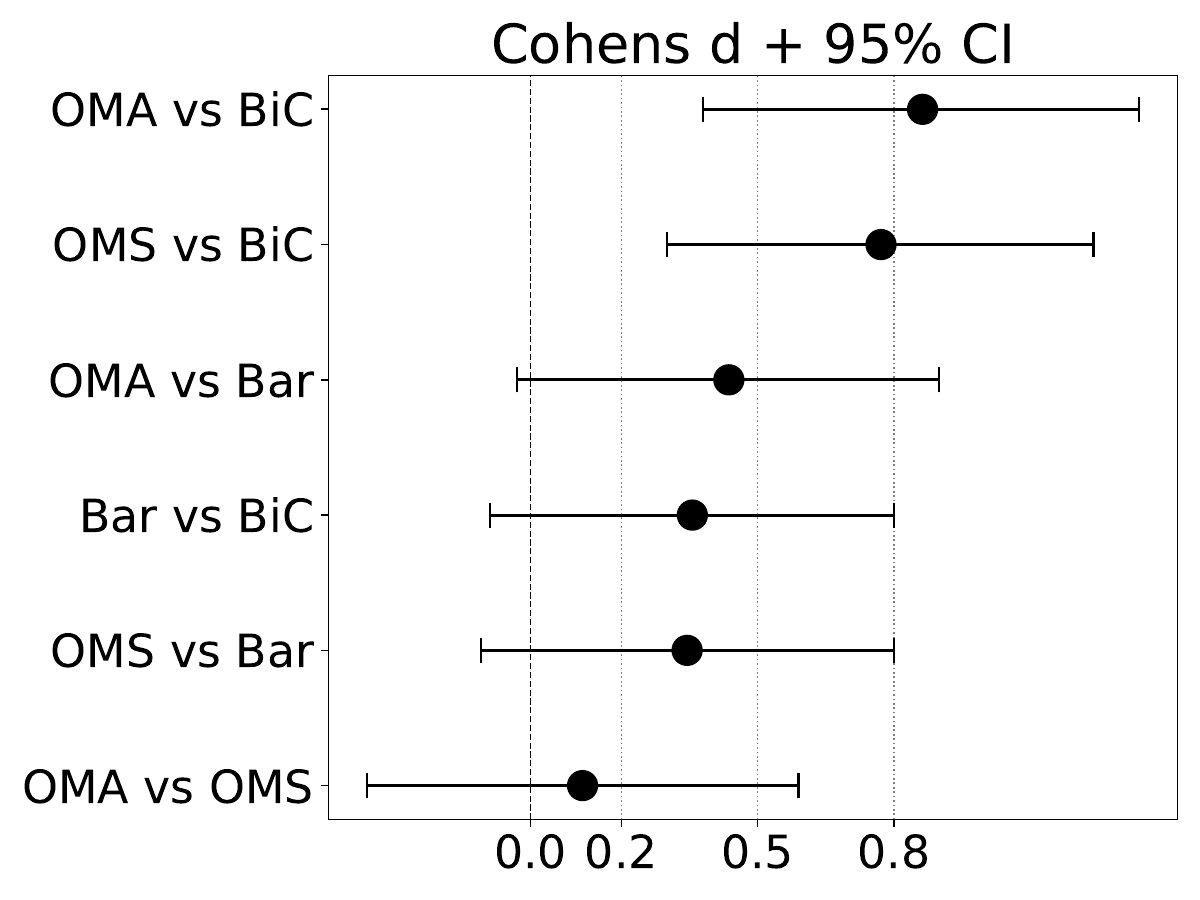}
    \label{fig:readability_understand}
\end{subfigure}

\caption{Comparative evaluation of the Understand Item of the PREVis Scale across the four tested encoding techniques.}
\label{fig:understand_app}
\end{figure*}

\begin{figure*}[hbt]
\centering

\begin{minipage}{\textwidth}
{\fontsize{20}{22}\selectfont\textbf {PREVis - Layout}} \label{sec:pre_lay}\\

\textbf{Rationale:} This item measures visual cleanliness and organization — whether the graphic looks cluttered or distracting.\\
\textbf{Hypothesis:} H2 \\
\textbf{Scoring:} Likert Scale (1-7) \\
\textbf{Links:} \\
\inlinevis{bc} \url{https://jorgeacostaupm.github.io/revisit/matrices_barchart/SHdPTGI3ZVZxS1EwU1pqOUdsS3V1QT09}, \\
\inlinevis{b} \url{https://jorgeacostaupm.github.io/revisit/matrices_bivariate/SHdPTGI3ZVZxS1EwU1pqOUdsS3V1QT09}, \\
\inlinevis{oms} \url{https://jorgeacostaupm.github.io/revisit/matrices_size/SHdPTGI3ZVZxS1EwU1pqOUdsS3V1QT09}, \\
\inlinevis{oma} \url{https://jorgeacostaupm.github.io/revisit/matrices_angle/SHdPTGI3ZVZxS1EwU1pqOUdsS3V1QT09}
\end{minipage}

\vspace{1em}
\begin{subfigure}[b]{\linewidth}
    \includegraphics[width=\linewidth]{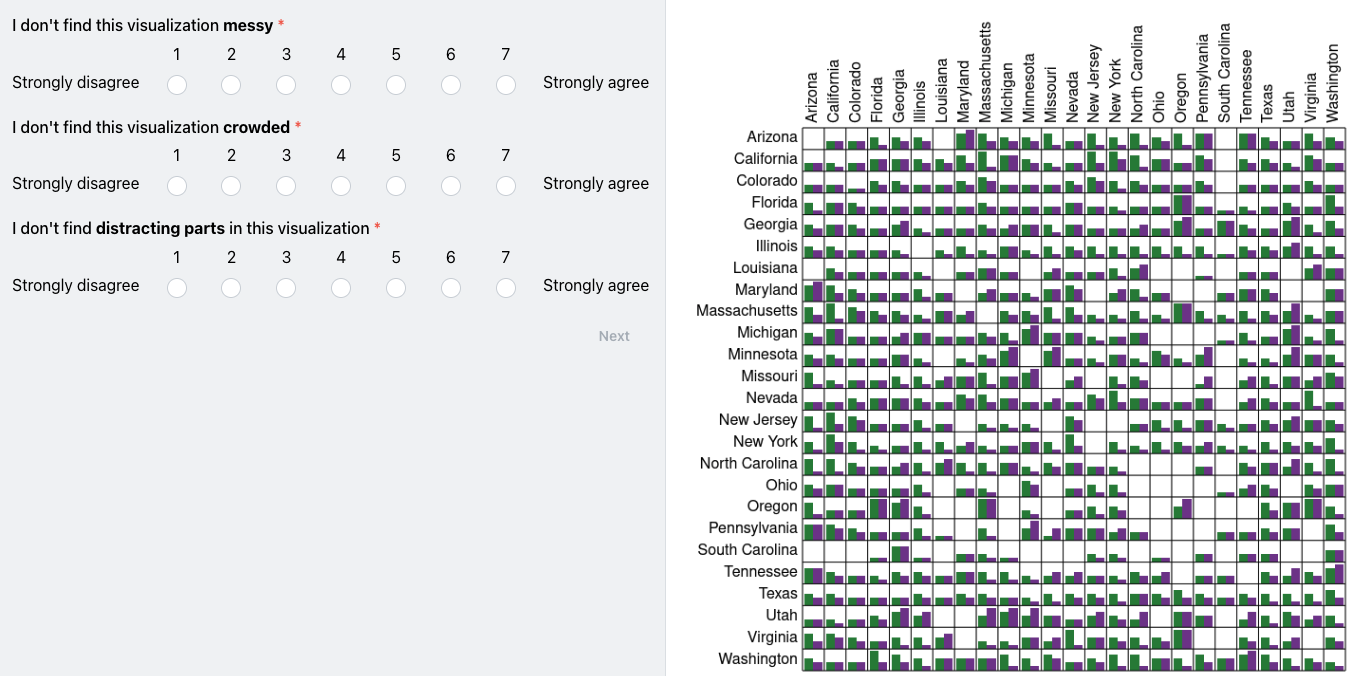}
\end{subfigure}
\vspace{1em}

\begin{subfigure}[b]{0.48\textwidth}
    \centering
        \includegraphics[width=0.48\textwidth]{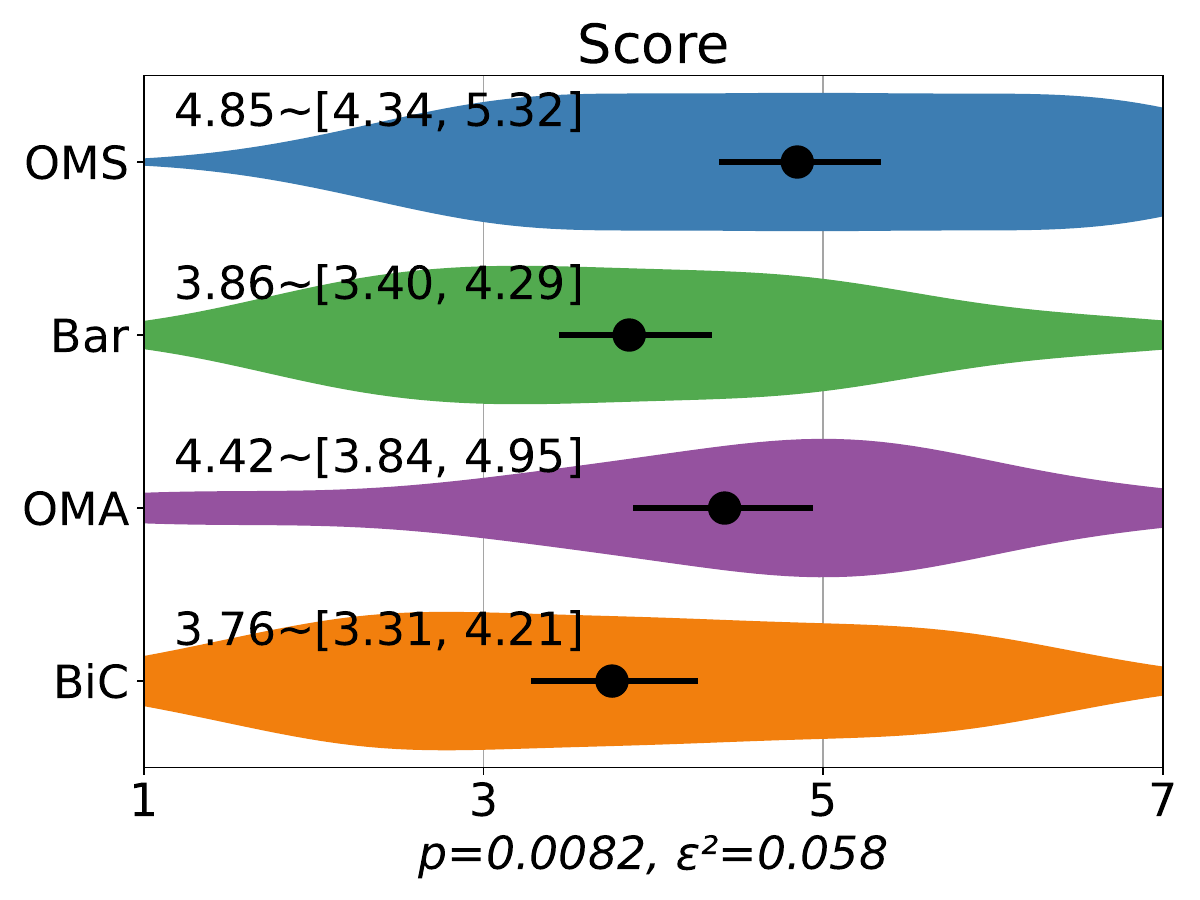}
    \hfill
    \includegraphics[width=0.48\textwidth]{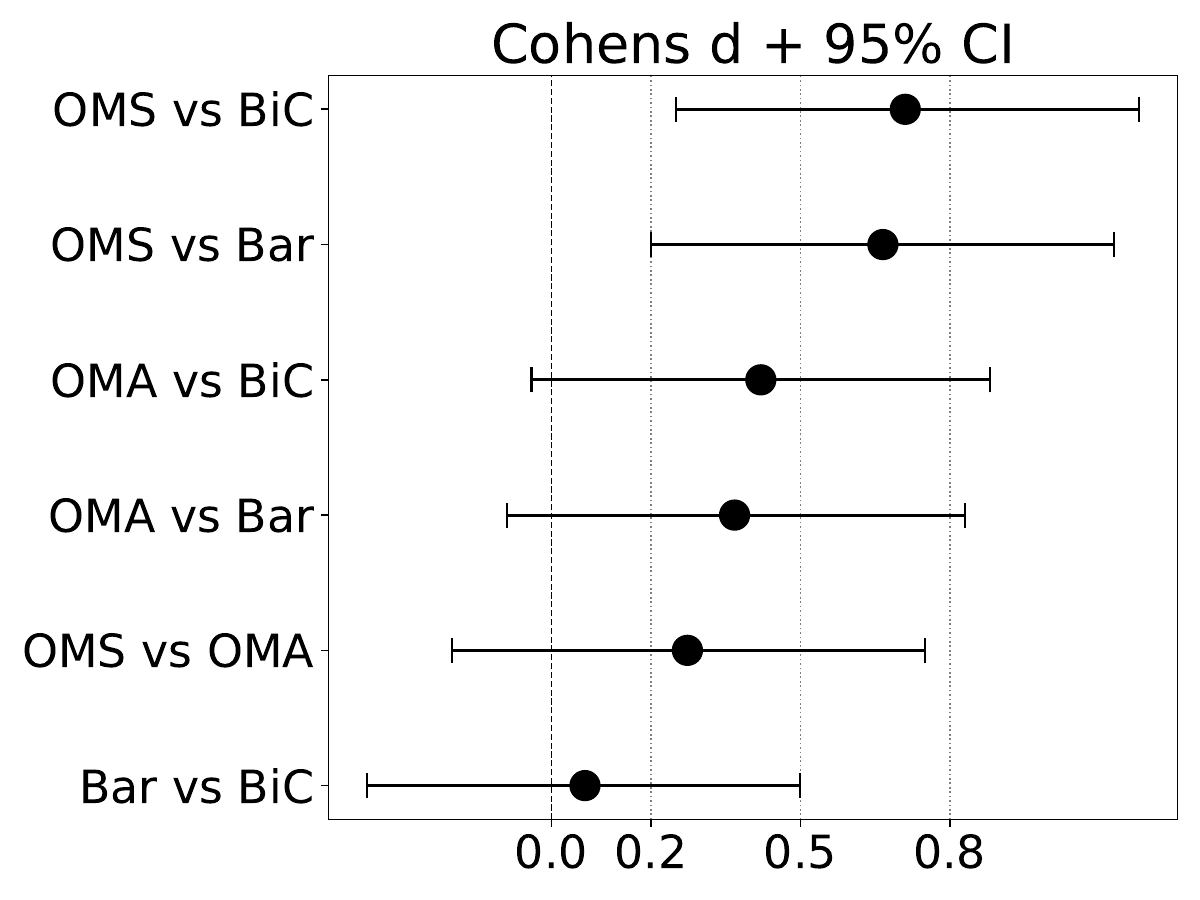}
    \label{fig:readability_layout}
\end{subfigure}

\caption{Comparative evaluation of the Layout Item of the PREVis Scale across the four tested encoding techniques.}
\label{fig:layout_app}
\end{figure*}

\begin{figure*}[hbt]
\centering

\begin{minipage}{\textwidth}
{\fontsize{20}{22}\selectfont\textbf {PREVis - Reading Features}} \label{sec:pre_fea}\\

\textbf{Rationale:} This item measures how easily users extract concrete information or values from the visualization.\\
\textbf{Hypothesis:} H2 \\
\textbf{Scoring:} Likert Scale (1-7) \\
\textbf{Links:} \\
\inlinevis{bc} \url{https://jorgeacostaupm.github.io/revisit/matrices_barchart/M3BGME5HMHhJNXZhblVLTGFGU0hoZz09}, \\
\inlinevis{b} \url{https://jorgeacostaupm.github.io/revisit/matrices_bivariate/M3BGME5HMHhJNXZhblVLTGFGU0hoZz09}, \\
\inlinevis{oms} \url{https://jorgeacostaupm.github.io/revisit/matrices_size/M3BGME5HMHhJNXZhblVLTGFGU0hoZz09}, \\
\inlinevis{oma} \url{https://jorgeacostaupm.github.io/revisit/matrices_angle/M3BGME5HMHhJNXZhblVLTGFGU0hoZz09}
\end{minipage}

\vspace{1em}
\begin{subfigure}[b]{\linewidth}
    \includegraphics[width=\linewidth]{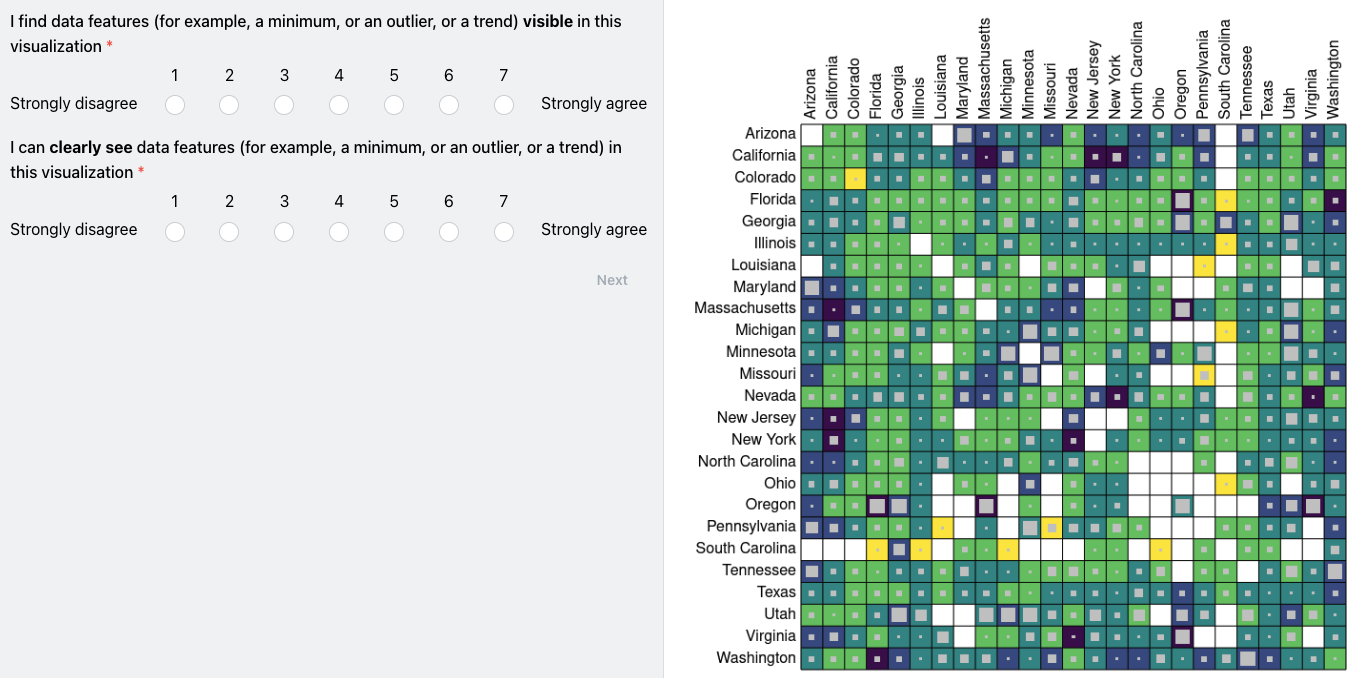}
\end{subfigure}
\vspace{1em}

\begin{subfigure}[b]{0.48\textwidth}
    \centering
        \includegraphics[width=0.48\textwidth]{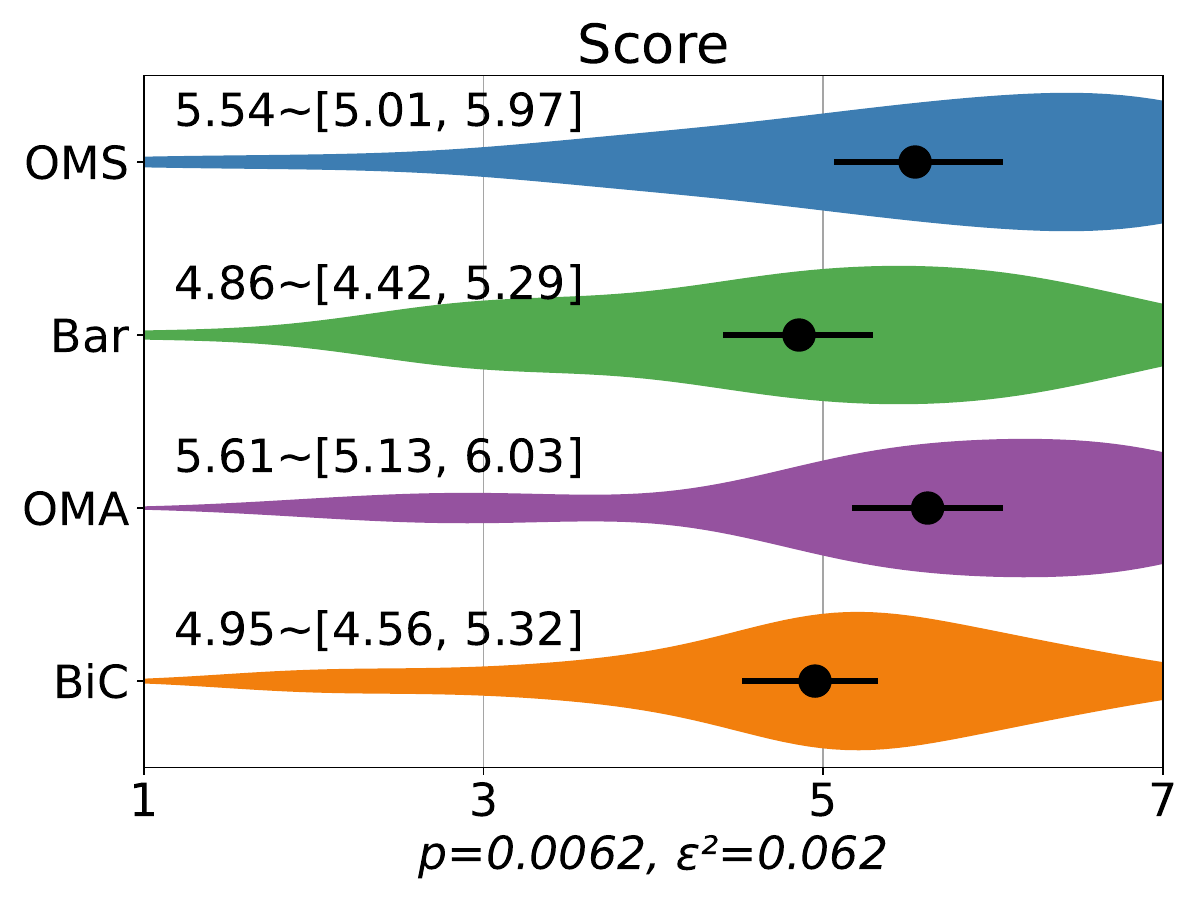}
    \hfill
    \includegraphics[width=0.48\textwidth]{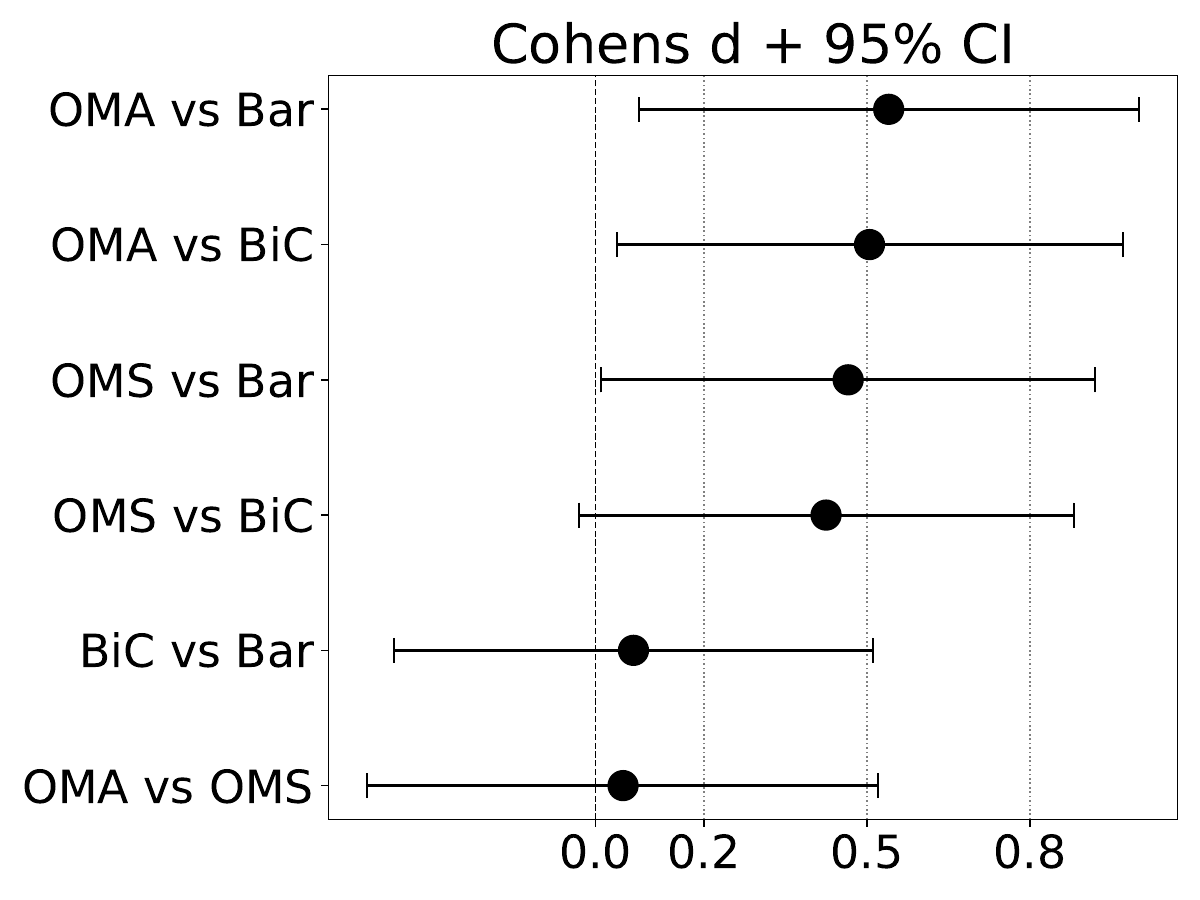}
    \label{fig:readability_features}
\end{subfigure}

\caption{Comparative evaluation of the Reading Features Item of the PREVis Scale across the four tested encoding techniques.}
\label{fig:features_app}
\end{figure*}

\begin{figure*}[hbt]
\centering

\begin{minipage}{\textwidth}
{\fontsize{20}{22}\selectfont\textbf {BeauVis: Aesthetics}} \label{sec:beau}\\

\textbf{Rationale:} This item measures the perceived aesthetics of visualizations, capturing users’ judgments of beauty, enjoy, visual quality, and overall aesthetic appeal.\\
\textbf{Hypothesis:} H3 \\
\textbf{Scoring:} Likert Scale (1 - 7) \\
\textbf{Links:} \\
\inlinevis{bc} \url{https://jorgeacostaupm.github.io/revisit/matrices_barchart/UzJVNzJseTBtdE1LY0x4dHMzNWVSUT09}, \\
\inlinevis{b} \url{https://jorgeacostaupm.github.io/revisit/matrices_bivariate/UzJVNzJseTBtdE1LY0x4dHMzNWVSUT09}, \\
\inlinevis{oms} \url{https://jorgeacostaupm.github.io/revisit/matrices_size/UzJVNzJseTBtdE1LY0x4dHMzNWVSUT09}, \\
\inlinevis{oma} \url{https://jorgeacostaupm.github.io/revisit/matrices_angle/UzJVNzJseTBtdE1LY0x4dHMzNWVSUT09}
\end{minipage}

\vspace{1em}
\begin{subfigure}[b]{\linewidth}
    \includegraphics[width=\linewidth]{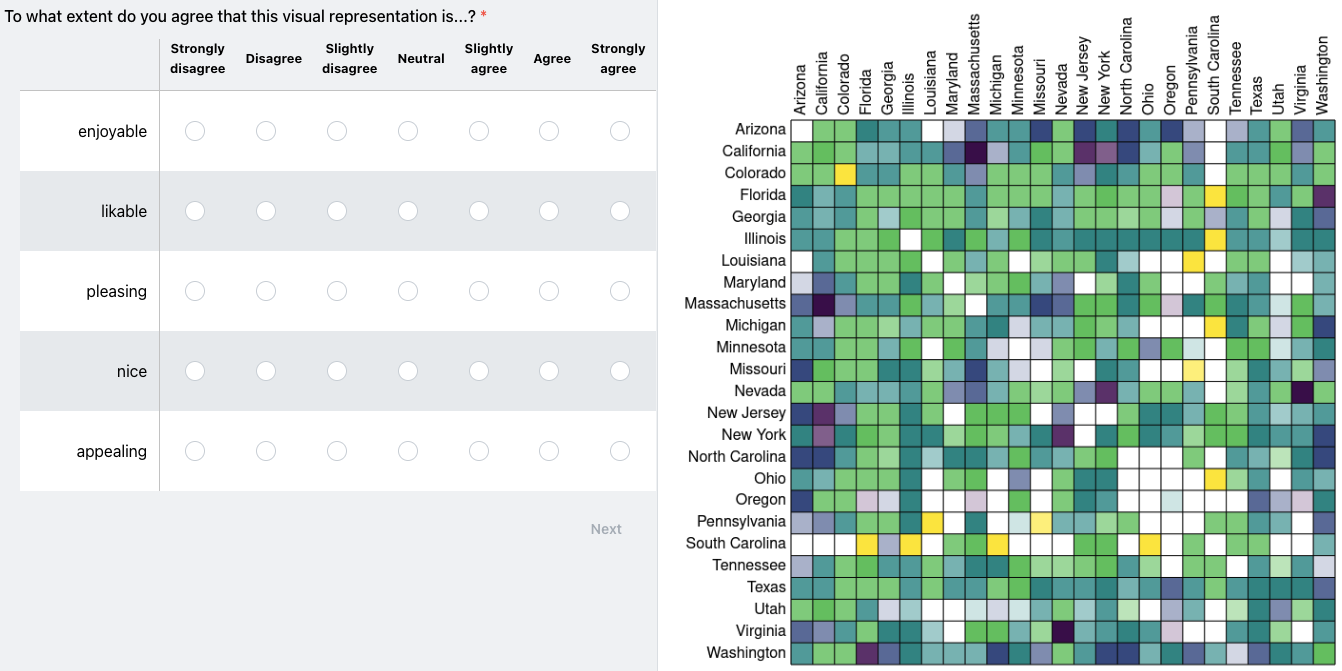}
\end{subfigure}
\vspace{1em}

\begin{subfigure}[b]{0.48\textwidth}
    \centering
        \includegraphics[width=0.48\textwidth]{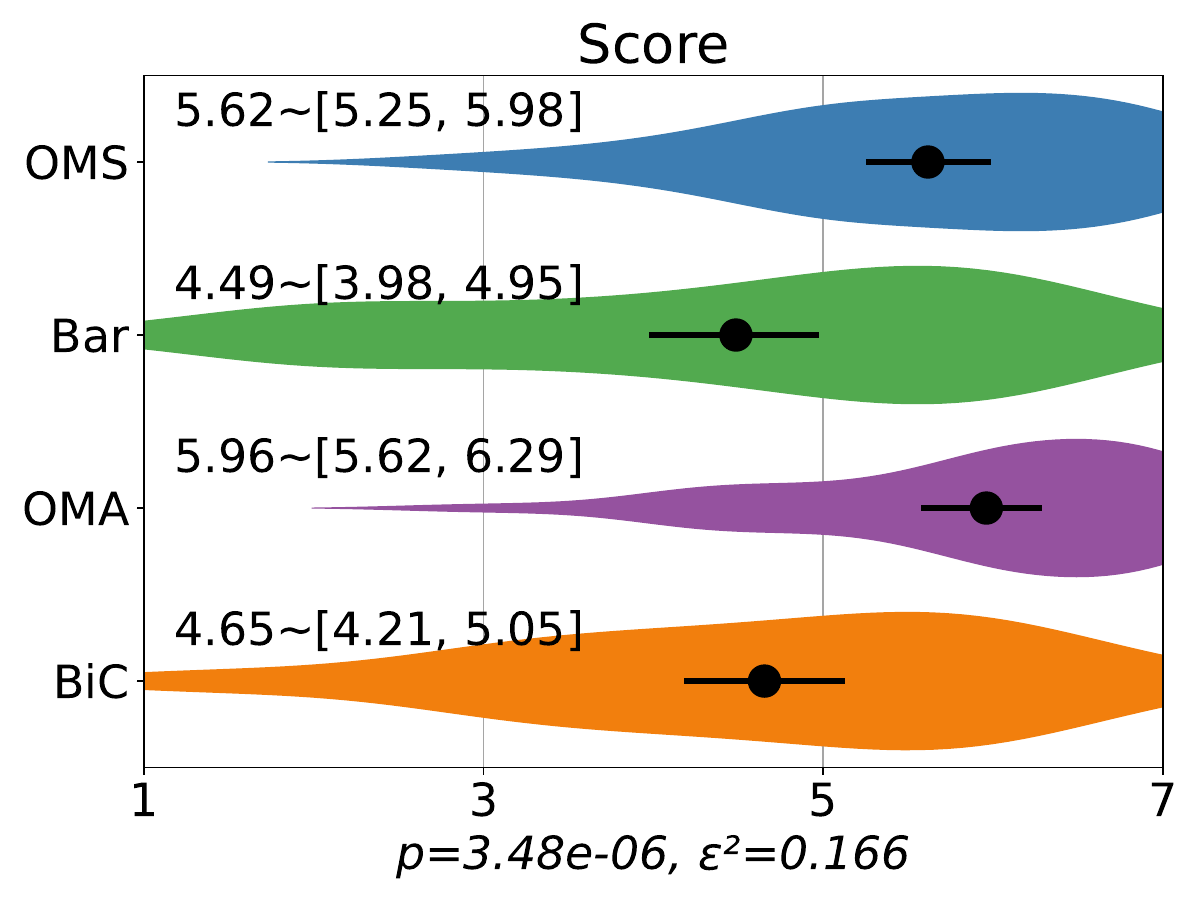}
    \hfill
    \includegraphics[width=0.48\textwidth]{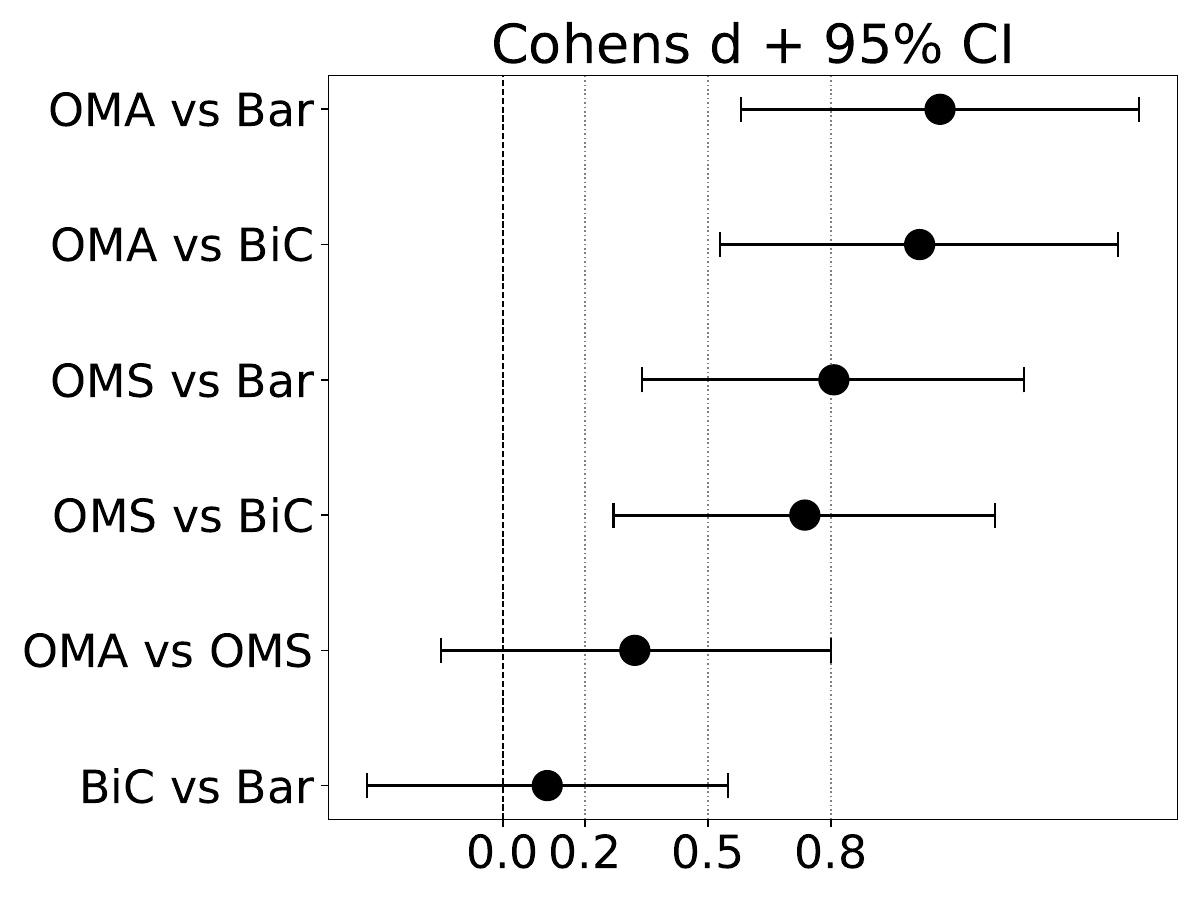}
    \caption{Aggregated BeauVis Scores.}
    \label{fig:readability_1}
\end{subfigure}

\caption{Comparative evaluation of perceived aesthetics score across the four tested encoding techniques.}
\end{figure*}

\fi

\end{document}